\RequirePackage{fix-cm}
\documentclass[aps,prd,twocolumn]{revtex4}

\usepackage{soul}
\RequirePackage{graphicx}
\RequirePackage{amsmath}
\RequirePackage{amssymb}
\RequirePackage{bm}
\RequirePackage{color}
\RequirePackage[pdfencoding=auto, psdextra]{hyperref}
\RequirePackage{verbatim}
\allowdisplaybreaks 
\usepackage{natbib} 
\usepackage{orcidlink} 
\newcommand{\ba}{\begin{eqnarray}}
\newcommand{\ea}{\end{eqnarray}}

\begin{document}
\bibliographystyle{unsrtnat} 

\title{ Radiative decays of the  $\Sigma_c$, $\Xi'_c$ and $\Omega_c$ charmed baryons }

\author{ A. Dávila-Rivera~\orcidlink{0009-0009-7499-5219}}
\affiliation{Facultad de Física, Universidad Autónoma de Zacatecas, Zacatecas, {C-580 98060}, {Mexico} }

\author{H. Garc{\'i}a-Tecocoatzi~\orcidlink{0000-0002-8767-9786}}
\affiliation{Tecnologico de Monterrey, Escuela de Ingenieria y Ciencias, General Ramon Corona 2514,
Zapopan 45138, Mexico}

\author{A. Ramirez-Morales~\orcidlink{0000-0001-8821-5708}}
\affiliation{Tecnologico de Monterrey, Escuela de Ingenieria y Ciencias, General Ramon Corona 2514,
Zapopan 45138, Mexico}

\author{Ailier Rivero-Acosta~\orcidlink{0000-0001-7651-3927}}\email[Corresponding author: ]{ailierrivero@gmail.com} 
\affiliation{Departamento de F\'isica, DCI, Campus Le\'on, Universidad de Guanajuato, Loma del Bosque 103, Lomas del Campestre, 37150, Le\'on, Guanajuato, Mexico}
\affiliation{Dipartimento di Fisica, Universit\`a di Genova, Via Dodecaneso 33, 16146 Genova, Italy}
\affiliation{INFN, Sezione di Genova, Via Dodecaneso 33, 16146 Genova, Italy}
\affiliation{Tecnologico de Monterrey, Escuela de Ingenieria y Ciencias, General Ramon Corona 2514,
Zapopan 45138, Mexico}

\author{E. Santopinto~\orcidlink{0000-0003-3942-6554}}
\affiliation{INFN, Sezione di Genova, Via Dodecaneso 33, 16146 Genova, Italy }

\author{Carlos Alberto Vaquera-Araujo~\orcidlink{0000-0001-8578-9263}} 
\affiliation{Secretar\'ia de Ciencia, Humanidades, Tecnolog\'ia e Innovaci\'on,  Insurgentes Sur 1582. Colonia Cr\'edito Constructor, Benito Ju\'arez, 03940, Ciudad de M\'exico, Mexico}
\affiliation{Departamento de F\'isica, DCI, Campus Le\'on, Universidad de
  Guanajuato, Loma del Bosque 103, Lomas del Campestre 37150, Le\'on, Guanajuato, Mexico}
\affiliation{Dual CP Institute of High Energy Physics, 28045, Colima, Mexico}

\begin{abstract}
In this work, we study the radiative decays of the $\Sigma_c$, $\Xi'_c$ and $\Omega_c$ charmed baryons, which belong to the flavor sextet ($\bf {6}_{\rm F}$), within the constituent quark model formalism. The electromagnetic transitions are calculated from the ground and $P$-wave states to ground states, as well as from the second shell states to both the ground and $P$-wave final states. These decays play a crucial role in confirming the existence of certain resonances when strong decays are not allowed. 
Moreover, electromagnetic decay widths are particularly useful for identifying resonances when states have the same mass and total decay width. 
A relevant case is the $\Omega_c (3327)$ state, whose branching ratios between the strong decay channels are comparable; thus, the radiative decay widths may help assign this state. We also make the assignment of the recently
discovered $\Xi'_c(2923)^{+}$ baryon, which is consistent with being the isospin partner of the $\Xi'_c(2923)^{0}$. This study presents, for the first time, the calculation of electromagnetic decays for $D_\rho$-wave states, $\rho-\lambda$ mixed states, and $\rho$-mode radially excited states in the charm sector. Throughout our calculations, we account for uncertainties arising from both experimental and model-dependent errors. 

\end{abstract}

\maketitle

\section{Introduction }  

In recent years, the spectroscopy of singly heavy baryons has advanced significantly, driven by the discovery of numerous excited states, improved mass and width measurements, and detailed studies of their decay channels~\cite{klempt2010baryon,cheng2022charmed}.  Substantial progress has also been made in the study of open charm and open bottom systems, supported by both spectroscopic models and experimental data~\cite{chen2017review}. Experimental efforts worldwide have played a central role, probing not only the mass spectra but also the production mechanisms in hadronic collisions~\cite{Chu:2016sjc,Wang:2022zgi}. These developments are reflected in the steadily growing number of singly charmed baryons listed in the Particle Data Group (PDG) compilation~\cite{ParticleDataGroup:2024cfk}.


The first evidence for a singly charmed baryon, 
identified as the $\Lambda_{c}^{+}$, was reported in 1975 by the Brookhaven National Laboratory (BNL group) in Ref.~\cite{Cazzoli:1975et}, with a mass of $2426$ MeV. This observation marked the starting point for the experimental study of charmed baryons. In 1976, the Fermilab Collaboration~\cite{Knapp:1976qw} reported the first evidence for the $\Sigma_c^{++}$ baryon with a mass of $2260~\mathrm{MeV}$, with a statistical significance of about 7$\sigma$. In 1980, the BEBC TST Neutrino Collaboration~\cite{BEBCTSTNeutrino:1980ktj} reported the first observation of the $\Sigma_c^+$, with a mass of $2457$ MeV. In 1983, the WA62 Collaboration~\cite{Biagi:1983en} reported the first observation of the ground state of the $\Xi_c$, with a mass of $2460~\mathrm{MeV}$. Two years later, in 1985, the same collaboration~\cite{Biagi:1984mu} provided the first evidence for the $\Omega_c^0$ baryon, with quark content $css$ and a mass of $2740~\mathrm{MeV}$. In 1993, Ref.~\cite{ARGUS:1993vtm} reported the observation of a charmed baryon with a mass of $2626.6$~MeV  ($90\%$ C.L.) in DORIS II at DESY. Theoretical estimates assign this mass to $P$-wave $\Lambda_c(2630)$ or $\Lambda_c(2640)$ states with $J^P = 1/2^-$ or $3/2^-$. In 1995, the CLEO Collaboration~\cite{CLEO:1994oxm} observed the $\Lambda_c(2595)^+$ and $\Lambda_c(2625)^+$ baryons, with masses of 2593 MeV and 2625 MeV, respectively. In 1997, Ref.~\cite{ARGUS:1997snv} studied and confirmed the existence of the excited charmed baryon $\Lambda_c(2595)^{+}$. In 1999, the CLEO Collaboration~\cite{CLEO:1999msf} presented evidence for the production of two new charmed baryon states, the $\Xi_{c}(2815)^{+}$ and $\Xi_{c}(2815)^{0}$ states with $J^P = 3/2^-$, which are charmed-strange analogs of the $\Lambda_{c}(2625)^{+}$. The following year, CLEO observed two new excited $\Lambda_c^+$ states with masses of about $3105~\mathrm{MeV}$ and $3221~\mathrm{MeV}$ (90\% CL)~\cite{CLEO:2000mbh}, and also provided evidence for two new charmed baryons,
the $J^P = 1/2^-$ $\Xi_{c}(2790)^0$ and $\Xi_{c}(2790)^+$ states, interpreted as the charmed-strange analogs of the $\Lambda_c(2595)^+$~\cite{CLEO:2000ibb}. In 2006, the Belle Collaboration~\cite{Belle:2006edu} reported the first observation of two charmed strange baryons $\Xi_c(snc)$, with masses of $2978.5$ MeV, and $3076.7$ MeV, respectively. They also performed a search for the isospin partner states and observed a significant signal at a mass of $3082.8$ MeV.
 
In 2008, the Belle Collaboration~\cite{Belle:2008yxs} reported a precise measurement of the masses of the $\Xi_c(2645)$ and $\Xi_c(2815)$ baryons, confirming previous observations by CLEO of the $\Xi_c(2815)$ baryon.

In 2011, the CDF Collaboration~\cite{CDF:2011zbc} measured the properties of several $\Lambda_c$ and $\Sigma_c$ baryons. They reported the following masses: $2453.90$~MeV for the $\Sigma_c(2455)^{++}$, $2453.74$~MeV for the $\Sigma_c(2455)^{0}$, $2517.19$~MeV for the $\Sigma_c(2520)^{++}$, $2519.34$~MeV for the $\Sigma_c(2520)^{0}$, $2592.25$~MeV for the $\Lambda_c(2595)^{+}$, and $2628.11$~MeV for the $\Lambda_c(2625)^{+}$ (all values at 90\% C.L.).

In 2017, the LHCb Collaboration~\cite{LHCb:2017uwr} reported the discovery of five new narrow excited $\Omega_c^0$ baryon states, namely, the $\Omega_c(3000)^0$ (with statistical significance of  $20.4\sigma$), the $\Omega_c(3050)^0$ ($20.4\sigma$), the $\Omega_c(3066)^0$ ($23.9\sigma$), the $\Omega_c(3090)^0$ ($21.1\sigma$), and the $\Omega_c(3119)^0$ ($10.4\sigma$). Additionally, a broad structure around
3188 MeV was seen. In the same year, the Belle Collaboration~\cite{Belle:2017ext}, observed five resonant states and confirmed four of the states announced by LHCb~\cite{LHCb:2017uwr}, the $\Omega_c(3000)^0$ (significance of 3.9$\sigma$), $\Omega_c(3050)^0$ (4.6$\sigma$), $\Omega_c(3066)^0$ (7.2$\sigma$), and $\Omega_c(3090)^0$ (5.7$\sigma$), while no significant signal was found for the $\Omega_c(3119)^0$. Belle also measured a signal excess at $3188$~MeV was also observed, corresponding to the $\Omega_c(3188)$ state reported by LHCb.

In 2020, the LHCb Collaboration~\cite{LHCb:2020iby} reported the observation of three excited neutral charmed baryons, $\Xi_c^0$. These states are the $\Xi_c(2923)^0$ with a mass of $2923.04$~MeV, the $\Xi_c(2939)^0$ with a mass of $2938.55$~MeV, and the $\Xi_c(2965)^0$ with a mass of $2964.88$~MeV.

The LHCb reported the first observation of the decay $\Omega_b^- \to \Xi_c^+ K^- \pi^-$, confirming four $\Omega_c^0$ states with significances above $5\sigma$~\cite{LHCb:2021ptx}. 
Subsequently, the analysis of the decay $B^- \to \Lambda_c^+ \bar{\Lambda}_c^- K^-$~\cite{LHCb:2022vns} resolved the $\Xi_c(2930)^0$ structure, previously observed by BABAR~\cite{BaBar:2007xtc} and Belle~\cite{Belle:2017jrt}, into two distinct states, $\Xi_c(2923)^0$ and $\Xi_c(2939)^0$, and also provided evidence for a new $\Xi_c(2880)^0$ with a local significance of $3.8\sigma$. 
More recently, LHCb~\cite{LHCb:2023sxp} observed two additional excited states, $\Omega_c(3185)^0$ and $\Omega_c(3327)^0$.

The discovery of new heavy-flavor baryons has further stimulated interest in this field.
In 2025 the LHCb Collaboration~\cite{LHCb:2025mge} observed four $\Xi_c^{'+}$ states with high significance. Their masses and decay widths were measured to be: $m[\Xi'_c(2815)^{+}]=2816.65$ MeV, $\Gamma[\Xi'_c(2815)^{+}]=2.07$ MeV,  $m[\Xi'_c(2923)^{+}]=2922.85$ MeV, $\Gamma[\Xi'_c(2923)^{+}]=5.3$ MeV,  
$m[\Xi'_c(2970)^{+}]=2968.6$ MeV, $\Gamma[\Xi'_c(2970)^{+}]=1.9$ MeV, 
$m[\Xi'_c(3080^{+}]=3076.8$ MeV, $\Gamma[\Xi'_c(3080)^{+}]=6.8$ MeV. The $\Xi'_c(2923)^{+}$ baryon was
discovered, being consistent with being the isospin partner of the previously observed $\Xi'_c(2923)^{0}$, and the first observation of the $\Xi'_c(3080)^{+}$ baryon was reported.



The spectroscopy of charmed baryons has been widely investigated in different theoretical frameworks. Early works employed QCD sum rules~\cite{Groote:1996em}, bootstrap quark models~\cite{Gerasyuta:1999pc}, and heavy quark symmetry corrections to $P$-wave states~\cite{Wang:2003zp}. Predictions for $\Xi_c$ and $\Xi_b$ masses were given in Ref.~\cite{Duraes:2007te}, while lattice QCD and chiral perturbation theory was used to compute spectra of singly, doubly, and triply charmed baryons~\cite{Brown:2014ena,Padmanath:2013bla}. $P$-wave excitations, including $\rho$- and $\lambda$-modes, were analyzed in Ref.~\cite{Chen:2015kpa}, and several studies assigned quantum numbers to newly observed $\Omega_c^0$ states~\cite{Padmanath:2017lng,Cheng:2017ove,Mao:2017wbz,Wang:2017vnc}. More recent lattice simulations and relativized quark models refined predictions for ground and excited states~\cite{Bahtiyar:2020uuj,Yang:2021lce,Pan:2023hwt,Yu:2022ymb,Wang:2023wii}, including mixing effects and unobserved resonances. In particular, Ref.~\cite{Li:2022xtj} studied strange single heavy baryons and provided preliminary quantum number assignments to recently observed $\Xi_c$ states such as $\Xi_{c}(3055)$, $\Xi_{c}(3080)$, $\Xi_{c}(2930)$, and others.

Strong decays of charmed baryons were investigated using Heavy Hadron Chiral Perturbation Theory~\cite{Pirjol:1997nh}, constituent and light-front quark models~\cite{Chiladze:1997ev,Tawfiq:1998nk,Ivanov:1998qe}, and nonrelativistic quark models with heavy-quark symmetry wavefunctions~\cite{Albertus:2005zy}. Decay patterns of $p$-wave states were studied using combined heavy quark and chiral symmetries~\cite{Cheng:2006dk}, while Bethe–Salpeter approaches were applied to $\Sigma_Q^{(*)} \to \Lambda_Q \pi$~\cite{Guo:2007qu}. More recently, the $^3P_0$ model was extensively used for excited $\Xi_c$ and $\Omega_c$ baryons~\cite{Ye:2017yvl,Zhao:2017fov}, with further analyses of $\rho$- and $\lambda$-mode mixing~\cite{Chen:2017sci}, and classification of higher excitations including $1D$- and $1F$-wave states~\cite{Luo:2023sne,Luo:2023sra}. Additional studies addressed the structure and decays of $2P$-wave states such as $\Lambda_c(2940)^+$ and $\Xi_c(3123)^+$~\cite{Yang:2023fsc,Yu:2023bxn}, with recent extensions to bottom baryons in the $\bar{3}_F$ representation~\cite{Wang:2024rai}.

Several interpretations of excited charmed baryons have been proposed. Molecular pictures have been suggested for $\Lambda_c(2593)^+$~\cite{Blechman:2003mq} and $\Lambda_c(2940)^+$~\cite{Ortega:2012cx}, while positive-parity assignments were proposed for $\Xi_c(2980)$ and $\Xi_c(3080)$~\cite{Cheng:2015naa}. Following the LHCb observation of five $\Omega_c^0$ states, quark potential and QCD sum rule analyses~\cite{Chen:2017gnu,Wang:2017zjw,Agaev:2017lip,Agaev:2017ywp} proposed spin–parity assignments as $1P$ negative-parity or radial excitations. Other works studied $\Xi_c(2930)$~\cite{Aliev:2018ube}, $\Xi_c(2923)^0$, $\Xi_c(2939)^0$, and $\Xi_c(2965)^0$ as $\lambda$-mode excitations~\cite{Lu:2020ivo}, and identified systematic patterns among $\bar{3}_F$ and $6_F$ states~\cite{Chen:2021eyk}. More recent studies applied unquenched quark models to $\Lambda_c(2910)^+$ and $\Lambda_c(2940)^+$~\cite{Zhang:2022pxc}, and discussed new excited $\Omega_c^0$ states at 3185 and 3327 MeV~\cite{Karliner:2023okv}.


The study of electromagnetic decays (EMDs) has steadily advanced over the years, expanding our knowledge of hadrons.  In 1999, the CLEO collaboration \cite{CLEO:1998wvk}, observed and studied the EMDs $\Xi^{'+}_c \rightarrow \Xi^{+}_c \gamma$ and $\Xi^{'0}_c \rightarrow \Xi^{0}_c \gamma$. These experimental results were compared with previous predictions (based on the quark model and QCD), marking a significant advance in the study of charmed baryons, as they represented the first experimental observation of radiative decays in excited charmed baryons.

In 2006, the BABAR collaboration \cite{BaBar:2006pve} measured the EMD $\Omega^{*}_c \rightarrow \Omega^{0}_c \gamma$, the first of its kind, with a statistical significance of 5.2$\sigma$. 
Subsequently, in 2008, the Belle collaboration \cite{Solovieva:2008fw} confirmed the observation of the decay $\Omega^{*0}_c \rightarrow \Omega^{0}_c \gamma$, where the excited state has a configuration $J^{P}=\frac{3}{2}^{+}$, and achieving a statistical significance of 6.4$\sigma$ for this decay.
In 2020, the Belle Collaboration~\cite{Belle:2020ozq} reported the first observations of EMDs from P-wave $\Xi_c$ states, namely $\Xi_c(2790)^{0} \rightarrow \Xi^{0}_c \gamma$ and $\Xi_c(2815)^{0} \rightarrow \Xi^{0}_c \gamma$, with statistical significances of $3.8\sigma$ and $8.6\sigma$, respectively.


Not only have experimental advances been achieved, but theoretical predictions of EMDs have also been made, allowing experimental studies to focus on specific decays. Some examples of the literature on theoretical investigations about EMDs of single
heavy baryons are given by the Refs.~\cite{Cheng:1992xi,Banuls:1999br,Jiang:2015xqa,Wang:2018cre,Zhu:1998ih,Wang:2010xfj,Wang:2009cd,Aliev:2009jt,Aliev:2011bm,Aliev:2014bma,Aliev:2016xvq,Bernotas:2013eia,Shah:2016nxi,Gandhi:2019xfw,Gandhi:2019bju,Yang:2019tst,Kim:2021xpp,Chow:1995nw,Gamermann:2010ga,Zhu:2000py,Luo:2025jpn,Luo:2025pzb,Bijker:2020tns,Ortiz-Pacheco:2023kjn,Garcia-Tecocoatzi:2025fxp,Wang:2017hej,Wang:2017kfr,Yao:2018jmc,Peng:2024pyl,Ivanov:1998wj,Ivanov:1999bk,Tawfiq:1999cf}. In most of these works, the analysis was performed only for ground states \cite{Cheng:1992xi,Banuls:1999br,Jiang:2015xqa,Wang:2018cre,Zhu:1998ih,Wang:2010xfj,Wang:2009cd,Aliev:2009jt,Aliev:2011bm,Aliev:2014bma,Aliev:2016xvq,Bernotas:2013eia,Shah:2016nxi,Gandhi:2019xfw,Gandhi:2019bju,Yang:2019tst,Kim:2021xpp}, only $P$-wave \cite{Chow:1995nw,Gamermann:2010ga,Zhu:2000py,Luo:2025jpn,Luo:2025pzb,Bijker:2020tns,Ortiz-Pacheco:2023kjn,Garcia-Tecocoatzi:2025fxp}, and for both ground and $P$-wave states \cite{Wang:2017hej,Wang:2017kfr,Yao:2018jmc,Peng:2024pyl,Ivanov:1998wj,Ivanov:1999bk,Tawfiq:1999cf} using different techniques.
In particular, chiral perturbation theory ($\chi$PT) was applied to the study of the EMDs of singly charmed baryons in \cite{Cheng:1992xi,Banuls:1999br,Jiang:2015xqa,Wang:2018cre}. Specifically, transitions from ground states to ground states were considered. 
Light-Cone QCD (LC-QCD) was used to study EMDs of ground states in \cite{Zhu:1998ih,Wang:2010xfj,Wang:2009cd,Aliev:2009jt,Aliev:2011bm,Aliev:2014bma,Aliev:2016xvq}. In \cite{Zhu:1998ih}, the authors studied the decays of the ground states corresponding to $\Sigma_{c(b)}$ baryons, and in \cite{Wang:2010xfj} the decay channels $\Xi_{Q}^{*} \to \Xi'_{Q} \gamma$ and $\Sigma_{Q}^{*} \to \Sigma_{Q} \gamma$, with $Q = b$ or $c$ were studied. Similarly, in Ref. \cite{Wang:2009cd}, the decay channel $\Omega_{Q}^{*} \to \Omega_{Q} \gamma$ was considered, and Ref. \cite{Aliev:2009jt} was devoted to the EMD of the ground sextet heavy-flavored spin-$3/2$ baryons to heavy-spin-$1/2$ baryons. The EMDs of ground-state spin-$3/2$ sextet heavy baryons to both sextet and anti-triplet heavy spin-$1/2$ baryons were calculated in \cite{Aliev:2011bm}. The decay channels $\Omega_{Q}^{*} \to \Omega_{Q} \gamma$ and $\Xi_{Q}^{*} \to \Xi'_{Q} \gamma$ were addressed in \cite{Aliev:2014bma}, and the EMDs of $\Sigma_{Q} \to \Lambda_{Q} \gamma$ and $\Xi'_{Q} \to \Xi_{Q} \gamma$ were investigated in \cite{Aliev:2016xvq}.

In \cite{Bernotas:2013eia}, the EMDs of the ground-state heavy baryons were calculated within the framework of a modified bag model.
The hypercentral constituent quark model (hCQM) was employed in Refs.~\cite{Shah:2016nxi,Gandhi:2019xfw,Gandhi:2019bju} to study the EMDs of singly charmed baryons. In 2016, Ref.~\cite{Shah:2016nxi} reported calculations of the EMD widths of $\Lambda_c^+$, $\Sigma_c^0$, $\Xi_c^0$, and $\Omega_c^0$. In 2019, the EMDs of the ground states of non-strange singly charmed baryons were investigated in Ref.~\cite{Gandhi:2019xfw}, while those of strange singly charmed baryons were analyzed in Ref.~\cite{Gandhi:2019bju}.

The chiral quark-soliton model ($\chi$QSM) was employed in~\cite{Yang:2019tst,Kim:2021xpp} to study EMDs of singly
charmed baryons. In 2019, Ref~\cite{Yang:2019tst} studied the transition magnetic moments and EMDs of the lowest-lying singly heavy baryon sextet. In 2021, in Ref~\cite{Kim:2021xpp} the authors investigated the EMDs of singly charmed baryons with spin $3/2$.

In 1995, in Ref.~\cite{Chow:1995nw}, the authors only studied the EMDs of $P$-wave excited $\Lambda_Q$ baryons in the chiral soliton model. The Coupled-Channel Approach (CCA)~\cite{Gamermann:2010ga} has been utilized to investigate the EMDs of singly charmed \(P\)-wave states, \(\Sigma_c(2792)^{\pm,0}\), \(\Xi_c(2790)^{+/0}\), \(\Xi_c(2970)^{+/0}\), and \(\Lambda_c(2595)\). LC-QCD was used to study EMDs of $P$-wave states in~\cite{Zhu:2000py,Luo:2025jpn,Luo:2025pzb}. In 2000, the EMDs of $P$-wave $\Lambda_{c}$ baryons into $\Sigma_c \gamma$ and $\Sigma_c^* \gamma$ channels were analyzed~\cite{Zhu:2000py}. In 2025, the EMDs of the $P$-wave charmed baryons baryons belonging to the flavor $\mathbf{\bar{3}}_{\rm F}$ representation were studied in Ref.~\cite{Luo:2025jpn}, while those of the $P$-wave charmed baryons in the $\mathbf{6}_{\rm F}$ representation were investigated in Ref.~\cite{Luo:2025pzb}.

Another approach used to study the EMDs of singly heavy baryons is the Constituent Quark Model (CQM)~\cite{Bijker:2020tns,Ortiz-Pacheco:2023kjn,Garcia-Tecocoatzi:2025fxp,Wang:2017hej,Wang:2017kfr,Yao:2018jmc,Peng:2024pyl}. In 2020, in Ref.~\cite{Bijker:2020tns}, the authors investigated the EMDs from $P$-wave to ground states of the $\Xi_{c/b}$ and $\Xi'_{c/b}$ baryons. In 2023, in Ref.~\cite{Ortiz-Pacheco:2023kjn} the EMDs from $P$-wave to ground states of singly heavy baryons were considered. In 2025, in Ref.~\cite{Garcia-Tecocoatzi:2025fxp} the authors analyzed the EMDs of the $\Xi_c(2790)^{+/0}$ and $\Xi_c(2815)^{+/0}$ baryons. They studied $P$-wave to ground state transitions using a non-relativistic Hamiltonian and ladder operators to evaluate the transition amplitudes.
The EMD of the low-lying $S$- and $P$-wave $\Lambda_{c(b)}$, $\Xi_{c(b)}$, $\Sigma_{c(b)}$, $\Xi'_{c(b)}$, and $\Omega_{c(b)}$ baryons were studied in \cite{Wang:2017hej,Wang:2017kfr}, where the Close-Copley replacement~\cite{Close:1970kt} was used to evaluate the transition amplitudes. This replacement, described in more detail in Refs.~\cite{Garcia-Tecocoatzi:2025fxp,Rivero-Acosta:2025drn}, relies on dimensional analysis to simplify the evaluation of the convective term of the non-relativistic electromagnetic interaction Hamiltonian. 

In 1999, the electromagnetic transitions between heavy baryon states using a relativistic three-quark model (RQM) were presented in \cite{Ivanov:1998wj,Ivanov:1999bk}, where the authors considered ground-state to ground-state transitions and some $P$-wave excited state to ground-state transitions.

Heavy quark symmetry was implemented in \cite{Tawfiq:1999cf} to analyze the EMD of the $S$-wave and $P$-wave singly heavy baryon states.

Up to now, there are only two works in the literature where a subset of the second shell charmed baryon states were studied using a CQM~\cite{Yao:2018jmc,Peng:2024pyl}. In both articles, the Close-Copley replacement~\cite{Close:1970kt} was implemented to evaluate the transition amplitudes. The study in \cite{Yao:2018jmc} focused on the EMDs of low-lying $D$-wave baryons, considering transitions from $D$-wave $\lambda$-mode excitations to $P$-wave $\lambda$-mode states of the $\Lambda_{c(b)}$, $\Xi_{c(b)}$, $\Sigma_{c(b)}$, $\Xi'_{c(b)}$, and $\Omega_{c(b)}$ baryons. 
In Ref.~\cite{Peng:2024pyl}, the EMDs of singly heavy baryons from both $D_{\lambda}$-wave and $\lambda$-mode radial excitations to ground and $P_{\lambda}$-wave final states were calculated.

In this study, we calculate the EMD widths of the $\Sigma_c$ with isospin 1, $\Xi'_c$ with isospin $1/2$, and $\Omega_c$ with isospin 0 charmed baryons, which belong to the flavor $\bf {6}_{\rm F}$-plet, within the constituent quark model formalism~\cite{Garcia-Tecocoatzi:2023btk,Garcia-Tecocoatzi:2025fxp,Rivero-Acosta:2025drn}. We evaluate the transition amplitudes analytically without introducing any further approximation to the nonrelativistic electromagnetic interaction Hamiltonian. We consider transitions from ground and $P$-wave states to ground states, as well as from all second-shell states to both ground and $P$-wave final states. This work presents the first calculation of the EMD of $D_\rho$-wave states, $\rho-\lambda$ mixed states, and $\rho-$ radially excited states for singly charmed baryons.
The paper is organized as follows: In Section~\ref{EMdecaywidths}, we briefly
present the quark model used to determine the mass spectra
and outline the formalism used to calculate the EMD widths. Section~\ref{Results} presents our results along with a discussion. Finally, our conclusions are summarized in Section~\ref{Conclusions}.

\section{Electromagnetic decay widths} 
\label{EMdecaywidths}

\begin{figure}[h]
    \centering
    \includegraphics[width=0.45\textwidth]{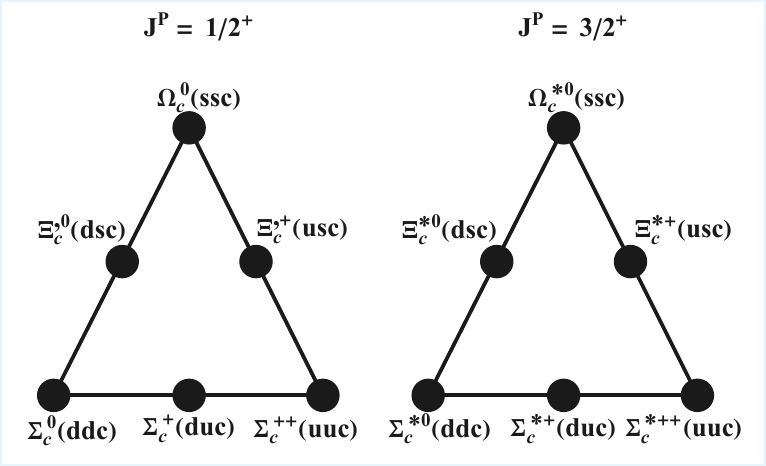}
	\caption{ The $SU_f(3)$ flavor sextet of the ground-state singly charmed baryons:  the states with ${\bf J}^P = {\bf \frac{1}{2}}^+$  are at the left side, and those with ${\bf J}^P = {\bf \frac{3}{2}}^+$   at the right side. }    
    \label{fig:sextet}
\end{figure}

In this section, we present the methodology used to calculate the electromagnetic decay widths of singly charmed baryons, specifically the $\Sigma_c$, $\Xi'_c$, and $\Omega_c$ states, belonging to the flavor sextet (see Fig.~\ref{fig:sextet}). 
We follow the formalism developed in Refs.~\cite{Garcia-Tecocoatzi:2023btk,Garcia-Tecocoatzi:2025fxp,Rivero-Acosta:2025drn,RiveroAcosta:2025nfh}, which focuses on the emission of left-handed photons in radiative transitions of the form $A \rightarrow A'\gamma$, where  $A$ and $A'$ denote the initial and final singly charmed baryons, respectively.


\subsection{Mass spectra} 
\label{Masses}

\begin{figure*}
\centering \includegraphics[scale=0.5] {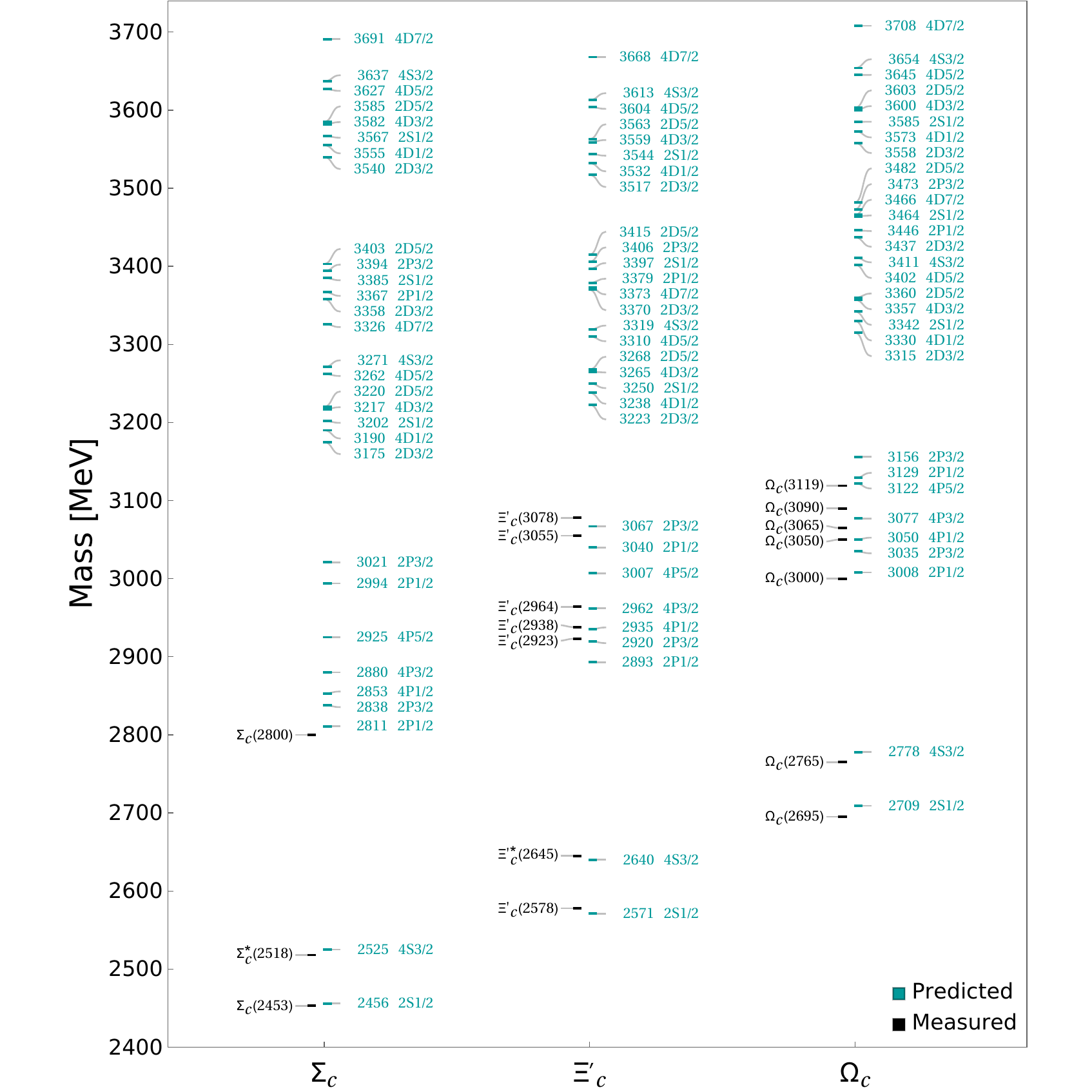}
\caption{ Comparison between the predicted masses of singly charmed baryon belonging to the ${\bf 6}_{\rm F}$-plet, as from Ref.~\cite{Garcia-Tecocoatzi:2022zrf}, 
with the experimental data from the PDG \cite{ParticleDataGroup:2024cfk}. The predicted masses are displayed in teal, while the experimental masses are reported in black \cite{ParticleDataGroup:2024cfk}. }
\label{fig:all-states}
\end{figure*}

In this work, we use the masses and assignments of singly charmed baryons obtained in Ref.~\cite{Garcia-Tecocoatzi:2022zrf}. The theoretical masses reproduce the trend observed in the available data reported in PDG \cite{ParticleDataGroup:2024cfk}, as shown on Fig.~\ref{fig:all-states}, where we present the spectra of singly charmed baryons computed using the three-quark model. These masses were calculated as the eigenvalues of the Hamiltonian introduced in Refs. \cite{Santopinto:2018ljf,Giachino:2020dsj}, which is given by 

\begin{eqnarray}
	H &=& H_{\rm h.o.}+P_{\rm S}\; {\bf S}^2_{\rm tot}
 + P_{\rm SL} \; {\bf S}_{\rm tot} \cdot {\bf L}_{\rm tot}+P_{\rm I}  \;  \bm{{\rm I}}^2+P_{\rm F}\; {\bf \hat{C}}_2,
 \nonumber
 \\
	\label{MassFormula}
\end{eqnarray}
where $H_{\rm h.o.}$ corresponds to the sum of the constituent masses and the harmonic oscillator Hamiltonian.
The symbols ${\bf S}_{\rm tot}, {\bf L}_{\rm tot}, \bm{{\rm I}}$ and  ${\bf \hat{C}}_2$ denote the spin, orbital angular momentum, isospin, and the $SU_f(3)$ Casimir operators, respectively, and they are weighted with the model parameters 
$P_{\rm S}, P_{\rm SL}, P_{\rm I}$, and $P_{\rm F}$ which are fitted to experimental data.

In the three-quark scheme, described in more detail in Refs.~\cite{Santopinto:2018ljf,Garcia-Tecocoatzi:2022zrf,Garcia-Tecocoatzi:2023btk,Giachino:2020dsj}, the  $H_{\rm h.o.}$ term  of Eq.  \ref{MassFormula} can be expressed using  the Jacobi coordinates 
$\boldsymbol{\rho} = (\boldsymbol{r}_1-\boldsymbol{r}_2)/\sqrt{2}$ and $\boldsymbol{\lambda}=(\boldsymbol{r}_1+\boldsymbol{r}_2-2\boldsymbol{r}_3)/\sqrt{6}$  and their conjugate momenta $\mathbf{p}_{\rho}$ and $ \mathbf{p}_{\lambda}$ in the following way:
\begin{eqnarray}
 H_{\rm h.o.}^{3q} =\sum_{i=1}^3m_i + \frac{\mathbf{p}_{\rho}^2}{2 m_{\rho}} 
+ \frac{\mathbf{p}_{\lambda}^2}{2 m_{\lambda}} 
+\frac{1}{2} m_{\rho} \omega^2_{\rho} \boldsymbol{\rho}^2   
+\frac{1}{2}  m_{\lambda} \omega^2_{\lambda} \boldsymbol{\lambda}^2 ,
	\nonumber \\
\label{eq:Hho}
\end{eqnarray}
where $m_{i}$ with $i=1,2$ are the light quark masses, $m_3$ is the charmed quark mass;
$m_\rho=(m_1+m_2)/2$, and $m_\lambda=3m_\rho m_3/(2m_\rho+m_3)$. The $\rho$- and  $\lambda$-oscillator frequencies are $\omega_{\rho(\lambda)}=\sqrt{\frac{3K_b}{m_{\rho(\lambda)}}}$, where $K_b$ is the harmonic oscillator constant. The quark masses and $K_b$ are model parameters fitted to experimental data (see Table I of ~\cite{Garcia-Tecocoatzi:2023btk} for the values used in this work).  Here the $\boldsymbol \rho$ coordinate describes the excitations within the light quark pair, while the $\boldsymbol \lambda$ coordinate describes the excitations between the light quark pair and the charm quark $c$. 

The eigenvalues of the Hamiltonian \ref{MassFormula}, proposed in Ref. \cite{Santopinto:2018ljf}, are given by the expression
\begin{eqnarray}
E^{3q}  &=& \sum_{i=1}^3m_i +  \omega_{\rho} n_{\rho} 
+ \omega_{\lambda} n_{\lambda} + P_{\rm S} \left[ S_{\rm tot}(S_{\rm tot}+1) \right]
\nonumber\\
&& + P_{\rm SL} \frac{1}{2} \Big[ J_{}(J_{}+1) - L_{\rm tot}(L_{\rm tot}+1) \nonumber\\
&& - S_{\rm tot}(S_{\rm tot}+1) \Big] 
+P_{\rm I}\left[ I(I+1)  \right]\nonumber \\
&& + P_{\rm F}\frac{1}{3} \left[ p(p+3)+q(q+3)+pq \right] ,
\label{MassFormula2}
\end{eqnarray} 
which is the formula used to obtain the mass spectrum of the singly charmed baryons used in the present work. Here, $ n_{\rho(\lambda)}= 2 k_{\rho(\lambda)}+l_{\rho(\lambda)}$ represents the quantum numbers of the harmonic oscillator, $k_{\rho(\lambda)}=0,1,...$ is the number of nodes (radial excitations) in the $\rho$($\lambda$) oscillators, $l_{\rho(\lambda)}=0,1,...$ is the orbital angular momentum of the $\rho$($\lambda$) oscillator. $J$ is the total angular momentum and $(p,q)$ are the Dynkin labels corresponding to the $SU(3)$ flavor representations, where the 
${\bf 6}_{\rm F}$-plet is characterized by $(p,q)=(2,0)$. In this work, we use $N=n_\rho+n_\lambda$, where the energy band $N=2$ corresponds second shell singly charmed baryons.

The singly charmed baryon $A$ can be described formally with the state: 
 \begin{eqnarray}
|\theta_c,\phi_A, k_A, J_A,{M_{J_A}}\rangle &
=& 
 | \theta_c \rangle \otimes |\phi_A, k_A ,J_A,{M_{J_A}}\rangle , \label{eq:statescomp} 
\end{eqnarray}
\noindent
where $| \theta_c \rangle =\frac{1}{\sqrt{6}}(|rgb \rangle  - | rbg \rangle + | gbr \rangle - | grb \rangle + | brg \rangle - | bgr \rangle)$ is the $SU_c(3)$ color singlet. $|\phi_A, k_A ,J_A,{M_{J_A}}\rangle$ represents the description of the baryon A, considering the flavor, spin, and spatial degrees of freedom, and is given by 

\begin{eqnarray}
|\phi_A, k_A ,J_A,{M_{J_A}}\rangle &
=& 
|\phi_A \rangle \otimes \sum_{M_{L},M_{S}} \langle L, M_L;S, M_S|J_A,{M_{J_A}}\rangle \nonumber \\
& \times &  \sum_{m_{l_\lambda},m_{l_\rho}} \langle l_\lambda, m_{l_\lambda}; l_\rho, m_{l_\rho}|L,M_L\rangle \nonumber \\
& \times &  \sum_{m_{S_{12}},m_{S_3}} \langle S_{12},m_{S_{12}};S_3,m_{S_3}|S,M_S\rangle \nonumber \\
& \times & \sum_{m_{S_1},m_{S_2}}  \langle S_1, m_{S_1};S_2,m_{S_2} |S_{12},m_{S_{12}} \rangle \nonumber\\
& \times & |S_1,m_{S_1}\rangle \otimes | S_2,m_{S_2}\rangle \otimes | S_3,m_{S_3}\rangle\nonumber  \\
& \otimes & |k_\rho, l_\rho, m_{l_\rho}, k_\lambda, l_\lambda, m_{l_\lambda} \rangle . \label{eq:states}  
\end{eqnarray}
\noindent
Here, $\phi_A$ stands for the flavor wave function, $k_A = k_\rho + k_\lambda$ is the total number of nodes, $J_A$ represents the total angular momentum, and $M_{J_A}$ is the total angular momentum projection. $|S_i, m_{S_i}\rangle$ denote the spin wave function of each quark ($i=1,2,3$), and $|k_\rho, l_\rho, m_{l_\rho}, k_\lambda, l_\lambda, m_{l_\lambda} \rangle$ is the harmonic-oscillator spatial baryon wave function, that can be expressed in terms of $\omega_{\rho}$ and $\omega_{\lambda}$ through the relations $\alpha^2_{\rho,\lambda}=\omega_{\rho,\lambda}m_{\rho,\lambda}$. The explicit form of the spatial baryon wave function in the momentum representation is given by 
\mbox{$\psi_{k_\rho,l_\rho,m_{l_\rho},k_\lambda,l_\lambda,m_{l_\lambda}}(\vec{p}_{\rho} ,\vec{p}_\lambda)=\langle \vec{p}_{\rho} ,\vec{p}_\lambda |k_\rho,l_\rho,m_{l_\rho},k_\lambda,l_\lambda,m_{l_\lambda}\rangle$.}

\subsection{Electromagnetic decay widths} 
\label{EMDecThe}

The electromagnetic interaction between photons and quarks is described at tree level by the Hamiltonian
\begin{equation}
H = -\sum_j e_j \, \bar{q}_j \gamma^\mu A_\mu q_j,
\end{equation}
where $e_j$, $q_j$, and $A_\mu$ are the charge, quark field, and electromagnetic field, respectively. Taking the nonrelativistic limit and keeping terms up to $\mathcal{O}(1/m_j)$, the interaction reduces to
\begin{eqnarray}
\mathcal{H}_{\rm em} &=& 2\sqrt{\frac{\pi}{k}}\sum^3_{j=1}\mu_j \left[ {\rm k} \, \mathbf{s}_{j,-} e^{-i \mathbf{k} \cdot \mathbf{r}_j} \right. \nonumber \\
&-& \left. \frac{1}{2}\left( \mathbf{p}_{j,-} e^{-i \mathbf{k} \cdot \mathbf{r}_j} + e^{-i \mathbf{k} \cdot \mathbf{r}_j} \mathbf{p}_{j,-} \right) \right],
\label{eq:Hem}
\end{eqnarray}
where $\mu_j = e_j/(2m_j)$, $\mathbf{s}_{j,-} = s_{j,x} - i s_{j,y}$, and $\mathbf{p}_{j,-} = p_{j,x} - i p_{j,y}$ are the magnetic moment, spin-ladder, and momentum-ladder operators of the $j$-th quark, respectively.

The partial decay width of a radiative transition $A \rightarrow A' \gamma$ is given by
\begin{equation}\label{gammaEM}
\Gamma_{\rm em} = \frac{\Phi}{(2\pi)^2} \frac{2}{2J_A+1} \sum_{M_{J_A}>0} |A_{M_{J_A}}|^2.
\end{equation}
Here the phase space factor is $\Phi = 4\pi (E_{A'}/m_A) {\rm k}^2$, with ${\rm k} = (m_A^2 - m_{A'}^2)/(2m_A)$ and $E_{A'} = \sqrt{m_{A'}^2 + {\rm k}^2}$. $A_{M_{J_A}}$ is the electromagnetic transition amplitude, 
\begin{equation}
A_{M_{J_A}} = \langle \phi_{A'}, k_{A'}, J_{A'}, M_{J_{A'}}-1 | \mathcal{H}_{\rm em} | \phi_{A}, k_{A}, J_{A}, M_{J_{A}} \rangle.
\end{equation}
 These baryon states diagonalize the Hamiltonian given in Eq.~\ref{MassFormula}. This Hamiltonian describes the charmed baryon spatial degrees of freedom in terms of two harmonic oscillators, using the Jacobi coordinates $\boldsymbol{\rho}$ and $\boldsymbol{\lambda}$, where the charmed baryon is modeled as a three-quark system. Namely, $\boldsymbol{\rho}$ represents the relative motion between the two light quarks, while $\boldsymbol{\lambda}$ describes the motion of the charm quark with respect to the center of mass of the light-quark pair in Ref.~\cite{Garcia-Tecocoatzi:2023btk, Garcia-Tecocoatzi:2022zrf, Santopinto:2018ljf}. Additionally, the Hamiltonian in Ref.~\cite{Santopinto:2018ljf} incorporates the contribution of spin, spin-orbit, isospin, and flavor degrees of freedom.
In accordance with Ref.~\cite{Rivero-Acosta:2025drn}, it is convenient to write the spatial sector, $\boldsymbol{\rho}(\boldsymbol{\lambda})$, of the charmed baryon states in terms of the basis $\left| l_{\lambda}, l_{\rho}, k_{\lambda}, k_{\rho} \right\rangle$, where $l_{\lambda,\rho}$ represents the orbital angular momenta and $k_{\lambda,\rho}$ denotes the number of nodes of the $\lambda$ and $\rho$ oscillators, hence $k_A$ becomes $k_A = k_\lambda + k_\rho$. In this context, we define the energy band as $N=n_\rho+n_\lambda$ where  $n_{\rho(\lambda)}= 2k_{\rho(\lambda)} + l_{\rho(\lambda)}$ represents the quantum numbers of the harmonic oscillator. 

In the analysis below, we will adopt the simplified notation  in terms of the number of nodes and the orbital angular momentum of the $\rho(\lambda)$ oscillators, $\left| l_{\lambda}, l_{\rho}, k_{\lambda}, k_{\rho} \right\rangle$ for the state of each baryon. Furthermore, to fully specify the baryon state, we also include the spectroscopic notation $^{2S+1}L_{x,J}$, where the subscript $x$ denotes the orbital excitation and takes one of the values $x = \lambda$, $\lambda\lambda$, $\rho$, $\rho\rho$, or $\lambda\rho$. Here, $\lambda$ corresponds to a single $\lambda$-mode excitation, $\lambda\lambda$ represents a double $\lambda$-mode excitation, $\rho$ indicates a single $\rho$-mode excitation, $\rho\rho$ signifies a double $\rho$-mode excitation, and $\lambda\rho$ denotes a mixed excitation. Additionally, we assign the total angular momentum and parity ${\bf J}^P$ and label the flavor multiplet with the symbol $\mathcal{F}$.\\

The $\mathcal{H}_{\rm em}$ is rewritten in terms of spin-flip and orbit-flip operators as
\begin{equation}
\mathcal{H}_{\rm em} = \sqrt{\frac{4\pi}{(2\pi)^3 {\rm k}}} \sum^3_{j=1} \mu_j \left[ {\rm k} \, \mathbf{s}_{j,-} \hat{U}_j - \frac{1}{2} \hat{T}_{j,-} \right],
\end{equation}
using 
${\rm k}\mathbf{ s}_{j,-}e^{-i \mathbf{k} \cdot \mathbf{ r}_j}\equiv 
{\rm k}\mathbf{ s}_{j,-} \hat{U}_j$ and  $\mathbf{p}_{j,-} \, e^{-i \mathbf{k} \cdot \mathbf{ r}_j} + e^{-i \mathbf{k} \cdot \mathbf{ r}_j} \, \mathbf{p}_{j,-}\equiv \hat{T}_{j,-} $. Details on the analytical evaluation of spatial matrix elements and baryon wave functions can be found in~\cite{Rivero-Acosta:2025drn}.


\begin{table*}[h!tp]
\caption{ Predicted electromagnetic decay widths (in keV) for $\Sigma_c(nnc)$ states with isospin 1 belonging to the flavor multiplet $\mathcal{F} = {\mathbf{ 6}}_{\rm F}$. The first column denotes the baryon name along with its predicted mass in both $S$-wave and $P$-wave configurations, corresponding to the $N = 0$ and $N = 1$ energy bands, respectively, where $N=n_\rho+n_\lambda$, taken from Ref.~\cite{Garcia-Tecocoatzi:2022zrf}. 
 The second column indicates the spin-parity $\mathbf{J}^{\rm P}$, while the third column shows the internal configuration of the state in the three-quark model, written as $\left| l_{\lambda}, l_{\rho}, k_{\lambda}, k_{\rho} \right\rangle$, where $l_{\lambda,\rho}$ are the orbital angular momenta and $k_{\lambda,\rho}$ are the number of nodes of the $\lambda$ and $\rho$ oscillators.
The fourth column provides the spectroscopic notation $^{2S+1}L_{x,J}$ associated with each state, where $x = \lambda$ or $\rho$ denotes the mode in which the orbital excitation occurs. 
Starting from the fifth column, the predicted electromagnetic decay widths are presented for each decay channel, computed using Eq.~\ref{gammaEM}. The decay products are listed at the top of each column. Zero values correspond to electromagnetic decays that are either forbidden by phase space or have widths too small to be displayed at this scale. Our results are compared with those from Refs.~\cite{Wang:2017kfr}, \cite{Ortiz-Pacheco:2023kjn}, and \cite{Luo:2025pzb}. The symbol ``$\cdots$'' indicates that no prediction is available for that particular state.}

\begin{center}
\scriptsize{
\begingroup
\setlength{\tabcolsep}{1.75pt} 
\renewcommand{\arraystretch}{1.35} 

\begin{tabular}{c c c c  c  c  p{1.0cm}  p{1.0cm}  p{1.0cm}  p{1.0cm}  p{1.0cm}  p{1.0cm}  p{1.0cm}  p{1.0cm}} \hline \hline
&    &    &    & $\Sigma^{++}_c \gamma$  & $\Sigma^{*++}_c \gamma$  & $\Sigma^{+}_c \gamma$  & $\Sigma^{*+}_c \gamma$  & $\Sigma^{0}_c \gamma$  & $\Sigma^{*0}_c \gamma$  & $\Lambda^{+}_c \gamma$ \\
$\mathbf{\Sigma_c(nnc)}$  & $\mathbf{J^P}$  & $\vert l_{\lambda}, l_{\rho}, k_{\lambda}, k_{\rho} \rangle$  & $^{2S+1}L_{x,J}$  & $^{2}S_{1/2}$  & $^{4}S_{3/2}$  & $^{2}S_{1/2}$  & $^{4}S_{3/2}$  & $^{2}S_{1/2}$  & $^{4}S_{3/2}$  & $^{2}S_{1/2}$  \\ \hline
 $N=0$  &  &  &  &  &  \\\
$\Sigma_c(2456)$  & $ \mathbf{\frac{1}{2}^+}$ & $\vert \,0\,,\,0\,,\,0\,,\,0 \,\rangle $ &$^{2}S_{1/2}$&$ 0$    &  0  &0  &0  &0  &0  &$ 168 _{- 62 }^{+ 76 }$   \\
 & &  &   & ... & ... & ... & ... & ... & ... & 80.6 & \cite{Wang:2017kfr} \\
 & &  &   & ... & ... & ... & ... & ... & ... & 87.2 & \cite{Ortiz-Pacheco:2023kjn} \\
$\Sigma_c(2525)$  & $ \mathbf{\frac{3}{2}^+}$ & $\vert \,0\,,\,0\,,\,0\,,\,0 \,\rangle $ &$^{4}S_{3/2}$&$ 4.2 _{- 3.3 }^{+ 6.4 }$   &  0   &  0  &0  &$ 2.8 _{- 2.3 }^{+ 3.4 }$    &  0  &$ 354 _{- 111 }^{+ 128 }$   \\
 & &  &   & 3.94 & ... & 0.004 & ... & 3.43 & ... & 373 & \cite{Wang:2017kfr} \\
  & &  &   & 2.1 & ... & 0 & ... & 1.8 & ... & 199.4 & \cite{Ortiz-Pacheco:2023kjn} \\
 $N=1$  &  &  &  &  &  \\
$\Sigma_c(2811)$  & $ \mathbf{\frac{1}{2}^-}$ & $\vert \,1\,,\,0\,,\,0\,,\,0 \,\rangle $ &$^{2}P_{\lambda,1/2}$&$ 160 _{- 42 }^{+ 45 }$    &  $ 7 _{- 3 }^{+ 4 }$    &  $ 2.6 _{- 2.1 }^{+ 3.4 }$    &  $ 0.6 _{- 0.2 }^{+ 0.3 }$    &  $ 249 _{- 31 }^{+ 25 }$    &  $ 1.1 _{- 0.6 }^{+ 0.8 }$    &  $ 198 _{- 64 }^{+ 78 }$   \\
 & &  &   & 283 & 3.04 & 1.6 & 0.31 & 205 & 0.39 & 48.3 & \cite{Wang:2017kfr} \\
 & &  &   & 67.1 & 3.4 & 4.6 & 0.3 & 156.1 & 0.6 & 110 & \cite{Ortiz-Pacheco:2023kjn} \\
$\Sigma_c(2853)$  & $ \mathbf{\frac{1}{2}^-}$ & $\vert \,1\,,\,0\,,\,0\,,\,0 \,\rangle $ &$^{4}P_{\lambda,1/2}$&$ 24 _{- 10 }^{+ 13 }$    &  $ 10 _{- 9 }^{+ 11 }$    &  $ 2.2 _{- 0.8 }^{+ 1.1 }$    &  $ 1.6 _{- 1.0 }^{+ 1.4 }$    &  $ 3.8 _{- 1.9 }^{+ 2.6 }$    &  $ 30 _{- 12 }^{+ 10 }$    &  $ 121 _{- 40 }^{+ 44 }$   \\
 & &  &   & 8.54 & 387 & 0.92 & 1.75 & 1.02 & 289 & 52.1 & \cite{Wang:2017kfr} \\
 & &  &   & 12.5 & 1.7 & 1.1 & 1.5 & 2.0 & 13.9 & 66.7 & \cite{Ortiz-Pacheco:2023kjn} \\
$\Sigma_c(2838)$  & $ \mathbf{\frac{3}{2}^-}$ & $\vert \,1\,,\,0\,,\,0\,,\,0 \,\rangle $ &$^{2}P_{\lambda,3/2}$&$ 964 _{- 239 }^{+ 259 }$    &  $ 10 _{- 4 }^{+ 5 }$    &  $ 15 _{- 9 }^{+ 10 }$    &  $ 0.9 _{- 0.3 }^{+ 0.4 }$    &  $ 594 _{- 70 }^{+ 72 }$    &  $ 1.6 _{- 0.8 }^{+ 1.1 }$    &  $ 226 _{- 70 }^{+ 78 }$   \\
 & &  &   & 210 & 14.7 & 4.64 & 1.55 & 245 & 1.82 & 87.3 & \cite{Wang:2017kfr} \\
 & &  &   & 551.8 & 5.6 & 4.9 & 0.5 & 432.2 & 0.9 & 132.9 & \cite{Ortiz-Pacheco:2023kjn} \\
$\Sigma_c(2880)$  & $ \mathbf{\frac{3}{2}^-}$ & $\vert \,1\,,\,0\,,\,0\,,\,0 \,\rangle $ &$^{4}P_{\lambda,3/2}$&$ 85 _{- 32 }^{+ 41 }$    &  $ 153 _{- 27 }^{+ 30 }$    &  $ 8 _{- 2 }^{+ 3 }$    &  $ 1.5 _{- 0.7 }^{+ 1.1 }$    &  $ 13 _{- 6 }^{+ 8 }$    &  $ 176 _{- 17 }^{+ 14 }$    &  $ 381 _{- 117 }^{+ 127 }$   \\
 & &  &   & 17.5 & 181 & 1.86 & 0.68 & 2.12 & 159 & 105 & \cite{Wang:2017kfr} \\
 & &  &   & 48.6 & 68.8 & 4.5 & 2 & 7.5 & 109.7 & 220.9 & \cite{Ortiz-Pacheco:2023kjn} \\
$\Sigma_c(2925)$  & $ \mathbf{\frac{5}{2}^-}$ & $\vert \,1\,,\,0\,,\,0\,,\,0 \,\rangle $ &$^{4}P_{\lambda,5/2}$&$ 79 _{- 27 }^{+ 37 }$    &  $ 995 _{- 270 }^{+ 278 }$    &  $ 7 _{- 2 }^{+ 2 }$    &  $ 11 _{- 8 }^{+ 10 }$    &  $ 12 _{- 5 }^{+ 7 }$    &  $ 647 _{- 82 }^{+ 87 }$    &  $ 288 _{- 85 }^{+ 89 }$   \\
 & &  &   & 13.6 & 168 & 1.46 & 0.89 & 1.64 & 160 & 59.4 & \cite{Wang:2017kfr} \\
 & &  &   & 49.7 & 450.3 & 4.7 & 2.3 & 7.4 & 335.1 & 179.8 & \cite{Ortiz-Pacheco:2023kjn} \\
  & &  &   & ... & $ 98^{+110}_{-85} $ & ... & $ 5.8^{+6.8}_{-4.6} $ & ... & $ 8.5^{+9.2}_{-6.8} $ & ... & \cite{Luo:2025pzb} \\
$\Sigma_c(2994)$  & $ \mathbf{\frac{1}{2}^-}$ & $\vert \,0\,,\,1\,,\,0\,,\,0 \,\rangle $ &$^{2}P_{\rho,1/2}$&$ 310 _{- 102 }^{+ 121 }$    &  $ 102 _{- 40 }^{+ 47 }$    &  $ 19 _{- 6 }^{+ 7 }$    &  $ 6 _{- 2 }^{+ 3 }$    &  $ 78 _{- 26 }^{+ 32 }$    &  $ 26 _{- 10 }^{+ 12 }$    &  $ 602 _{- 70 }^{+ 55 }$   \\
 & &  &   & ... & ... & ... & ... & ... & ... & ... & \cite{Wang:2017kfr} \\
 & &  &   & 178.1 & 56.6 & 10.1 & 3.2 & 49.1 & 15.6 & 618.2 & \cite{Ortiz-Pacheco:2023kjn} \\
  & &  &   & $ 1100^{+2200}_{-1000} $ & $ 390^{+1000}_{-390} $ & $ 17^{+33}_{-16} $ & $ 6.2^{+14.6}_{-6.2} $ & $ 290^{+540}_{-290} $ & $ 99^{+250}_{-99} $ & ... & \cite{Luo:2025pzb} \\
$\Sigma_c(3021)$  & $ \mathbf{\frac{3}{2}^-}$ & $\vert \,0\,,\,1\,,\,0\,,\,0 \,\rangle $ &$^{2}P_{\rho,3/2}$&$ 357 _{- 111 }^{+ 121 }$    &  $ 123 _{- 45 }^{+ 54 }$    &  $ 22 _{- 7 }^{+ 8 }$    &  $ 8 _{- 3 }^{+ 3 }$    &  $ 89 _{- 28 }^{+ 31 }$    &  $ 31 _{- 11 }^{+ 14 }$    &  $ 562 _{- 66 }^{+ 63 }$   \\
 & &  &   & ... & ... & ... & ... & ... & ... & ... & \cite{Wang:2017kfr} \\
 & &  &   & 215.5 & 72.2 & 12.2 & 4.1 & 59.4 & 19.9 & 624.6 & \cite{Ortiz-Pacheco:2023kjn} \\
  & &  &   & $ 680^{+1700}_{-660} $ & $ 590^{+1400}_{-570} $ & $ 10^{+19.6}_{-9.1} $ & $ 9.2^{+21.4}_{-8.9} $ & $ 170^{+310}_{-160} $ & $ 150^{+350}_{-140} $ & ... & \cite{Luo:2025pzb} \\
\hline \hline
\end{tabular}
\endgroup
}
\end{center}
\label{sigmasEM}
\end{table*}

\begin{table*}[h!tp]
\caption{Same as table \ref{sigmasEM}, but for $\Xi'_c$ states with isospin $1/2$.}
\begin{center}
\scriptsize{
\begingroup
\setlength{\tabcolsep}{1.75pt} 
\renewcommand{\arraystretch}{1.35} 

\begin{tabular}{c c c c  p{1.0cm}  p{1.0cm}  p{1.0cm}  p{1.0cm}  p{1.0cm}  p{1.0cm}  p{1.0cm}  p{1.0cm}  p{1.0cm}} \hline \hline
&    &    &    & $\Xi_c^{+} \gamma$  & $\Xi_c^{0} \gamma$  & $\Xi_c'^{+} \gamma$  & $\Xi_c^{*+} \gamma$  & $\Xi_c'^{0} \gamma$  & $\Xi_c^{*0} \gamma$ \\
$\mathbf{\Xi'_c(snc)}$  & $\mathbf{J^P}$  & $\vert l_{\lambda}, l_{\rho}, k_{\lambda}, k_{\rho} \rangle$  & $^{2S+1}L_{x,J}$  & $^2S_{1/2}$  & $^2S_{1/2}$  & $^2S_{1/2}$  & $^4S_{3/2}$  & $^2S_{1/2}$  & $^4S_{3/2}$  \\ \hline
 $N=0$  &  &  &  &  &  \\\
$\Xi'_c(2571)$  & $ \mathbf{\frac{1}{2}^+}$ & $\vert \,0\,,\,0\,,\,0\,,\,0 \,\rangle $ &$^{2}S_{1/2}$&$ 24 _{- 12 }^{+ 19 }$    &  $ 0.5 _{- 0.3 }^{+ 0.5 }$    &  0  &0  &0  &0 \\
 & &  &   & 42.3 & 0 & ... & ... & ... & ... & \cite{Wang:2017kfr} \\
 & &  &   & 20.6 & 0.4 & ... & ... & ... & ... & \cite{Ortiz-Pacheco:2023kjn} \\
$\Xi'_c(2640)$  & $ \mathbf{\frac{3}{2}^+}$ & $\vert \,0\,,\,0\,,\,0\,,\,0 \,\rangle $ &$^{4}S_{3/2}$&$ 99 _{- 37 }^{+ 48 }$    &  $ 2.1 _{- 1.0 }^{+ 1.5 }$    &  $ 0.2 _{- 0.1 }^{+ 0.2 }$    &  0  &$ 2.0 _{- 1.3 }^{+ 1.9 }$    &  0 \\
 & &  &   & 139 & 0 & 0.004 & ... & 3.03 & ... & \cite{Wang:2017kfr} \\
 & &  &   & 74.2 & 1.6 & 0.1 & ... & 1.4 & ... & \cite{Ortiz-Pacheco:2023kjn} \\ 
$N=1$  &  &  &  &  &  \\
$\Xi'_c(2893)$  & $ \mathbf{\frac{1}{2}^-}$ & $\vert \,1\,,\,0\,,\,0\,,\,0 \,\rangle $ &$^{2}P_{\lambda,1/2}$&$ 59 _{- 18 }^{+ 24 }$    &  $ 1.2 _{- 0.6 }^{+ 0.8 }$    &  $ 3.6 _{- 3.5 }^{+ 5.1 }$    &  $ 0.5 _{- 0.2 }^{+ 0.2 }$    &  $ 253 _{- 14 }^{+ 15 }$    &  $ 0.3 _{- 0.1 }^{+ 0.2 }$   \\
 & &  &   & 46.4 & 0 & 0.03 & 1.61 & 472 & 1.0 & \cite{Wang:2017kfr} \\
 & &  &   & 37.9 & 0.8 & 0 & 0.3 & 158.3 & 0.1 & \cite{Ortiz-Pacheco:2023kjn} \\
$\Xi'_c(2935)$  & $ \mathbf{\frac{1}{2}^-}$ & $\vert \,1\,,\,0\,,\,0\,,\,0 \,\rangle $ &$^{4}P_{\lambda,1/2}$&$ 41 _{- 15 }^{+ 18 }$    &  $ 0.9 _{- 0.4 }^{+ 0.6 }$    &  $ 2.4 _{- 0.8 }^{+ 1.1 }$    &  $ 0.2 _{- 0.2 }^{+ 0.6 }$    &  $ 1.2 _{- 0.6 }^{+ 0.8 }$    &  $ 40 _{- 6 }^{+ 5 }$   \\
 & &  &   & 14.5 & 0 & 0.33 & 0.16 & 0.2 & 125 & \cite{Wang:2017kfr} \\
 & &  &   & 25.2 & 0.5 & 1.3 & 0.1 & 0.6 & 18.5 & \cite{Ortiz-Pacheco:2023kjn} \\
$\Xi'_c(2920)$  & $ \mathbf{\frac{3}{2}^-}$ & $\vert \,1\,,\,0\,,\,0\,,\,0 \,\rangle $ &$^{2}P_{\lambda,3/2}$&$ 73 _{- 23 }^{+ 26 }$    &  $ 1.6 _{- 0.7 }^{+ 0.9 }$    &  $ 49 _{- 25 }^{+ 28 }$    &  $ 0.8 _{- 0.3 }^{+ 0.3 }$    &  $ 449 _{- 36 }^{+ 36 }$    &  $ 0.4 _{- 0.2 }^{+ 0.2 }$   \\
 & &  &   & 46.1 & 0 & 12.1 & 1.59 & 302 & 1.05 & \cite{Wang:2017kfr} \\
 & &  &   & 50.2 & 1.1 & 18.7 & 0.5 & 339.3 & 0.3 & \cite{Ortiz-Pacheco:2023kjn} \\
$\Xi'_c(2962)$  & $ \mathbf{\frac{3}{2}^-}$ & $\vert \,1\,,\,0\,,\,0\,,\,0 \,\rangle $ &$^{4}P_{\lambda,3/2}$&$ 140 _{- 45 }^{+ 53 }$    &  $ 3.0 _{- 1.4 }^{+ 1.9 }$    &  $ 9 _{- 3 }^{+ 3 }$    &  $ 6 _{- 4 }^{+ 5 }$    &  $ 4.5 _{- 2.1 }^{+ 2.6 }$    &  $ 173 _{- 9 }^{+ 8 }$   \\
 & &  &   & 54.6 & 0 & 2.06 & 1.64 & 1.21 & 187 & \cite{Wang:2017kfr} \\
 & &  &   & 90.6 & 1.9 & 5.2 & 0.9 & 2.5 & 108 & \cite{Ortiz-Pacheco:2023kjn} \\
$\Xi'_c(3007)$  & $ \mathbf{\frac{5}{2}^-}$ & $\vert \,1\,,\,0\,,\,0\,,\,0 \,\rangle $ &$^{4}P_{\lambda,5/2}$&$ 120 _{- 38 }^{+ 44 }$    &  $ 2.5 _{- 1.2 }^{+ 1.5 }$    &  $ 9 _{- 3 }^{+ 3 }$    &  $ 45 _{- 27 }^{+ 30 }$    &  $ 4.3 _{- 2.0 }^{+ 2.3 }$    &  $ 486 _{- 45 }^{+ 48 }$   \\
 & &  &   & 32.0 & 0 & 1.63 & 2.35 & 0.93 & 192 & \cite{Wang:2017kfr} \\
 & &  &   & 83.4 & 1.8 & 5.8 & 14.1 & 2.7 & 248.2 & \cite{Ortiz-Pacheco:2023kjn} \\
  & &  &   & ... & ... & ... & $19^{+41}_{-19}$ & ... & $44^{+100}_{-44}$  & \cite{Luo:2025pzb} \\
$\Xi'_c(3040)$  & $ \mathbf{\frac{1}{2}^-}$ & $\vert \,0\,,\,1\,,\,0\,,\,0 \,\rangle $ &$^{2}P_{\rho,1/2}$&$ 852 _{- 66 }^{+ 68 }$    &  $ 18 _{- 5 }^{+ 5 }$    &  $ 24 _{- 9 }^{+ 11 }$    &  $ 7 _{- 2 }^{+ 3 }$    &  $ 32 _{- 10 }^{+ 11 }$    &  $ 9 _{- 3 }^{+ 4 }$   \\
  & &  &   & ... & ... & ... & ... & ... & ... & \cite{Wang:2017kfr} \\
 & &  &   & 709.5 & 15 & 12.7 & 3.5 & 20.7 & 5.7 & \cite{Ortiz-Pacheco:2023kjn} \\
  & &  &   & ... & ... & $44^{+97}_{-43}$ & $15^{+42}_{-15}$ & $25^{+50}_{-25}$ & $22^{+60}_{-22}$  & \cite{Luo:2025pzb} \\
$\Xi'_c(3067)$  & $ \mathbf{\frac{3}{2}^-}$ & $\vert \,0\,,\,1\,,\,0\,,\,0 \,\rangle $ &$^{2}P_{\rho,3/2}$&$ 826 _{- 59 }^{+ 60 }$    &  $ 17 _{- 5 }^{+ 5 }$    &  $ 29 _{- 10 }^{+ 12 }$    &  $ 8 _{- 3 }^{+ 4 }$    &  $ 39 _{- 11 }^{+ 13 }$    &  $ 12 _{- 4 }^{+ 4 }$   \\
 & &  &   & ... & ... & ... & ... & ... & ... & \cite{Wang:2017kfr} \\
 & &  &   & 760.5 & 16.1 & 16.4 & 4.8 & 26.8 & 7.8 & \cite{Ortiz-Pacheco:2023kjn} \\
  & &  &   & ... & ... & $130^{+260}_{-130}$ & $57^{+160}_{-57}$ & $71^{+140}_{-70}$ & $62^{+170}_{-62}$  & \cite{Luo:2025pzb} \\
\hline \hline
\end{tabular}
\endgroup
}
\end{center}
\label{cascadesprimeEM}
\end{table*}


\begin{table}[h!tp]

\caption{Same as Table \ref{sigmasEM}, but for $\Omega_{c}$ states with isospin 0. The results are compared with those of Refs.~\cite{Wang:2017hej,Ortiz-Pacheco:2023kjn,Luo:2025pzb}. }
\begin{center}
\scriptsize{
\begingroup
\setlength{\tabcolsep}{1.75pt} 
\renewcommand{\arraystretch}{1.35} 

\begin{tabular}{c c c c  p{1.0cm}  p{1.0cm}  p{1.0cm}  p{1.0cm}  p{1.0cm}} \hline \hline 
&    &    &    & $\Omega_{c} \gamma$  & $\Omega^{*}_{c} \gamma$ \\
$\mathbf{\Omega_c(ssc)}$  & $\mathbf{J^P}$  & $\vert l_{\lambda}, l_{\rho}, k_{\lambda}, k_{\rho} \rangle$  & $^{2S+1}L_{x,J}$  & $^2S_{1/2}$  & $^4S_{3/2}$  \\ \hline 
 $N=0$  &  &  &  &  &  \\
$\Omega_c(2709)$  & $ \mathbf{\frac{1}{2}^+}$ & $\vert \,0\,,\,0\,,\,0\,,\,0 \,\rangle $ &$^{2}S_{1/2}$&0  &0 \\
$\Omega_c(2778)$  & $ \mathbf{\frac{3}{2}^+}$ & $\vert \,0\,,\,0\,,\,0\,,\,0 \,\rangle $ &$^{4}S_{3/2}$&$ 1.4 _{- 1.0 }^{+ 1.6 }$    &  0 \\
 & &  &   & 0.89 & ... & \cite{Wang:2017hej} \\
  & &  &   & 1 & ... & \cite{Ortiz-Pacheco:2023kjn} \\
 $N=1$  &  &  &  &  &  \\
$\Omega_c(3008)$  & $ \mathbf{\frac{1}{2}^-}$ & $\vert \,1\,,\,0\,,\,0\,,\,0 \,\rangle $ &$^{2}P_{\lambda,1/2}$&$ 212 _{- 14 }^{+ 15 }$    &  $ 0.1 _{- 0.0 }^{+ 0.0 }$   \\
 & &  &   & 200-360 & $\sim$10 & \cite{Wang:2017hej} \\
 & &  &  & 135.7 & 0 & \cite{Ortiz-Pacheco:2023kjn} \\
$\Omega_c(3050)$  & $ \mathbf{\frac{1}{2}^-}$ & $\vert \,1\,,\,0\,,\,0\,,\,0 \,\rangle $ &$^{4}P_{\lambda,1/2}$&$ 0.3 _{- 0.2 }^{+ 0.2 }$    &  $ 37 _{- 4 }^{+ 4 }$   \\
 & &  &  & ... & ... & \cite{Wang:2017hej} \\
 & &  &   & 0.2 & 17.5 & \cite{Ortiz-Pacheco:2023kjn} \\
$\Omega_c(3035)$  & $ \mathbf{\frac{3}{2}^-}$ & $\vert \,1\,,\,0\,,\,0\,,\,0 \,\rangle $ &$^{2}P_{\lambda,3/2}$&$ 320 _{- 22 }^{+ 22 }$    &  $ 0.1 _{- 0.1 }^{+ 0.1 }$   \\
 & &  &   & 350 & 0.568 & \cite{Wang:2017hej} \\
 & &  &   & 251.6 & 0.1 & \cite{Ortiz-Pacheco:2023kjn} \\
$\Omega_c(3077)$  & $ \mathbf{\frac{3}{2}^-}$ & $\vert \,1\,,\,0\,,\,0\,,\,0 \,\rangle $ &$^{4}P_{\lambda,3/2}$&$ 1.3 _{- 0.7 }^{+ 0.7 }$    &  $ 143 _{- 8 }^{+ 8 }$   \\
 & &  &  & 1.12 & 330 & \cite{Wang:2017hej} \\
 & &  &   & 0.7 & 90.9 & \cite{Ortiz-Pacheco:2023kjn} \\
$\Omega_c(3122)$  & $ \mathbf{\frac{5}{2}^-}$ & $\vert \,1\,,\,0\,,\,0\,,\,0 \,\rangle $ &$^{4}P_{\lambda,5/2}$&$ 1.3 _{- 0.7 }^{+ 0.7 }$    &  $ 346 _{- 27 }^{+ 26 }$   \\
 & &  &   & 0.1 & 180 & \cite{Wang:2017hej} \\
 & &  &   & 0.7 & 177 & \cite{Ortiz-Pacheco:2023kjn} \\
 & &  &   & ... & $5.2^{+14.6}_{-5.2}$ & \cite{Luo:2025pzb} \\
$\Omega_c(3129)$  & $ \mathbf{\frac{1}{2}^-}$ & $\vert \,0\,,\,1\,,\,0\,,\,0 \,\rangle $ &$^{2}P_{\rho,1/2}$&$ 12 _{- 4 }^{+ 4 }$    &  $ 3.0 _{- 1.0 }^{+ 1.1 }$   \\
 & &  &   & ... & ... & \cite{Wang:2017hej} \\
 & &  &   & 8.1 & 1.9 & \cite{Ortiz-Pacheco:2023kjn} \\
 & &  &  & $76^{+170}_{-76}$ & $12^{+36}_{-12}$ & \cite{Luo:2025pzb} \\
$\Omega_c(3156)$  & $ \mathbf{\frac{3}{2}^-}$ & $\vert \,0\,,\,1\,,\,0\,,\,0 \,\rangle $ &$^{2}P_{\rho,3/2}$&$ 16 _{- 4 }^{+ 4 }$    &  $ 4.1 _{- 1.2 }^{+ 1.3 }$   \\
 & &  &   & ... & ... & \cite{Wang:2017hej} \\
 & &  &   & 11 & 2.9 & \cite{Ortiz-Pacheco:2023kjn} \\
 & &  &  & $20^{+43}_{-20}$ & $12^{+38}_{-12}$ & \cite{Luo:2025pzb} \\
\hline \hline
\end{tabular}
\endgroup
}
\end{center}
\label{omegasEM}
\end{table}

\subsection{Uncertainties}

In this work, we report an uncertainty for each electromagnetic decay width calculation. The sources of uncertainty arise from the masses of the charmed baryons involved in the electromagnetic transition $A \rightarrow A'\gamma$, which originate from the experimentally measured masses reported by the PDG~\cite{ParticleDataGroup:2024cfk}. Additionally, we include a model uncertainty to account for the approximate description of the charmed baryon masses in Eq.~(\ref{MassFormula2}). These uncertainties are propagated through a Monte Carlo simulation for error estimation. Specifically, each baryon mass is sampled from a Gaussian distribution whose mean corresponds to either the experimental PDG value or the theoretical prediction from Ref.~\cite{Garcia-Tecocoatzi:2022zrf}, and whose standard deviation is given by the corresponding experimentally reported uncertainty plus the model uncertainty. For each set of sampled masses, the electromagnetic decay width is computed. By repeating this procedure $10^3$ times, we obtain a distribution for each decay width. The mean of each distribution is taken as the central value, and the 68\% confidence-level uncertainty is estimated using quantiles. The fitting and error-propagation procedures are carried out using \texttt{MINUIT}~\cite{James:1975dr} and \texttt{NUMPY}~\cite{Harris:2020xlr}.

\section{Results and Discussion}
\label{Results}



\begin{table*}[htp!]
\caption{ Predicted electromagnetic decay widths (in keV) for transitions from second shell $\Sigma_c^{+}(snc)$ states with isospin 1 belonging to the flavor multiplet $\mathcal{F} = {\bf 6}_{\rm F}$ to $\Sigma_c^{+}(snc)$ states. The first column lists the baryon name along with its predicted mass in the $N = 2$ energy band, where $N = n_\rho + n_\lambda$, as calculated in Ref.~\cite{Garcia-Tecocoatzi:2022zrf}. 
The second column displays the spin-parity $\mathbf{J}^{\rm P}$, while the third column shows the internal configuration $\left| l_{\lambda}, l_{\rho}, k_{\lambda}, k_{\rho} \right\rangle$ in the three-quark model, where $l_{\lambda,\rho}$ are the orbital angular momenta and $k_{\lambda,\rho}$ denote the number of radial nodes of the $\lambda$ and $\rho$ oscillators. The fourth column presents the spectroscopic notation $^{2S+1}L_{x,J}$ for each state, where the subscript $x$ specifies the orbital excitation and can take the values $x = \lambda$, $\lambda\lambda$, $\rho$, $\rho\rho$, or $\lambda\rho$. 
Starting from the fifth column, the electromagnetic decay widths are shown, computed using Eq.~\ref{gammaEM}. Each column corresponds to a specific decay channel; the final-state particles are indicated at the top of each column, and their spectroscopic notation $^{2S+1}L_{x,J}$ is provided in the second row. Zero values correspond to decay widths that are either forbidden by phase space or too small to be shown at this scale. Our results are compared with those in Refs.~\cite{Yao:2018jmc, Peng:2024pyl}. The symbol ``$\cdots$'' indicates that no prediction is available for the corresponding state in Refs.~\cite{Yao:2018jmc} and~\cite{Peng:2024pyl}. }
\begin{center}
\scriptsize{
\begingroup
\setlength{\tabcolsep}{1.75pt} 
\renewcommand{\arraystretch}{1.35} 

\begin{tabular}{c c c c  p{1.0cm}  p{1.0cm}  p{1.0cm}  p{1.0cm}  p{1.0cm}  p{1.0cm}  p{1.0cm}  p{1.0cm}  p{1.0cm}  p{1.0cm}  p{1.0cm}  p{1.0cm}} \hline \hline
&    &    &    & $\Sigma_{c}^{+} \gamma$  & $\Sigma_{c}^{*+} \gamma$  & $\Sigma_{c}^{+} \gamma$  & $\Sigma_{c}^{*+} \gamma$  & $\Sigma_{c}^{+} \gamma$  & $\Sigma_{c}^{*+} \gamma$  & $\Sigma_{c}^{*+} \gamma$  & $\Sigma_{c}^{+} \gamma$  & $\Sigma_{c}^{+} \gamma$ \\
$\mathbf{\Sigma_c(nnc)}$  & $\mathbf{J^P}$  & $\vert l_{\lambda}, l_{\rho}, k_{\lambda}, k_{\rho} \rangle$  & $^{2S+1}L_{x,J}$  & $^2S_{1/2}$  & $^4S_{3/2}$  & $^2P_{\lambda,1/2}$  & $^4P_{\lambda,1/2}$  & $^2P_{\lambda,3/2}$  & $^4P_{\lambda,3/2}$  & $^4P_{\lambda,5/2}$  & $^2P_{\rho,1/2}$  & $^2P_{\rho,3/2}$  \\ \hline
 $N=2$  &  &  &  &  &  \\\ 
$\Sigma_c(3175)$  & $ \mathbf{\frac{3}{2}^+}$ & $\vert \,2\,,\,0\,,\,0\,,\,0 \,\rangle $ &$^{2}D_{\lambda\lambda,3/2}$&$ 13 _{- 2 }^{+ 3 }$    &  $ 0.5 _{- 0.3 }^{+ 0.4 }$  &  $ 2.9 _{- 1.7 }^{+ 3.3 }$    &  $ 2.8 _{- 1.4 }^{+ 2.0 }$    &  $ 2.9 _{- 1.5 }^{+ 2.0 }$    &  $ 0.6 _{- 0.3 }^{+ 0.4 }$    &  0  &0  &0  \\
  &  &  & & $...$  & $...$  & $1.5$ & $0.55$  & $1.82$  & $1.27$  & $0.28$  & $...$  & $...$  & \cite{Yao:2018jmc} \\
$\Sigma_c(3220)$  & $ \mathbf{\frac{5}{2}^+}$ & $\vert \,2\,,\,0\,,\,0\,,\,0 \,\rangle $ &$^{2}D_{\lambda\lambda,5/2}$&$ 63 _{- 9 }^{+ 9 }$    &  $ 0.7 _{- 0.4 }^{+ 0.5 }$  &  $ 14 _{- 6 }^{+ 8 }$    &  $ 0.4 _{- 0.2 }^{+ 0.3 }$    &  $ 8 _{- 6 }^{+ 7 }$    &  $ 1.6 _{- 0.7 }^{+ 0.8 }$    &  $ 0.9 _{- 0.5 }^{+ 0.6 }$    &  0  &0   \\
  &  &  & & $...$  & $...$  & $2.27$ & $0.03$  & $0.81$  & $0.51$  & $1.36$ & $...$  & $...$  & \cite{Yao:2018jmc} \\
$\Sigma_c(3190)$  & $ \mathbf{\frac{1}{2}^+}$ & $\vert \,2\,,\,0\,,\,0\,,\,0 \,\rangle $ &$^{4}D_{\lambda\lambda,1/2}$&$ 0.8 _{- 0.4 }^{+ 0.5 }$    &  $ 4.2 _{- 2.1 }^{+ 2.2 }$    &  $ 3.0 _{- 1.4 }^{+ 2.1 }$    &  $ 2.8 _{- 2.7 }^{+ 5.3 }$    &  0  &$ 0.8 _{- 0.7 }^{+ 0.9 }$    &  $ 0.2 _{- 0.1 }^{+ 0.2 }$    &  0  &0    \\
  &  &  & & $...$  & $...$ & $2.56$ & $0.97$  & $0.2$  & $1.26$  & $0.1$ & $...$  & $...$  & \cite{Yao:2018jmc} \\
 &  &  & & 0.1  & 56  &  $2.4$   & $...$  & $0.8$  & $1$  & $0.3$  & $...$  & $...$  & \cite{Peng:2024pyl} \\
$\Sigma_c(3217)$  & $ \mathbf{\frac{3}{2}^+}$ & $\vert \,2\,,\,0\,,\,0\,,\,0 \,\rangle $ &$^{4}D_{\lambda\lambda,3/2}$&$ 1.8 _{- 1.0 }^{+ 1.1 }$    &  $ 9 _{- 1 }^{+ 1 }$  &  $ 9 _{- 3 }^{+ 4 }$    &  $ 6 _{- 3 }^{+ 5 }$    &  $ 0.4 _{- 0.2 }^{+ 0.3 }$    &  $ 1.1 _{- 1.0 }^{+ 1.7 }$    &  $ 0.2 _{- 0.1 }^{+ 0.2 }$    &  0  &0    \\
  &  &  & & $...$  & $...$ & $2.58$ & $0.45$  & $0.06$  & $0.9$  & $0.42$ & $...$  & $...$  & \cite{Yao:2018jmc} \\
$\Sigma_c(3262)$  & $ \mathbf{\frac{5}{2}^+}$ & $\vert \,2\,,\,0\,,\,0\,,\,0 \,\rangle $ &$^{4}D_{\lambda\lambda,5/2}$&$ 2.8 _{- 1.5 }^{+ 1.6 }$    &  $ 18 _{- 2 }^{+ 2 }$ &  $ 4.7 _{- 1.4 }^{+ 1.7 }$    &  $ 5 _{- 3 }^{+ 3 }$    &  $ 9 _{- 3 }^{+ 4 }$    &  $ 6 _{- 3 }^{+ 5 }$    &  $ 0.5 _{- 0.5 }^{+ 0.8 }$    &  0  &0   \\
  &  &  & & $...$  & $...$ & $0.8$ & $0.14$  & $0.52$  & $0.47$  & $0.35$ & $...$  & $...$  & \cite{Yao:2018jmc} \\
$\Sigma_c(3326)$  & $ \mathbf{\frac{7}{2}^+}$ & $\vert \,2\,,\,0\,,\,0\,,\,0 \,\rangle $ &$^{4}D_{\lambda\lambda,7/2}$&$ 2.0 _{- 1.1 }^{+ 1.1 }$    &  $ 76 _{- 11 }^{+ 10 }$ &  $ 0.1 _{- 0.0 }^{+ 0.1 }$    &  $ 0.6 _{- 0.4 }^{+ 0.6 }$    &  $ 15 _{- 5 }^{+ 5 }$    &  $ 8 _{- 4 }^{+ 6 }$    &  $ 10 _{- 8 }^{+ 10 }$    &  0  &0   \\
  &  &  & & $...$  & $...$ & $0.001$ & $0.005$  & $0.42$  & $0.17$  & $0.55$ & $...$  & $...$  & \cite{Yao:2018jmc} \\
 &  &  & & 0.1  & 21.5 &  $0$   & $...$  & $0.4$  & $0.3$  & $0.5$  & $...$  & $...$  & \cite{Peng:2024pyl} \\
$\Sigma_c(3202)$  & $ \mathbf{\frac{1}{2}^+}$ & $\vert \,0\,,\,0\,,\,1\,,\,0 \,\rangle $ &$^{2}S_{1/2}$&$ 22 _{- 9 }^{+ 10 }$    &  $ 1.5 _{- 0.8 }^{+ 1.0 }$  &  $ 57 _{- 27 }^{+ 32 }$    &  $ 0.1 _{- 0.0 }^{+ 0.0 }$    &  $ 66 _{- 26 }^{+ 33 }$    &  $ 0.2 _{- 0.1 }^{+ 0.1 }$    &  $ 4.3 _{- 2.3 }^{+ 3.2 }$    &  0  &0    \\
 &  &  & & 3.3  & 0.8 &  $0$   & $...$  & $0.1$  & $1$  & $0.1$  & $...$  & $...$  & \cite{Peng:2024pyl} \\
$\Sigma_c(3271)$  & $ \mathbf{\frac{3}{2}^+}$ & $\vert \,0\,,\,0\,,\,1\,,\,0 \,\rangle $ &$^{4}S_{3/2}$&$ 5 _{- 3 }^{+ 3 }$    &  $ 25 _{- 9 }^{+ 10 }$  &  $ 24 _{- 6 }^{+ 7 }$    &  $ 2.5 _{- 2.2 }^{+ 2.6 }$    &  $ 58 _{- 18 }^{+ 20 }$    &  $ 3.3 _{- 2.5 }^{+ 3.1 }$    &  $ 21 _{- 11 }^{+ 14 }$    &  0  &0    \\
 &  &  & & 0.7  & 3.5  &  $0.1$   & $...$  & $0$  & $0.3$  & $0.8$  & $...$  & $...$  &  \cite{Peng:2024pyl} \\
$\Sigma_c(3567)$  & $ \mathbf{\frac{1}{2}^+}$ & $\vert \,0\,,\,0\,,\,0\,,\,1 \,\rangle $ &$^{2}S_{1/2}$&$ 47 _{- 5 }^{+ 4 }$    &  $ 6 _{- 1 }^{+ 1 }$  &  $ 19 _{- 12 }^{+ 18 }$    &  $ 0.1 _{- 0.0 }^{+ 0.0 }$    &  $ 112 _{- 61 }^{+ 85 }$    &  $ 0.2 _{- 0.1 }^{+ 0.2 }$    &  $ 4.3 _{- 2.8 }^{+ 4.3 }$    &  $ 58 _{- 18 }^{+ 16 }$    &  $ 158 _{- 54 }^{+ 52 }$     \\
$\Sigma_c(3637)$  & $ \mathbf{\frac{3}{2}^+}$ & $\vert \,0\,,\,0\,,\,0\,,\,1 \,\rangle $ &$^{4}S_{3/2}$&$ 12 _{- 1 }^{+ 1 }$    &  $ 41 _{- 5 }^{+ 3 }$    &  $ 14 _{- 7 }^{+ 9 }$    &  $ 0.3 _{- 0.1 }^{+ 0.0 }$    &  $ 35 _{- 19 }^{+ 24 }$    &  $ 3.5 _{- 1.4 }^{+ 1.8 }$    &  $ 67 _{- 38 }^{+ 56 }$    &  $ 67 _{- 16 }^{+ 10 }$    &  $ 189 _{- 52 }^{+ 37 }$      \\
$\Sigma_c(3358)$  & $ \mathbf{\frac{3}{2}^+}$ & $\vert \,1\,,\,1\,,\,0\,,\,0 \,\rangle $ &$^{2}D_{\lambda\rho,3/2}$&$ 22 _{- 7 }^{+ 7 }$    &  $ 9 _{- 3 }^{+ 3 }$  &  $ 3.0 _{- 0.8 }^{+ 0.8 }$    &  $ 10 _{- 3 }^{+ 4 }$    &  $ 2.6 _{- 0.5 }^{+ 0.4 }$    &  $ 3.4 _{- 1.4 }^{+ 1.9 }$    &  $ 0.2 _{- 0.1 }^{+ 0.2 }$    &  $ 1.5 _{- 0.8 }^{+ 1.1 }$    &  $ 1.3 _{- 0.8 }^{+ 1.1 }$    \\
$\Sigma_c(3403)$  & $ \mathbf{\frac{5}{2}^+}$ & $\vert \,1\,,\,1\,,\,0\,,\,0 \,\rangle $ &$^{2}D_{\lambda\rho,5/2}$&$ 25 _{- 7 }^{+ 6 }$    &  $ 11 _{- 3 }^{+ 3 }$  &  $ 8 _{- 2 }^{+ 2 }$    &  $ 2.1 _{- 1.0 }^{+ 1.4 }$    &  $ 4.8 _{- 1.1 }^{+ 1.0 }$    &  $ 5 _{- 2 }^{+ 2 }$    &  $ 3.1 _{- 1.1 }^{+ 1.1 }$    &  $ 0.6 _{- 0.4 }^{+ 0.6 }$    &  $ 4.6 _{- 2.5 }^{+ 3.6 }$  \\
$\Sigma_c(3367)$  & $ \mathbf{\frac{1}{2}^-}$ & $\vert \,1\,,\,1\,,\,0\,,\,0 \,\rangle $ &$^{2}P_{\lambda\rho,1/2}$&0  &0  &  0  &$ 14 _{- 5 }^{+ 7 }$    &  0  &$ 9 _{- 4 }^{+ 5 }$    &  $ 1.7 _{- 0.8 }^{+ 1.0 }$    &  $ 1.6 _{- 1.6 }^{+ 2.2 }$    &  $ 0.6 _{- 0.5 }^{+ 0.6 }$   \\
$\Sigma_c(3394)$  & $ \mathbf{\frac{3}{2}^-}$ & $\vert \,1\,,\,1\,,\,0\,,\,0 \,\rangle $ &$^{2}P_{\lambda\rho,3/2}$&0  &0 &  $ 13 _{- 4 }^{+ 5 }$    &  $ 5 _{- 2 }^{+ 2 }$    &  $ 6 _{- 2 }^{+ 2 }$    &  $ 6 _{- 2 }^{+ 3 }$    &  $ 2.0 _{- 0.9 }^{+ 1.1 }$    &  $ 0.1 _{- 0.1 }^{+ 0.2 }$    &  $ 2.4 _{- 2.0 }^{+ 2.4 }$   \\
$\Sigma_c(3385)$  & $ \mathbf{\frac{1}{2}^+}$ & $\vert \,1\,,\,1\,,\,0\,,\,0 \,\rangle $ &$^{2}S_{\lambda\rho,1/2}$&0  &0   &$ 2.1 _{- 0.4 }^{+ 0.2 }$    &  $ 1.1 _{- 0.1 }^{+ 0.1 }$    &  $ 4.2 _{- 0.5 }^{+ 0.3 }$    &  $ 0.2 _{- 0.0 }^{+ 0.0 }$    &  $ 1.7 _{- 0.4 }^{+ 0.3 }$    &  $ 2.6 _{- 2.3 }^{+ 2.9 }$    &  $ 3.5 _{- 2.0 }^{+ 2.5 }$  \\
$\Sigma_c(3540)$  & $ \mathbf{\frac{3}{2}^+}$ & $\vert \,0\,,\,2\,,\,0\,,\,0 \,\rangle $ &$^{2}D_{\rho\rho,3/2}$&$ 4.9 _{- 0.2 }^{+ 0.3 }$    &  $ 2.1 _{- 0.4 }^{+ 0.3 }$ &  $ 5 _{- 1 }^{+ 0 }$    &  $ 0.2 _{- 0.1 }^{+ 0.2 }$    &  $ 3.7 _{- 0.9 }^{+ 0.8 }$    &  0  &$ 0.2 _{- 0.1 }^{+ 0.2 }$    &  $ 8 _{- 3 }^{+ 4 }$    &  $ 9 _{- 4 }^{+ 4 }$     \\
$\Sigma_c(3585)$  & $ \mathbf{\frac{5}{2}^+}$ & $\vert \,0\,,\,2\,,\,0\,,\,0 \,\rangle $ &$^{2}D_{\rho\rho,5/2}$&$ 54 _{- 6 }^{+ 4 }$    &  $ 2.3 _{- 0.4 }^{+ 0.2 }$  &  $ 10 _{- 4 }^{+ 4 }$    &  $ 0.3 _{- 0.2 }^{+ 0.2 }$    &  $ 36 _{- 14 }^{+ 15 }$    &  0  &$ 0.3 _{- 0.2 }^{+ 0.3 }$    &  $ 19 _{- 7 }^{+ 7 }$    &  $ 14 _{- 5 }^{+ 5 }$    \\
$\Sigma_c(3555)$  & $ \mathbf{\frac{1}{2}^+}$ & $\vert \,0\,,\,2\,,\,0\,,\,0 \,\rangle $ &$^{4}D_{\rho\rho,1/2}$&$ 2.3 _{- 0.3 }^{+ 0.2 }$    &  $ 0.3 _{- 0.3 }^{+ 0.8 }$  &  $ 0.3 _{- 0.2 }^{+ 0.2 }$    &  $ 2.0 _{- 0.4 }^{+ 0.2 }$    &  $ 0.5 _{- 0.3 }^{+ 0.4 }$    &  $ 1.7 _{- 0.4 }^{+ 0.2 }$    &  $ 0.6 _{- 0.3 }^{+ 0.2 }$    &  $ 17 _{- 7 }^{+ 8 }$    &  0   \\
$\Sigma_c(3582)$  & $ \mathbf{\frac{3}{2}^+}$ & $\vert \,0\,,\,2\,,\,0\,,\,0 \,\rangle $ &$^{4}D_{\rho\rho,3/2}$&$ 4.7 _{- 0.5 }^{+ 0.3 }$    &  $ 9 _{- 1 }^{+ 1 }$  &  $ 0.7 _{- 0.4 }^{+ 0.5 }$    &  $ 3.1 _{- 0.7 }^{+ 0.6 }$    &  $ 1.2 _{- 0.7 }^{+ 1.0 }$    &  $ 0.1 _{- 0.0 }^{+ 0.0 }$    &  $ 3.5 _{- 1.5 }^{+ 1.8 }$    &  $ 42 _{- 16 }^{+ 17 }$    &  $ 2.4 _{- 1.0 }^{+ 1.3 }$    \\
$\Sigma_c(3627)$  & $ \mathbf{\frac{5}{2}^+}$ & $\vert \,0\,,\,2\,,\,0\,,\,0 \,\rangle $ &$^{4}D_{\rho\rho,5/2}$&$ 6 _{- 0 }^{+ 0 }$    &  $ 14 _{- 1 }^{+ 1 }$   &  $ 1.2 _{- 0.6 }^{+ 0.8 }$    &  $ 1.2 _{- 0.1 }^{+ 0.1 }$    &  $ 2.1 _{- 1.2 }^{+ 1.4 }$    &  $ 6 _{- 2 }^{+ 2 }$    &  $ 5 _{- 2 }^{+ 2 }$    &  $ 18 _{- 6 }^{+ 5 }$    &  $ 38 _{- 13 }^{+ 13 }$    \\
$\Sigma_c(3691)$  & $ \mathbf{\frac{7}{2}^+}$ & $\vert \,0\,,\,2\,,\,0\,,\,0 \,\rangle $ &$^{4}D_{\rho\rho,7/2}$&$ 4.1 _{- 0.3 }^{+ 0.2 }$    &  $ 53 _{- 7 }^{+ 6 }$  &  $ 1.1 _{- 0.5 }^{+ 0.6 }$    &  $ 11 _{- 5 }^{+ 5 }$    &  $ 1.9 _{- 0.9 }^{+ 1.1 }$    &  $ 8 _{- 3 }^{+ 2 }$    &  $ 33 _{- 13 }^{+ 14 }$    &  $ 1.0 _{- 0.6 }^{+ 0.9 }$    &  $ 48 _{- 15 }^{+ 12 }$   \\
\hline \hline
\end{tabular}

\endgroup
}
\end{center}
\label{sigmas+_EM}
\end{table*}

\begin{table*}[htp!]
\caption{Same as table \ref{sigmas+_EM}, but for transitions from second shell $\Sigma_c^{+}(snc)$ states with isospin 1  to $\Lambda_c^{+}$ states.}
\begin{center}
\scriptsize{
\begingroup
\setlength{\tabcolsep}{1.75pt} 
\renewcommand{\arraystretch}{1.35} 

\begin{tabular}{c c c c  p{1.0cm}  p{1.0cm}  p{1.0cm}  p{1.0cm}  p{1.0cm}  p{1.0cm}  p{1.0cm}  p{1.0cm}  p{1.0cm}  p{1.0cm}  p{1.0cm}} \hline \hline
&    &    &    & $\Lambda_{c}^{+} \gamma$ & $\Lambda_{c}^{+} \gamma$  & $\Lambda_{c}^{+} \gamma$  & $\Lambda_{c}^{+} \gamma$  & $\Lambda_{c}^{+} \gamma$  & $\Lambda_{c}^{+} \gamma$  & $\Lambda_{c}^{+} \gamma$  & $\Lambda_{c}^{+} \gamma$ \\
$\mathbf{\Sigma_c(nnc)}$  & $\mathbf{J^P}$  & $\vert l_{\lambda}, l_{\rho}, k_{\lambda}, k_{\rho} \rangle$  & $^{2S+1}L_{x,J}$  & $^2S_{1/2}$  & $^2P_{\lambda,1/2}$  & $^2P_{\lambda,3/2}$  & $^2P_{\rho,1/2}$  & $^4P_{\rho,1/2}$  & $^2P_{\rho,3/2}$  & $^4P_{\rho,3/2}$  & $^4P_{\rho,5/2}$  \\ \hline
 $N=2$  &  &  &  &  &  \\\ 
$\Sigma_c(3175)$  & $ \mathbf{\frac{3}{2}^+}$ & $\vert \,2\,,\,0\,,\,0\,,\,0 \,\rangle $ &$^{2}D_{\lambda\lambda,3/2}$& $ 111 _{- 36 }^{+ 36 }$    &  $79 _{- 23 }^{+ 25 }$    &  $ 82 _{- 23 }^{+ 25 }$    &  $ 11 _{- 5 }^{+ 7 }$    &  0  &$ 2.1 _{- 1.0 }^{+ 1.3 }$    &  0  &0 \\
  &  &  & & $...$  &  $17.53$  & $64.6$  & $...$ &  $...$   & $...$  & $...$  & $...$ & \cite{Yao:2018jmc} \\
$\Sigma_c(3220)$  & $ \mathbf{\frac{5}{2}^+}$ & $\vert \,2\,,\,0\,,\,0\,,\,0 \,\rangle $ &$^{2}D_{\lambda\lambda,5/2}$& $ 123 _{- 37 }^{+ 36 }$    &  $ 184 _{- 52 }^{+ 56 }$    &  $ 138 _{- 37 }^{+ 38 }$    &  $ 4.6 _{- 1.9 }^{+ 2.7 }$    &  0  &$ 25 _{- 13 }^{+ 19 }$    &  0  &0 \\
  &  &  & & $...$  & $25.36$  & $42.95$  & $...$ &  $...$   & $...$  & $...$  & $...$ & \cite{Yao:2018jmc} \\
$\Sigma_c(3190)$  & $ \mathbf{\frac{1}{2}^+}$ & $\vert \,2\,,\,0\,,\,0\,,\,0 \,\rangle $ &$^{4}D_{\lambda\lambda,1/2}$&  $ 57 _{- 19 }^{+ 20 }$    &  $ 161 _{- 54 }^{+ 58 }$    &  $ 0.1 _{- 0.1 }^{+ 0.1 }$    &  0  &$ 4.1 _{- 2.2 }^{+ 3.1 }$    &  $ 0.1 _{- 0.0 }^{+ 0.1 }$    &  $ 3.0 _{- 1.9 }^{+ 2.9 }$    &  $ 0.2 _{- 0.1 }^{+ 0.1 }$   \\
  &  &  &  & $...$  & $24.7$  & $21.2$  & $...$ &  $...$  & $...$  & $...$  & $...$ & \cite{Yao:2018jmc} \\
 &  &  &  & 11   & 38.7  & 19.7  & $...$ &  $...$  & $...$  & $...$  & $...$ & \cite{Peng:2024pyl} \\
$\Sigma_c(3217)$  & $ \mathbf{\frac{3}{2}^+}$ & $\vert \,2\,,\,0\,,\,0\,,\,0 \,\rangle $ &$^{4}D_{\lambda\lambda,3/2}$& $ 122 _{- 38 }^{+ 37 }$  &$ 410 _{- 123 }^{+ 134 }$    &  $ 24 _{- 8 }^{+ 10 }$    &  $ 0.1 _{- 0.1 }^{+ 0.2 }$    &  $ 6 _{- 3 }^{+ 4 }$    &  $ 0.2 _{- 0.1 }^{+ 0.3 }$    &  $ 1.0 _{- 0.6 }^{+ 0.9 }$    &  $ 1.3 _{- 0.9 }^{+ 1.5 }$   \\
  &  &  &  & $...$  & $88.8$  & $7.71$ & $...$ &  $...$   & $...$  & $...$  & $...$ & \cite{Yao:2018jmc} \\
$\Sigma_c(3262)$  & $ \mathbf{\frac{5}{2}^+}$ & $\vert \,2\,,\,0\,,\,0\,,\,0 \,\rangle $ &$^{4}D_{\lambda\lambda,5/2}$&  $ 180 _{- 51 }^{+ 48 }$ &$ 170 _{- 42 }^{+ 44 }$    &  $ 364 _{- 98 }^{+ 104 }$    &  $ 0.4 _{- 0.2 }^{+ 0.4 }$    &  $ 5 _{- 2 }^{+ 3 }$    &  $ 0.5 _{- 0.4 }^{+ 0.7 }$    &  $ 7 _{- 4 }^{+ 5 }$    &  $ 2.5 _{- 1.4 }^{+ 2.2 }$   \\
  &  &  & & $...$  & $31.3$  & $53.9$  & $...$ &  $...$   & $...$  & $...$  & $...$ & \cite{Yao:2018jmc} \\
$\Sigma_c(3326)$  & $ \mathbf{\frac{7}{2}^+}$ & $\vert \,2\,,\,0\,,\,0\,,\,0 \,\rangle $ &$^{4}D_{\lambda\lambda,7/2}$&  $ 124 _{- 32 }^{+ 27 }$  &$ 12 _{- 6 }^{+ 9 }$    &  $ 456 _{- 107 }^{+ 103 }$    &  $ 0.6 _{- 0.4 }^{+ 0.7 }$    &  $ 3.5 _{- 2.1 }^{+ 3.6 }$    &  $ 0.9 _{- 0.6 }^{+ 0.9 }$    &  $ 12 _{- 6 }^{+ 8 }$    &  $ 24 _{- 14 }^{+ 21 }$   \\
  &  &  &  & $...$  & $0.35$  & $45.4$  & $...$ &  $...$   & $...$  & $...$  & $...$ & \cite{Yao:2018jmc} \\
 &  &  & & 14.6  &  0.3  & 53  & $...$ &  $...$  & $...$  & $...$  & $...$ & \cite{Peng:2024pyl} \\
$\Sigma_c(3202)$  & $ \mathbf{\frac{1}{2}^+}$ & $\vert \,0\,,\,0\,,\,1\,,\,0 \,\rangle $ &$^{2}S_{1/2}$& $ 296 _{- 93 }^{+ 88 }$  &$ 542 _{- 116 }^{+ 106 }$    &  $ 1491 _{- 348 }^{+ 336 }$    &  $ 0.7 _{- 0.5 }^{+ 0.4 }$    &  0  &$ 29 _{- 19 }^{+ 32 }$    &  0  &$ 0.2 _{- 0.1 }^{+ 0.4 }$   \\
 &  &  & & 117.8  &  11.7  & 19.5  & $...$ &  $...$  & $...$  & $...$  & $...$ & \cite{Peng:2024pyl} \\
$\Sigma_c(3271)$  & $ \mathbf{\frac{3}{2}^+}$ & $\vert \,0\,,\,0\,,\,1\,,\,0 \,\rangle $ &$^{4}S_{3/2}$&  $ 336 _{- 95 }^{+ 86 }$  &$ 607 _{- 94 }^{+ 68 }$    &  $ 1738 _{- 317 }^{+ 251 }$    &  $ 4.9 _{- 3.1 }^{+ 5.9 }$    &  $ 5 _{- 2 }^{+ 3 }$    &  $ 10 _{- 6 }^{+ 12 }$    &  $ 8 _{- 4 }^{+ 6 }$    &  $ 14 _{- 10 }^{+ 20 }$   \\
 &  &  &  & 151.2  &  13.3  & 23.9  & $...$ &  $...$  & $...$  & $...$  & $...$ & \cite{Peng:2024pyl} \\
$\Sigma_c(3567)$  & $ \mathbf{\frac{1}{2}^+}$ & $\vert \,0\,,\,0\,,\,0\,,\,1 \,\rangle $ &$^{2}S_{1/2}$  &  $ 310 _{- 34 }^{+ 24 }$  &  $ 626 _{- 258 }^{+ 289 }$    &  $ 1672 _{- 731 }^{+ 805 }$    &  $ 2999 _{- 224 }^{+ 158 }$    &  $ 1.5 _{- 0.2 }^{+ 0.1 }$    &  $ 940 _{- 138 }^{+ 68 }$    &  $ 7 _{- 1 }^{+ 1 }$    &  $ 230 _{- 48 }^{+ 31 }$   \\
$\Sigma_c(3637)$  & $ \mathbf{\frac{3}{2}^+}$ & $\vert \,0\,,\,0\,,\,0\,,\,1 \,\rangle $ &$^{4}S_{3/2}$& $ 293 _{- 49 }^{+ 37 }$   &  $ 792 _{- 300 }^{+ 293 }$    &  $ 2164 _{- 865 }^{+ 893 }$    &  $ 183 _{- 35 }^{+ 19 }$    &  $ 448 _{- 54 }^{+ 40 }$    &  $ 568 _{- 89 }^{+ 42 }$    &  $ 570 _{- 54 }^{+ 39 }$    &  $ 696 _{- 127 }^{+ 61 }$   \\
$\Sigma_c(3358)$  & $ \mathbf{\frac{3}{2}^+}$ & $\vert \,1\,,\,1\,,\,0\,,\,0 \,\rangle $ &$^{2}D_{\lambda\rho,3/2}$&  $ 233 _{- 35 }^{+ 28 }$   &  $ 209 _{- 36 }^{+ 46 }$    &  $ 158 _{- 30 }^{+ 31 }$    &  $ 33 _{- 11 }^{+ 12 }$    &  $ 103 _{- 45 }^{+ 50 }$    &  $ 29 _{- 9 }^{+ 8 }$    &  $ 37 _{- 18 }^{+ 23 }$    &  $ 2.6 _{- 1.3 }^{+ 1.9 }$   \\
$\Sigma_c(3403)$  & $ \mathbf{\frac{5}{2}^+}$ & $\vert \,1\,,\,1\,,\,0\,,\,0 \,\rangle $ &$^{2}D_{\lambda\rho,5/2}$&  $ 209 _{- 38 }^{+ 31 }$  &  $ 73 _{- 21 }^{+ 20 }$    &  $ 290 _{- 26 }^{+ 33 }$    &  $ 84 _{- 28 }^{+ 31 }$    &  $ 22 _{- 11 }^{+ 16 }$    &  $ 53 _{- 16 }^{+ 16 }$    &  $ 57 _{- 24 }^{+ 28 }$    &  $ 35 _{- 14 }^{+ 16 }$   \\
$\Sigma_c(3367)$  & $ \mathbf{\frac{1}{2}^-}$ & $\vert \,1\,,\,1\,,\,0\,,\,0 \,\rangle $ &$^{2}P_{\lambda\rho,1/2}$  &$ 15 _{- 1 }^{+ 1 }$   &  $ 290 _{- 58 }^{+ 52 }$    &  $ 14 _{- 7 }^{+ 8 }$    &  0  &$ 144 _{- 64 }^{+ 76 }$    &  0  &$ 99 _{- 44 }^{+ 59 }$    &  $ 18 _{- 9 }^{+ 12 }$   \\
$\Sigma_c(3394)$  & $ \mathbf{\frac{3}{2}^-}$ & $\vert \,1\,,\,1\,,\,0\,,\,0 \,\rangle $ &$^{2}P_{\lambda\rho,3/2}$ &$ 14 _{- 1 }^{+ 1 }$   &  $ 71 _{- 12 }^{+ 11 }$    &  $ 224 _{- 50 }^{+ 48 }$    &  $ 135 _{- 51 }^{+ 56 }$    &  $ 55 _{- 23 }^{+ 30 }$    &  $ 60 _{- 23 }^{+ 26 }$    &  $ 68 _{- 30 }^{+ 37 }$    &  $ 22 _{- 11 }^{+ 13 }$   \\
$\Sigma_c(3385)$  & $ \mathbf{\frac{1}{2}^+}$ & $\vert \,1\,,\,1\,,\,0\,,\,0 \,\rangle $ &$^{2}S_{\lambda\rho,1/2}$ &0  &  $ 437 _{- 25 }^{+ 22 }$    &  $ 225 _{- 15 }^{+ 12 }$    &  $ 24 _{- 4 }^{+ 3 }$    &  $ 12 _{- 3 }^{+ 2 }$    &  $ 50 _{- 10 }^{+ 7 }$    &  $ 2.4 _{- 0.7 }^{+ 0.5 }$    &  $ 19 _{- 7 }^{+ 6 }$   \\
$\Sigma_c(3540)$  & $ \mathbf{\frac{3}{2}^+}$ & $\vert \,0\,,\,2\,,\,0\,,\,0 \,\rangle $ &$^{2}D_{\rho\rho,3/2}$  &  $ 125 _{- 12 }^{+ 8 }$   &  $ 39 _{- 17 }^{+ 18 }$    &  $ 70 _{- 32 }^{+ 36 }$    &  $ 185 _{- 70 }^{+ 93 }$    &  $ 137 _{- 37 }^{+ 31 }$    &  $ 97 _{- 20 }^{+ 16 }$    &  $ 51 _{- 18 }^{+ 18 }$    &  $ 4.2 _{- 1.5 }^{+ 1.8 }$   \\
$\Sigma_c(3585)$  & $ \mathbf{\frac{5}{2}^+}$ & $\vert \,0\,,\,2\,,\,0\,,\,0 \,\rangle $ &$^{2}D_{\rho\rho,5/2}$&  $ 122 _{- 16 }^{+ 11 }$   &  $ 46 _{- 19 }^{+ 20 }$    &  $ 84 _{- 37 }^{+ 39 }$    &  $ 626 _{- 132 }^{+ 100 }$    &  $ 22 _{- 7 }^{+ 8 }$    &  $ 1207 _{- 220 }^{+ 185 }$    &  $ 70 _{- 19 }^{+ 16 }$    &  $ 51 _{- 14 }^{+ 11 }$   \\
$\Sigma_c(3555)$  & $ \mathbf{\frac{1}{2}^+}$ & $\vert \,0\,,\,2\,,\,0\,,\,0 \,\rangle $ &$^{4}D_{\rho\rho,1/2}$&  $ 62 _{- 8 }^{+ 5 }$    &  $ 21 _{- 10 }^{+ 10 }$    &  $ 37 _{- 18 }^{+ 20 }$    &  $ 80 _{- 17 }^{+ 12 }$    &  $ 13 _{- 13 }^{+ 48 }$    &  $ 0.4 _{- 0.4 }^{+ 0.9 }$    &  $ 185 _{- 12 }^{+ 10 }$    &  $ 23 _{- 3 }^{+ 2 }$   \\
$\Sigma_c(3582)$  & $ \mathbf{\frac{3}{2}^+}$ & $\vert \,0\,,\,2\,,\,0\,,\,0 \,\rangle $ &$^{4}D_{\rho\rho,3/2}$&  $ 122 _{- 17 }^{+ 11 }$   &  $ 46 _{- 20 }^{+ 21 }$    &  $ 83 _{- 37 }^{+ 40 }$    &  $ 194 _{- 36 }^{+ 24 }$    &  $ 588 _{- 34 }^{+ 34 }$    &  $ 15 _{- 4 }^{+ 5 }$    &  $ 6 _{- 6 }^{+ 20 }$    &  $ 54 _{- 9 }^{+ 8 }$   \\
$\Sigma_c(3627)$  & $ \mathbf{\frac{5}{2}^+}$ & $\vert \,0\,,\,2\,,\,0\,,\,0 \,\rangle $ &$^{4}D_{\rho\rho,5/2}$& $ 160 _{- 26 }^{+ 19 }$   &  $ 72 _{- 28 }^{+ 28 }$    &  $ 133 _{- 54 }^{+ 57 }$    &  $ 72 _{- 10 }^{+ 6 }$    &  $ 289 _{- 47 }^{+ 27 }$    &  $ 163 _{- 25 }^{+ 18 }$    &  $ 882 _{- 78 }^{+ 70 }$    &  $ 19 _{- 3 }^{+ 10 }$   \\
$\Sigma_c(3691)$  & $ \mathbf{\frac{7}{2}^+}$ & $\vert \,0\,,\,2\,,\,0\,,\,0 \,\rangle $ &$^{4}D_{\rho\rho,7/2}$&  $ 94 _{- 20 }^{+ 17 }$    &   $ 54 _{- 19 }^{+ 17 }$    &  $ 101 _{- 37 }^{+ 35 }$    &  $ 9 _{- 4 }^{+ 4 }$    &  $ 75 _{- 28 }^{+ 31 }$    &  $ 176 _{- 18 }^{+ 11 }$    &  $ 489 _{- 101 }^{+ 76 }$    &  $ 1598 _{- 283 }^{+ 221 }$   \\
\hline \hline
\end{tabular}

\endgroup
}
\end{center}
\label{sigmas_lambdas_EM}
\end{table*}

\begin{table*}[h!tp]
\caption{Same as table \ref{sigmas+_EM}, but for $\Sigma_{c}^{++}$ with isospin 1.}
\begin{center}
\scriptsize{
\begingroup
\setlength{\tabcolsep}{1.75pt} 
\renewcommand{\arraystretch}{1.35} 

\begin{tabular}{c c c c  p{1.0cm}  p{1.0cm}  p{1.0cm}  p{1.0cm}  p{1.0cm}  p{1.0cm}  p{1.0cm}  p{1.0cm}  p{1.0cm}  p{1.0cm}} \hline \hline
&    &    &    & $\Sigma_{c}^{++} \gamma$  & $\Sigma_{c}^{*++} \gamma$   & $\Sigma_{c}^{++} \gamma$  & $\Sigma_{c}^{++} \gamma$  & $\Sigma_{c}^{++} \gamma$  & $\Sigma_{c}^{++} \gamma$  & $\Sigma_{c}^{++} \gamma$  & $\Sigma_{c}^{++} \gamma$  & $\Sigma_{c}^{++} \gamma$  &  \\
$\mathbf{\Sigma_c(nnc)}$  & $\mathbf{J^P}$  & $\vert l_{\lambda}, l_{\rho}, k_{\lambda}, k_{\rho} \rangle$  & $^{2S+1}L_{x,J}$  & $^2S_{1/2}$  & $^4S_{3/2}$  & $^2P_{\lambda,1/2}$  & $^4P_{\lambda,1/2}$  & $^2P_{\lambda,3/2}$  & $^4P_{\lambda,3/2}$  & $^4P_{\lambda,5/2}$  & $^2P_{\rho,1/2}$  & $^2P_{\rho,3/2}$  &   \\ \hline
 $N=2$  &  &  &  &  &  \\\ 
$\Sigma_c(3175)$  & $ \mathbf{\frac{3}{2}^+}$ & $\vert \,2\,,\,0\,,\,0\,,\,0 \,\rangle $ &$^{2}D_{\lambda\lambda,3/2}$&$ 87 _{- 5 }^{+ 6 }$    &  $ 11 _{- 5 }^{+ 6 }$  &  $ 421 _{- 81 }^{+ 86 }$    &  $ 31 _{- 16 }^{+ 24 }$    &  $ 36 _{- 9 }^{+ 8 }$    &  $ 7 _{- 4 }^{+ 5 }$    &  $ 0.4 _{- 0.3 }^{+ 0.4 }$    &  0  &0  &  \\
  &  &  & & $...$  & $...$ &  $231.9$ & $5.09$  & $78.5$  & $11.9$  & $2.66$ & $...$  & $...$  &  \cite{Yao:2018jmc} \\
$\Sigma_c(3220)$  & $ \mathbf{\frac{5}{2}^+}$ & $\vert \,2\,,\,0\,,\,0\,,\,0 \,\rangle $ &$^{2}D_{\lambda\lambda,5/2}$&$ 715 _{- 158 }^{+ 162 }$    &  $ 14 _{- 6 }^{+ 8 }$  &  $ 215 _{- 88 }^{+ 110 }$    &  $ 4.4 _{- 2.2 }^{+ 3.2 }$    &  $ 1280 _{- 270 }^{+ 252 }$    &  $ 18 _{- 8 }^{+ 12 }$    &  $ 10 _{- 5 }^{+ 7 }$    &  0  &0  &  \\
  &  &  & & $...$  & $...$ & $58.4$ & $0.29$  & $34.9$  & $4.73$  & $12.93$ & $...$  & $...$  &  \cite{Yao:2018jmc} \\
$\Sigma_c(3190)$  & $ \mathbf{\frac{1}{2}^+}$ & $\vert \,2\,,\,0\,,\,0\,,\,0 \,\rangle $ &$^{4}D_{\lambda\lambda,1/2}$&$ 17 _{- 8 }^{+ 9 }$    &  $ 20 _{- 14 }^{+ 14 }$  &  $ 33 _{- 16 }^{+ 23 }$    &  $ 255 _{- 123 }^{+ 123 }$    &  $ 0.2 _{- 0.1 }^{+ 0.0 }$    &  $ 78 _{- 25 }^{+ 30 }$    &  $ 3.6 _{- 2.4 }^{+ 4.1 }$    &  0  &0  & \\
  &  &  & & $...$  & $...$ & $15.01$ & $136.5$  & $1.93$  & $94.6$  & $10.1$ & $...$  & $...$  &   \cite{Yao:2018jmc} \\
 &  &  & & 2.5  & 562.5 & $254.1$  & $...$  & $101.8$  & $46$  & $5.6$  & $...$  & -  &  \cite{Peng:2024pyl} \\
$\Sigma_c(3217)$  & $ \mathbf{\frac{3}{2}^+}$ & $\vert \,2\,,\,0\,,\,0\,,\,0 \,\rangle $ &$^{4}D_{\lambda\lambda,3/2}$&$ 38 _{- 16 }^{+ 19 }$    &  $ 90 _{- 18 }^{+ 23 }$  &  $ 95 _{- 40 }^{+ 50 }$    &  $ 398 _{- 90 }^{+ 91 }$    &  $ 4.7 _{- 2.1 }^{+ 3.1 }$    &  $ 119 _{- 45 }^{+ 45 }$    &  $ 15 _{- 5 }^{+ 6 }$    &  0  &0  &  \\
  &  &  & & $...$  & $...$ & $23.6$ & $52.7$  & $0.59$  & $121.8$  & $32.3$ & $...$  & $...$  &  \cite{Yao:2018jmc} \\
$\Sigma_c(3262)$  & $ \mathbf{\frac{5}{2}^+}$ & $\vert \,2\,,\,0\,,\,0\,,\,0 \,\rangle $ &$^{4}D_{\lambda\lambda,5/2}$&$ 61 _{- 25 }^{+ 27 }$    &  $ 183 _{- 34 }^{+ 38 }$  &  $ 49 _{- 18 }^{+ 22 }$    &  $ 104 _{- 49 }^{+ 59 }$    &  $ 93 _{- 36 }^{+ 44 }$    &  $ 759 _{- 165 }^{+ 155 }$    &  $ 61 _{- 21 }^{+ 22 }$    &  0  &0  &  \\
  &  &  & & $...$  & $...$ & $7.4$ & $9.79$  & $4.9$  & $48.5$  & $49.4$ & $...$  & $...$  &  \cite{Yao:2018jmc} \\
$\Sigma_c(3326)$  & $ \mathbf{\frac{7}{2}^+}$ & $\vert \,2\,,\,0\,,\,0\,,\,0 \,\rangle $ &$^{4}D_{\lambda\lambda,7/2}$&$ 47 _{- 18 }^{+ 19 }$    &  $ 800 _{- 174 }^{+ 164 }$ &  $ 1.0 _{- 0.7 }^{+ 1.1 }$    &  $ 10 _{- 6 }^{+ 10 }$    &  $ 152 _{- 55 }^{+ 68 }$    &  $ 164 _{- 78 }^{+ 95 }$    &  $ 1509 _{- 357 }^{+ 334 }$    &  $ 0.1 _{- 0.1 }^{+ 0.2 }$    &  $ 0.1 _{- 0.1 }^{+ 0.3 }$    &  \\
  &  &  & & $...$  & $...$ & $0.017$ & $1.3$  & $3.94$  & $13.5$  & $23.3$ & $...$  & $...$  &   \cite{Yao:2018jmc} \\
 &  &  & & 3.2  & 186.8 & $0.3$  & $...$  & $6.2$  & $2.4$  & $221.1$  & $...$  & $...$  &  \cite{Peng:2024pyl} \\
$\Sigma_c(3202)$  & $ \mathbf{\frac{1}{2}^+}$ & $\vert \,0\,,\,0\,,\,1\,,\,0 \,\rangle $ &$^{2}S_{1/2}$&$ 376 _{- 147 }^{+ 177 }$    &  $ 32 _{- 14 }^{+ 17 }$   &  $ 1979 _{- 606 }^{+ 696 }$    &  $ 0.7 _{- 0.4 }^{+ 0.4 }$    &  $ 381 _{- 206 }^{+ 289 }$    &  $ 2.4 _{- 1.2 }^{+ 1.5 }$    &  $ 48 _{- 25 }^{+ 36 }$    &  $ 0.1 _{- 0.1 }^{+ 0.1 }$    &  $ 0.1 _{- 0.1 }^{+ 0.2 }$    &  \\
 &  &  & & 55.6  & 5.8 & $29.6$  & $...$  & $12.4$  & $145.4$  & $1.2$  & $...$  & $...$  & \cite{Peng:2024pyl} \\
$\Sigma_c(3271)$  & $ \mathbf{\frac{3}{2}^+}$ & $\vert \,0\,,\,0\,,\,1\,,\,0 \,\rangle $ &$^{4}S_{3/2}$&$ 116 _{- 48 }^{+ 51 }$    &  $ 357 _{- 140 }^{+ 163 }$  &  $ 239 _{- 72 }^{+ 85 }$    &  $ 327 _{- 85 }^{+ 88 }$    &  $ 594 _{- 210 }^{+ 246 }$    &  $ 386 _{- 100 }^{+ 101 }$    &  $ 184 _{- 123 }^{+ 192 }$    &  $ 0.5 _{- 0.5 }^{+ 1.1 }$    &  $ 0.7 _{- 0.6 }^{+ 1.7 }$    &    \\
 &  &  & & 24.3  & 50.9  & $10.6$  & $...$  & $40.7$  & $23$  & $162$  & $...$  & $...$ & \cite{Peng:2024pyl} \\
$\Sigma_c(3567)$  & $ \mathbf{\frac{1}{2}^+}$ & $\vert \,0\,,\,0\,,\,0\,,\,1 \,\rangle $ &$^{2}S_{1/2}$&$ 749 _{- 91 }^{+ 58 }$    &  $ 89 _{- 15 }^{+ 11 }$  &  $ 304 _{- 195 }^{+ 288 }$    &  $ 0.8 _{- 0.5 }^{+ 0.7 }$    &  $ 1785 _{- 961 }^{+ 1333 }$    &  $ 2.9 _{- 1.7 }^{+ 2.5 }$    &  $ 70 _{- 43 }^{+ 68 }$    &  $ 931 _{- 270 }^{+ 242 }$    &  $ 2539 _{- 864 }^{+ 801 }$    &   \\
$\Sigma_c(3637)$  & $ \mathbf{\frac{3}{2}^+}$ & $\vert \,0\,,\,0\,,\,0\,,\,1 \,\rangle $ &$^{4}S_{3/2}$&$ 192 _{- 13 }^{+ 11 }$    &  $ 660 _{- 80 }^{+ 50 }$  &  $ 219 _{- 116 }^{+ 141 }$    &  $ 5 _{- 1 }^{+ 1 }$    &  $ 558 _{- 303 }^{+ 402 }$    &  $ 56 _{- 23 }^{+ 28 }$    &  $ 1064 _{- 616 }^{+ 835 }$    &  $ 1073 _{- 240 }^{+ 158 }$    &  $ 3032 _{- 800 }^{+ 586 }$    &   \\
$\Sigma_c(3358)$  & $ \mathbf{\frac{3}{2}^+}$ & $\vert \,1\,,\,1\,,\,0\,,\,0 \,\rangle $ &$^{2}D_{\lambda\rho,3/2}$&$ 360 _{- 111 }^{+ 105 }$    &  $ 149 _{- 50 }^{+ 54 }$  &  $ 48 _{- 12 }^{+ 13 }$    &  $ 151 _{- 54 }^{+ 59 }$    &  $ 42 _{- 9 }^{+ 6 }$    &  $ 55 _{- 23 }^{+ 30 }$    &  $ 3.8 _{- 1.8 }^{+ 2.9 }$    &  $ 286 _{- 49 }^{+ 47 }$    &  $ 78 _{- 21 }^{+ 24 }$    &   \\
$\Sigma_c(3403)$  & $ \mathbf{\frac{5}{2}^+}$ & $\vert \,1\,,\,1\,,\,0\,,\,0 \,\rangle $ &$^{2}D_{\lambda\rho,5/2}$&$ 400 _{- 113 }^{+ 103 }$    &  $ 172 _{- 55 }^{+ 52 }$ &  $ 127 _{- 36 }^{+ 36 }$    &  $ 34 _{- 16 }^{+ 21 }$    &  $ 77 _{- 17 }^{+ 14 }$    &  $ 84 _{- 30 }^{+ 32 }$    &  $ 50 _{- 17 }^{+ 17 }$    &  $ 18 _{- 9 }^{+ 14 }$    &  $ 324 _{- 60 }^{+ 56 }$    &   \\
$\Sigma_c(3367)$  & $ \mathbf{\frac{1}{2}^-}$ & $\vert \,1\,,\,1\,,\,0\,,\,0 \,\rangle $ &$^{2}P_{\lambda\rho,1/2}$&0  &0  &0  &$ 216 _{- 84 }^{+ 105 }$    &  0  &$ 147 _{- 57 }^{+ 72 }$    &  $ 27 _{- 12 }^{+ 17 }$    &  $ 411 _{- 96 }^{+ 86 }$    &  $ 68 _{- 20 }^{+ 20 }$    &   \\
$\Sigma_c(3394)$  & $ \mathbf{\frac{3}{2}^-}$ & $\vert \,1\,,\,1\,,\,0\,,\,0 \,\rangle $ &$^{2}P_{\lambda\rho,3/2}$&0  &0  &$ 206 _{- 66 }^{+ 74 }$    &  $ 84 _{- 30 }^{+ 36 }$    &  $ 91 _{- 32 }^{+ 34 }$    &  $ 101 _{- 38 }^{+ 47 }$    &  $ 33 _{- 13 }^{+ 17 }$    &  $ 94 _{- 24 }^{+ 20 }$    &  $ 387 _{- 91 }^{+ 86 }$    &   \\
$\Sigma_c(3385)$  & $ \mathbf{\frac{1}{2}^+}$ & $\vert \,1\,,\,1\,,\,0\,,\,0 \,\rangle $ &$^{2}S_{\lambda\rho,1/2}$&0  &0 &$ 33 _{- 6 }^{+ 3 }$    &  $ 17 _{- 2 }^{+ 1 }$    &  $ 68 _{- 10 }^{+ 5 }$    &  $ 3.3 _{- 0.5 }^{+ 0.3 }$    &  $ 27 _{- 7 }^{+ 5 }$    &  $ 348 _{- 107 }^{+ 108 }$    &  $ 184 _{- 58 }^{+ 59 }$    &   \\
$\Sigma_c(3540)$ & $ \mathbf{\frac{3}{2}^+}$ & $\vert \,0\,,\,2\,,\,0\,,\,0 \,\rangle $ &$^{2}D_{\rho\rho,3/2}$ & $ 78 _{- 4 }^{+ 5 }$    &  $ 34 _{- 6 }^{+ 5 }$  &  $ 86 _{- 13 }^{+ 8 }$    &  $ 3.0 _{- 1.8 }^{+ 2.8 }$    &  $ 59 _{- 15 }^{+ 13 }$    &  $ 0.5 _{- 0.3 }^{+ 0.4 }$    &  $ 2.9 _{- 1.9 }^{+ 3.4 }$    &  $ 129 _{- 50 }^{+ 54 }$    &  $ 138 _{- 55 }^{+ 57 }$    &    \\
$\Sigma_c(3585)$  & $ \mathbf{\frac{5}{2}^+}$ & $\vert \,0\,,\,2\,,\,0\,,\,0 \,\rangle $ &$^{2}D_{\rho\rho,5/2}$&$ 862 _{- 93 }^{+ 65 }$    &  $ 36 _{- 6 }^{+ 4 }$ &  $ 159 _{- 58 }^{+ 68 }$    &  $ 4.3 _{- 2.5 }^{+ 3.6 }$    &  $ 573 _{- 220 }^{+ 237 }$    &  $ 0.7 _{- 0.4 }^{+ 0.6 }$    &  $ 4.5 _{- 2.7 }^{+ 4.5 }$    &  $ 300 _{- 106 }^{+ 116 }$    &  $ 228 _{- 83 }^{+ 82 }$    &    \\
$\Sigma_c(3555)$  & $ \mathbf{\frac{1}{2}^+}$ & $\vert \,0\,,\,2\,,\,0\,,\,0 \,\rangle $ &$^{4}D_{\rho\rho,1/2}$&$ 37 _{- 5 }^{+ 3 }$    &  $ 4.6 _{- 4.6 }^{+ 13.3 }$  &  $ 4.6 _{- 2.7 }^{+ 4.0 }$    &  $ 32 _{- 6 }^{+ 3 }$    &  $ 8 _{- 5 }^{+ 7 }$    &  $ 27 _{- 6 }^{+ 4 }$    &  $ 10 _{- 5 }^{+ 2 }$    &  $ 264 _{- 108 }^{+ 124 }$    &  $ 0.3 _{- 0.3 }^{+ 0.3 }$    &  \\
$\Sigma_c(3582)$  & $ \mathbf{\frac{3}{2}^+}$ & $\vert \,0\,,\,2\,,\,0\,,\,0 \,\rangle $ &$^{4}D_{\rho\rho,3/2}$&$ 75 _{- 9 }^{+ 5 }$    &  $ 148 _{- 15 }^{+ 13 }$  &  $ 11 _{- 6 }^{+ 8 }$    &  $ 49 _{- 12 }^{+ 10 }$    &  $ 18 _{- 11 }^{+ 15 }$    &  $ 0.9 _{- 0.1 }^{+ 0.2 }$    &  $ 56 _{- 24 }^{+ 29 }$    &  $ 673 _{- 253 }^{+ 263 }$    &  $ 38 _{- 16 }^{+ 20 }$    &    \\
$\Sigma_c(3627)$  & $ \mathbf{\frac{5}{2}^+}$ & $\vert \,0\,,\,2\,,\,0\,,\,0 \,\rangle $ &$^{4}D_{\rho\rho,5/2}$&$ 104 _{- 8 }^{+ 6 }$    &  $ 229 _{- 13 }^{+ 12 }$ &  $ 19 _{- 10 }^{+ 13 }$    &  $ 19 _{- 2 }^{+ 2 }$    &  $ 34 _{- 18 }^{+ 24 }$    &  $ 104 _{- 34 }^{+ 33 }$    &  $ 85 _{- 33 }^{+ 39 }$    &  $ 282 _{- 92 }^{+ 83 }$    &  $ 600 _{- 213 }^{+ 214 }$    &   \\
$\Sigma_c(3691)$  & $ \mathbf{\frac{7}{2}^+}$ & $\vert \,0\,,\,2\,,\,0\,,\,0 \,\rangle $ &$^{4}D_{\rho\rho,7/2}$&$ 66 _{- 5 }^{+ 4 }$    &  $ 842 _{- 122 }^{+ 88 }$ &  $ 17 _{- 8 }^{+ 10 }$    &  $ 178 _{- 75 }^{+ 79 }$    &  $ 30 _{- 15 }^{+ 18 }$    &  $ 135 _{- 41 }^{+ 37 }$    &  $ 523 _{- 207 }^{+ 230 }$    &  $ 16 _{- 9 }^{+ 13 }$    &  $ 761 _{- 228 }^{+ 199 }$    &   \\
\hline \hline
\end{tabular}

\endgroup
}
\end{center}
\label{sigmas_charged_++_EM}
\end{table*}

\begin{table*}[h!tp]
\caption{Same as table \ref{sigmas+_EM}, but for $\Sigma_c^{0}$ states with isospin 1.}
\begin{center}
\scriptsize{
\begingroup
\setlength{\tabcolsep}{1.75pt} 
\renewcommand{\arraystretch}{1.35} 

\begin{tabular}{c c c c   p{1.0cm}  p{1.0cm}  p{1.0cm}  p{1.0cm}  p{1.0cm}  p{1.0cm}  p{1.0cm}  p{1.0cm}  p{1.0cm}  p{1.0cm}  p{1.0cm}  p{1.0cm}} \hline \hline
&    &    &     & $\Sigma_{c}^{0} \gamma$  & $\Sigma_{c}^{*0} \gamma$  & $\Sigma_{c}^{0} \gamma$  & $\Sigma_{c}^{0} \gamma$  & $\Sigma_{c}^{0} \gamma$  & $\Sigma_{c}^{0} \gamma$  & $\Sigma_{c}^{0} \gamma$  & $\Sigma_{c}^{0} \gamma$  & $\Sigma_{c}^{0} \gamma$ \\
$\mathbf{\Sigma_c(nnc)}$  & $\mathbf{J^P}$  & $\vert l_{\lambda}, l_{\rho}, k_{\lambda}, k_{\rho} \rangle$  & $^{2S+1}L_{x,J}$  &  $^2S_{1/2}$  & $^4S_{3/2}$  & $^2P_{\lambda,1/2}$&$^4P_{\lambda,1/2}$  & $^2P_{\lambda,3/2}$  & $^4P_{\lambda,3/2}$  & $^4P_{\lambda,5/2}$  & $^2P_{\rho,1/2}$  & $^2P_{\rho,3/2}$  \\ \hline
 $N=2$  &  &  &  &  &  \\\ 
$\Sigma_c(3175)$  & $ \mathbf{\frac{3}{2}^+}$ & $\vert \,2\,,\,0\,,\,0\,,\,0 \,\rangle $ &$^{2}D_{\lambda\lambda,3/2}$&  $ 10 _{- 3 }^{+ 4 }$    &  $ 3.3 _{- 1.3 }^{+ 1.6 }$    &  $ 496 _{- 35 }^{+ 30 }$    &  $ 4.9 _{- 2.7 }^{+ 3.9 }$    &  $ 48 _{- 6 }^{+ 5 }$    &  $ 1.1 _{- 0.6 }^{+ 1.0 }$    &  $ 0.1 _{- 0.0 }^{+ 0.1 }$    &  0  &0 \\
  &  &  &  & $...$  & $...$ &  $164.2$  & $0.6$  & 38.6 &  1.42  & 0.33  & $...$  & $...$ &  \cite{Yao:2018jmc} \\
$\Sigma_c(3220)$  & $ \mathbf{\frac{5}{2}^+}$ & $\vert \,2\,,\,0\,,\,0\,,\,0 \,\rangle $ &$^{2}D_{\lambda\lambda,5/2}$&  $ 120 _{- 43 }^{+ 41 }$    &  $ 4.4 _{- 1.6 }^{+ 2.0 }$    &   $ 50 _{- 21 }^{+ 28 }$    &  $ 0.7 _{- 0.4 }^{+ 0.6 }$    &  $ 937 _{- 88 }^{+ 73 }$    &  $ 2.8 _{- 1.5 }^{+ 2.0 }$    &  $ 1.6 _{- 0.9 }^{+ 1.4 }$    &  0  &0 \\
  &  &  &  & $...$  & $...$ & $-$  & $-$  & - &  -  & -  & $...$  & $...$ &  \cite{Yao:2018jmc} \\
$\Sigma_c(3190)$  & $ \mathbf{\frac{1}{2}^+}$ & $\vert \,2\,,\,0\,,\,0\,,\,0 \,\rangle $ &$^{4}D_{\lambda\lambda,1/2}$&     $ 5 _{- 2 }^{+ 3 }$    &  $ 0.3 _{- 0.3 }^{+ 0.8 }$    &  $ 5 _{- 3 }^{+ 4 }$    &  $ 358 _{- 77 }^{+ 55 }$    &  0  &$ 49 _{- 10 }^{+ 11 }$    &  $ 1.1 _{- 0.7 }^{+ 1.3 }$    &  0  &0 \\
  &  &  &   & $...$  & $...$ &  $0.45$  & $94.3$  & 0.24 &  55.9  & 6.72  & $...$  & $...$ &  \cite{Yao:2018jmc} \\
 &  &  & & 0.8  &  76.6 &  165.4  & $...$ &  71  & 23.6  & $1.7$  & $...$ & $...$ & \cite{Peng:2024pyl} \\
$\Sigma_c(3217)$  & $ \mathbf{\frac{3}{2}^+}$ & $\vert \,2\,,\,0\,,\,0\,,\,0 \,\rangle $ &$^{4}D_{\lambda\lambda,3/2}$ &   $ 12 _{- 4 }^{+ 5 }$    &  $ 15 _{- 6 }^{+ 8 }$    &  $ 14 _{- 7 }^{+ 10 }$    &  $ 314 _{- 22 }^{+ 22 }$    &  $ 0.7 _{- 0.4 }^{+ 0.6 }$    &  $ 163 _{- 29 }^{+ 20 }$    &  $ 10 _{- 2 }^{+ 2 }$    &  0  &0 \\
  &  &  &  & $...$  & $...$ &  $2.71$  & $38.61$  & 0.08 &  84.22  & 19.6  & $...$  & $...$ &  \cite{Yao:2018jmc} \\
$\Sigma_c(3262)$  & $ \mathbf{\frac{5}{2}^+}$ & $\vert \,2\,,\,0\,,\,0\,,\,0 \,\rangle $ &$^{4}D_{\lambda\lambda,5/2}$&      $ 20 _{- 6 }^{+ 7 }$    &  $ 28 _{- 11 }^{+ 11 }$    &   $ 7 _{- 3 }^{+ 4 }$    &  $ 32 _{- 14 }^{+ 19 }$    &  $ 14 _{- 6 }^{+ 9 }$    &  $ 601 _{- 50 }^{+ 44 }$    &  $ 83 _{- 12 }^{+ 9 }$    &  0  &0 \\
  &  &  &  & $...$  & $...$ &  $0.87$  & $5.64$  & 0.6 & 44.2  & 35  & $...$  & $...$ &  \cite{Yao:2018jmc} \\
$\Sigma_c(3326)$  & $ \mathbf{\frac{7}{2}^+}$ & $\vert \,2\,,\,0\,,\,0\,,\,0 \,\rangle $ &$^{4}D_{\lambda\lambda,7/2}$&     $ 16 _{- 5 }^{+ 5 }$    &  $ 120 _{- 44 }^{+ 42 }$    &  $ 0.2 _{- 0.2 }^{+ 0.3 }$    &  $ 2.8 _{- 1.6 }^{+ 2.8 }$    &  $ 22 _{- 10 }^{+ 12 }$    &  $ 51 _{- 20 }^{+ 29 }$    &  $ 1096 _{- 129 }^{+ 120 }$    &  0  &0 \\
  &  &  & &  $...$  & $...$ & $0.004$  & $1$  & 0.49 &  8.22  & 30.5  & $...$  & $...$ &  \cite{Yao:2018jmc} \\
 &  &  & &   1.1  &  19.5 &  0.1  & $...$ & 1.6  & 0.2  & $213.3$ & $...$ & $...$ & \cite{Peng:2024pyl} \\
$\Sigma_c(3202)$  & $ \mathbf{\frac{1}{2}^+}$ & $\vert \,0\,,\,0\,,\,1\,,\,0 \,\rangle $ &$^{2}S_{1/2}$&    $ 100 _{- 38 }^{+ 44 }$    &  $ 10 _{- 4 }^{+ 5 }$    &  $ 870 _{- 172 }^{+ 198 }$    &  $ 0.1 _{- 0.1 }^{+ 0.1 }$    &  $ 13 _{- 13 }^{+ 32 }$    &  $ 0.4 _{- 0.2 }^{+ 0.3 }$    &  $ 8 _{- 4 }^{+ 6 }$    &  0  &0 \\
 &  &  & &    14.6  &  0.4 &  28.5  & $...$ &  11.5  & 103.0  & $0.2$  & $...$ & $...$ & \cite{Peng:2024pyl} \\
$\Sigma_c(3271)$  & $ \mathbf{\frac{3}{2}^+}$ & $\vert \,0\,,\,0\,,\,1\,,\,0 \,\rangle $ &$^{4}S_{3/2}$&    $ 38 _{- 12 }^{+ 13 }$    &  $ 78 _{- 35 }^{+ 40 }$    &   $ 33 _{- 14 }^{+ 16 }$    &  $ 225 _{- 29 }^{+ 28 }$    &  $ 84 _{- 38 }^{+ 45 }$    &  $ 271 _{- 33 }^{+ 33 }$    &  $ 22 _{- 18 }^{+ 36 }$    &  $ 0.1 _{- 0.1 }^{+ 0.3 }$    &  $ 0.2 _{- 0.2 }^{+ 0.4 }$   \\
 &  &  & &   10.8  &  11.5 &  8.9  & $...$ &  39.3  & 14.4  & $121.1$  & $...$ & $...$ & \cite{Peng:2024pyl} \\
$\Sigma_c(3567)$  & $ \mathbf{\frac{1}{2}^+}$ & $\vert \,0\,,\,0\,,\,0\,,\,1 \,\rangle $ &$^{2}S_{1/2}$&     $ 187 _{- 22 }^{+ 14 }$    &  $ 22 _{- 4 }^{+ 3 }$    &  $ 75 _{- 48 }^{+ 68 }$    &  $ 0.2 _{- 0.1 }^{+ 0.2 }$    &  $ 448 _{- 248 }^{+ 320 }$    &  $ 0.7 _{- 0.4 }^{+ 0.6 }$    &  $ 17 _{- 11 }^{+ 18 }$    &  $ 233 _{- 72 }^{+ 60 }$    &  $ 633 _{- 222 }^{+ 202 }$   \\
$\Sigma_c(3637)$  & $ \mathbf{\frac{3}{2}^+}$ & $\vert \,0\,,\,0\,,\,0\,,\,1 \,\rangle $ &$^{4}S_{3/2}$&   $ 48 _{- 3 }^{+ 3 }$    &  $ 165 _{- 20 }^{+ 12 }$    &  $ 55 _{- 28 }^{+ 35 }$    &  $ 1.3 _{- 0.3 }^{+ 0.2 }$    &  $ 140 _{- 75 }^{+ 97 }$    &  $ 14 _{- 6 }^{+ 7 }$    &  $ 268 _{- 159 }^{+ 218 }$    &  $ 269 _{- 60 }^{+ 40 }$    &  $ 757 _{- 203 }^{+ 151 }$   \\
$\Sigma_c(3358)$  & $ \mathbf{\frac{3}{2}^+}$ & $\vert \,1\,,\,1\,,\,0\,,\,0 \,\rangle $ &$^{2}D_{\lambda\rho,3/2}$&  $ 90 _{- 28 }^{+ 26 }$    &  $ 37 _{- 13 }^{+ 14 }$    &  $ 12 _{- 3 }^{+ 3 }$    &  $ 38 _{- 13 }^{+ 14 }$    &  $ 10 _{- 2 }^{+ 2 }$    &  $ 14 _{- 6 }^{+ 8 }$    &  $ 1.0 _{- 0.4 }^{+ 0.7 }$    &  $ 270 _{- 15 }^{+ 12 }$    &  $ 48 _{- 6 }^{+ 7 }$   \\
$\Sigma_c(3403)$  & $ \mathbf{\frac{5}{2}^+}$ & $\vert \,1\,,\,1\,,\,0\,,\,0 \,\rangle $ &$^{2}D_{\lambda\rho,5/2}$&    $ 100 _{- 28 }^{+ 25 }$    &  $ 43 _{- 14 }^{+ 14 }$    &  $ 32 _{- 9 }^{+ 9 }$    &  $ 8 _{- 4 }^{+ 5 }$    &  $ 19 _{- 4 }^{+ 4 }$    &  $ 21 _{- 7 }^{+ 8 }$    &  $ 12 _{- 4 }^{+ 4 }$    &  $ 8 _{- 4 }^{+ 5 }$    &  $ 336 _{- 17 }^{+ 17 }$   \\
$\Sigma_c(3367)$  & $ \mathbf{\frac{1}{2}^-}$ & $\vert \,1\,,\,1\,,\,0\,,\,0 \,\rangle $ &$^{2}P_{\lambda\rho,1/2}$&  0  &0  &  0  &$ 54 _{- 21 }^{+ 25 }$    &  0  &$ 37 _{- 14 }^{+ 18 }$    &  $ 7 _{- 3 }^{+ 4 }$    &  $ 320 _{- 34 }^{+ 29 }$    &  $ 46 _{- 8 }^{+ 8 }$   \\
$\Sigma_c(3394)$  & $ \mathbf{\frac{3}{2}^-}$ & $\vert \,1\,,\,1\,,\,0\,,\,0 \,\rangle $ &$^{2}P_{\lambda\rho,3/2}$&  0  &0  &  $ 51 _{- 16 }^{+ 20 }$    &  $ 21 _{- 8 }^{+ 9 }$    &  $ 23 _{- 8 }^{+ 8 }$    &  $ 25 _{- 9 }^{+ 11 }$    &  $ 8 _{- 3 }^{+ 5 }$    &  $ 92 _{- 10 }^{+ 9 }$    &  $ 280 _{- 31 }^{+ 29 }$   \\
$\Sigma_c(3385)$  & $ \mathbf{\frac{1}{2}^+}$ & $\vert \,1\,,\,1\,,\,0\,,\,0 \,\rangle $ &$^{2}S_{\lambda\rho,1/2}$&  0  &0  &  $ 8 _{- 2 }^{+ 1 }$    &  $ 4.2 _{- 0.5 }^{+ 0.3 }$    &  $ 17 _{- 2 }^{+ 1 }$    &  $ 0.8 _{- 0.1 }^{+ 0.1 }$    &  $ 7 _{- 2 }^{+ 1 }$    &  $ 241 _{- 40 }^{+ 41 }$    &  $ 97 _{- 17 }^{+ 18 }$   \\
$\Sigma_c(3540)$ & $ \mathbf{\frac{3}{2}^+}$ & $\vert \,0\,,\,2\,,\,0\,,\,0 \,\rangle $ &$^{2}D_{\rho\rho,3/2}$ &   $ 20 _{- 1 }^{+ 1 }$    &  $ 9 _{- 2 }^{+ 1 }$    &   $ 21 _{- 4 }^{+ 2 }$    &  $ 0.8 _{- 0.5 }^{+ 0.7 }$    &  $ 15 _{- 4 }^{+ 3 }$    &  $ 0.1 _{- 0.1 }^{+ 0.1 }$    &  $ 0.7 _{- 0.5 }^{+ 0.8 }$    &  $ 32 _{- 12 }^{+ 13 }$    &  $ 34 _{- 13 }^{+ 14 }$   \\
$\Sigma_c(3585)$  & $ \mathbf{\frac{5}{2}^+}$ & $\vert \,0\,,\,2\,,\,0\,,\,0 \,\rangle $ &$^{2}D_{\rho\rho,5/2}$&   $ 216 _{- 23 }^{+ 16 }$    &  $ 9 _{- 1 }^{+ 1 }$    &  $ 40 _{- 15 }^{+ 17 }$    &  $ 1.1 _{- 0.6 }^{+ 0.9 }$    &  $ 143 _{- 57 }^{+ 57 }$    &  $ 0.2 _{- 0.1 }^{+ 0.1 }$    &  $ 1.1 _{- 0.7 }^{+ 1.1 }$    &  $ 75 _{- 28 }^{+ 29 }$    &  $ 57 _{- 21 }^{+ 20 }$   \\
$\Sigma_c(3555)$  & $ \mathbf{\frac{1}{2}^+}$ & $\vert \,0\,,\,2\,,\,0\,,\,0 \,\rangle $ &$^{4}D_{\rho\rho,1/2}$&  $ 9 _{- 1 }^{+ 1 }$  &  $ 1.1 _{- 1.1 }^{+ 3.3 }$  &  $ 1.2 _{- 0.7 }^{+ 1.0 }$    &  $ 8 _{- 1 }^{+ 1 }$    &  $ 1.9 _{- 1.2 }^{+ 1.8 }$    &  $ 7 _{- 2 }^{+ 1 }$    &  $ 2.5 _{- 1.2 }^{+ 0.6 }$    &  $ 66 _{- 28 }^{+ 30 }$    &  $ 0.1 _{- 0.1 }^{+ 0.1 }$   \\
$\Sigma_c(3582)$  & $ \mathbf{\frac{3}{2}^+}$ & $\vert \,0\,,\,2\,,\,0\,,\,0 \,\rangle $ &$^{4}D_{\rho\rho,3/2}$ &  $ 19 _{- 2 }^{+ 1 }$  &  $ 37 _{- 4 }^{+ 3 }$  & $ 2.7 _{- 1.6 }^{+ 2.1 }$    &  $ 12 _{- 3 }^{+ 2 }$    &  $ 4.6 _{- 2.6 }^{+ 3.9 }$    &  $ 0.2 _{- 0.0 }^{+ 0.0 }$    &  $ 14 _{- 6 }^{+ 7 }$    &  $ 169 _{- 62 }^{+ 66 }$    &  $ 10 _{- 4 }^{+ 5 }$   \\
$\Sigma_c(3627)$  & $ \mathbf{\frac{5}{2}^+}$ & $\vert \,0\,,\,2\,,\,0\,,\,0 \,\rangle $ &$^{4}D_{\rho\rho,5/2}$&     $ 26 _{- 2 }^{+ 2 }$    &  $ 57 _{- 3 }^{+ 3 }$    &  $ 4.9 _{- 2.6 }^{+ 3.1 }$    &  $ 4.8 _{- 0.5 }^{+ 0.4 }$    &  $ 8 _{- 4 }^{+ 6 }$    &  $ 26 _{- 8 }^{+ 8 }$    &  $ 21 _{- 8 }^{+ 10 }$    &  $ 70 _{- 22 }^{+ 22 }$    &  $ 150 _{- 51 }^{+ 52 }$   \\
$\Sigma_c(3691)$  & $ \mathbf{\frac{7}{2}^+}$ & $\vert \,0\,,\,2\,,\,0\,,\,0 \,\rangle $ &$^{4}D_{\rho\rho,7/2}$&   $ 16 _{- 1 }^{+ 1 }$    &  $ 210 _{- 30 }^{+ 22 }$  &  $ 4.2 _{- 2.1 }^{+ 2.5 }$    &  $ 44 _{- 18 }^{+ 21 }$    &  $ 8 _{- 4 }^{+ 5 }$    &  $ 34 _{- 10 }^{+ 9 }$    &  $ 131 _{- 53 }^{+ 54 }$    &  $ 3.9 _{- 2.3 }^{+ 3.4 }$    &  $ 190 _{- 57 }^{+ 48 }$   \\
\hline \hline
\end{tabular}

\endgroup
}
\end{center}
\label{sigmas_zero_EM}
\end{table*}


\begin{turnpage}
\begin{table*}[htp]
\caption{Same as table \ref{sigmas+_EM}, but for $\Xi_{c}^{'0}$ states with isospin $1/2$.}
\scriptsize{
\begingroup
\setlength{\tabcolsep}{1.75pt} 
\renewcommand{\arraystretch}{1.35} 

\begin{tabular}{c c c c  p{1.0cm}  p{1.0cm}  p{1.0cm}  p{1.0cm}  p{1.0cm}  p{1.0cm}  p{1.0cm}  p{1.0cm}  p{1.0cm}  p{1.0cm}  p{1.0cm}  p{1.0cm}  p{1.0cm}  p{1.0cm}  p{1.0cm}  p{1.0cm}  p{1.0cm}  p{1.0cm}  p{1.0cm}  p{1.0cm}} \hline \hline
&    &    &    & $\Xi_{c}^{0} \gamma$  & $\Xi'^{0}_{c} \gamma$  & $\Xi^{*0}_{c} \gamma$  & $\Xi^{0}_{c} \gamma$  & $\Xi^{0}_{c} \gamma$  & $\Xi^{0}_{c} \gamma$  & $\Xi^{0}_{c} \gamma$  & $\Xi^{0}_{c} \gamma$  & $\Xi^{0}_{c} \gamma$  & $\Xi^{0}_{c} \gamma$  & $\Xi'^{0}_{c} \gamma$  & $\Xi^{*0}_{c} \gamma$  & $\Xi'^{0}_{c} \gamma$  & $\Xi^{*0}_{c} \gamma$  & $\Xi^{*0}_{c} \gamma$  & $\Xi'^{0}_{c} \gamma$  & $\Xi'^{0}_{c} \gamma$ \\
$\mathbf{\Xi'_c(snc)}$  & $\mathbf{J^P}$  & $\vert l_{\lambda}, l_{\rho}, k_{\lambda}, k_{\rho} \rangle$  & $^{2S+1}L_{x,J}$  & $^2S_{1/2}$  & $^2S_{1/2}$  & $^4S_{3/2}$  & $^2P_{\lambda,1/2}$  & $^2P_{\lambda,3/2}$  & $^2P_{\rho,1/2}$  & $^4P_{\rho,1/2}$  & $^2P_{\rho,3/2}$  & $^4P_{\rho,3/2}$  & $^4P_{\rho,5/2}$  & $^2P_{\lambda,1/2}$  & $^4P_{\lambda,1/2}$  & $^2P_{\lambda,3/2}$  & $^4P_{\lambda,3/2}$  & $^4P_{\lambda,5/2}$  & $^2P_{\rho,1/2}$  & $^2P_{\rho,3/2}$  \\ \hline
\hline
 $N=2$  &  &  &  &  &  \\\ 
$\Xi'_c(3223)$  & $ \mathbf{\frac{3}{2}^+}$ & $\vert \,2\,,\,0\,,\,0\,,\,0 \,\rangle $ &$^{2}D_{\lambda\lambda,3/2}$&$ 0.9 _{- 0.4 }^{+ 0.6 }$    &  $ 5 _{- 2 }^{+ 2 }$    &  $ 1.1 _{- 0.3 }^{+ 0.4 }$    &  $ 0.6 _{- 0.2 }^{+ 0.3 }$    &  $ 0.6 _{- 0.2 }^{+ 0.3 }$    &  $ 0.1 _{- 0.0 }^{+ 0.0 }$    &  0  &0  &0  &0  &$ 473 _{- 25 }^{+ 24 }$    &  $ 1.4 _{- 0.7 }^{+ 1.0 }$    &  $ 51 _{- 3 }^{+ 3 }$    &  $ 0.3 _{- 0.1 }^{+ 0.2 }$    &  0  &0  &0 \\
  &  &  & & $...$  & $...$ & $...$  & $0$ & $0$  & $...$  & $...$  & $...$ & $...$  & $...$  & $64.2$  & $0.69$  & 29.2 &  0.54  & 0.12  & $...$  & $...$ &  \cite{Yao:2018jmc} \\
$\Xi'_c(3268)$  & $ \mathbf{\frac{5}{2}^+}$ & $\vert \,2\,,\,0\,,\,0\,,\,0 \,\rangle $ &$^{2}D_{\lambda\lambda,5/2}$&$ 1.1 _{- 0.5 }^{+ 0.6 }$    &  $ 56 _{- 21 }^{+ 21 }$    &  $ 1.6 _{- 0.5 }^{+ 0.5 }$    &  $ 1.5 _{- 0.6 }^{+ 0.8 }$    &  $ 1.1 _{- 0.5 }^{+ 0.6 }$    &  0  &0  &$ 0.1 _{- 0.1 }^{+ 0.1 }$    &  0  &0  &$ 19 _{- 6 }^{+ 8 }$    &  $ 0.2 _{- 0.1 }^{+ 0.2 }$    &  $ 762 _{- 56 }^{+ 57 }$    &  $ 0.8 _{- 0.4 }^{+ 0.5 }$    &  $ 0.4 _{- 0.2 }^{+ 0.3 }$    &  0  &0 \\
 &  &  & & $...$  & $...$ & $...$  & $0$ & $0$  & $...$  & $...$  & $...$ & $...$  & $...$  & $6.38$  & $0.04$  & 50.1 &  0.23  & 0.61  & $...$  & $...$ &  \cite{Yao:2018jmc} \\
$\Xi'_c(3238)$  & $ \mathbf{\frac{1}{2}^+}$ & $\vert \,2\,,\,0\,,\,0\,,\,0 \,\rangle $ &$^{4}D_{\lambda\lambda,1/2}$&$ 0.5 _{- 0.2 }^{+ 0.3 }$    &  $ 2.1 _{- 0.7 }^{+ 1.0 }$    &  $ 1.4 _{- 0.8 }^{+ 1.0 }$    &  $ 1.2 _{- 0.6 }^{+ 0.8 }$    &  0  &0  &0  &0  &0  &0  &$ 1.6 _{- 0.8 }^{+ 1.3 }$    &  $ 373 _{- 30 }^{+ 23 }$    &  0  &$ 35 _{- 6 }^{+ 6 }$    &  $ 0.3 _{- 0.2 }^{+ 0.4 }$    &  0  &0 \\
 &  &  & & $...$  & $...$ & $...$  & $0$ & $0$  & $...$  & $...$  & $...$ & $...$  & $...$   & $0.03$  & 184.92 &  0.08  & 42 &  5.15  & $...$  & $...$ &  \cite{Yao:2018jmc} \\
&  &  & & 0 & 0.4 & 32.4  & $0.2$ & $0.1$  & $...$  & $...$  & $...$  & $...$  & $...$  & 137.3  & $...$ &  57.6  & 16.2  & $0.6$  & $...$ & $...$ & \cite{Peng:2024pyl} \\
$\Xi'_c(3265)$  & $ \mathbf{\frac{3}{2}^+}$ & $\vert \,2\,,\,0\,,\,0\,,\,0 \,\rangle $ &$^{4}D_{\lambda\lambda,3/2}$&$ 1.1 _{- 0.5 }^{+ 0.7 }$    &  $ 4.9 _{- 1.5 }^{+ 1.9 }$    &  $ 6 _{- 3 }^{+ 3 }$    &  $ 3.3 _{- 1.6 }^{+ 2.1 }$    &  $ 0.2 _{- 0.1 }^{+ 0.1 }$    &  0  &0  &0  &0  &0  &$ 4.8 _{- 2.2 }^{+ 3.3 }$    &  $ 268 _{- 19 }^{+ 18 }$    &  $ 0.2 _{- 0.1 }^{+ 0.1 }$    &  $ 169 _{- 11 }^{+ 9 }$    &  $ 7 _{- 1 }^{+ 1 }$    &  0  &0 \\
 &  &  & & $...$  & $...$ & $...$  & $0$ & $0$  & $...$  & $...$  & $...$ & $...$  & $...$  & $0.3$  & $74.7$  & 0.03 &  66.24  & 14  & $...$  & $...$ &  \cite{Yao:2018jmc} \\
$\Xi'_c(3310)$  & $ \mathbf{\frac{5}{2}^+}$ & $\vert \,2\,,\,0\,,\,0\,,\,0 \,\rangle $ &$^{4}D_{\lambda\lambda,5/2}$&$ 1.7 _{- 0.9 }^{+ 1.0 }$    &  $ 9 _{- 2 }^{+ 3 }$    &  $ 12 _{- 5 }^{+ 5 }$    &  $ 1.6 _{- 0.7 }^{+ 0.9 }$    &  $ 3.2 _{- 1.4 }^{+ 1.8 }$    &  0  &0  &0  &0  &0  &$ 2.6 _{- 1.2 }^{+ 1.4 }$    &  $ 14 _{- 5 }^{+ 6 }$    &  $ 4.8 _{- 2.2 }^{+ 2.7 }$    &  $ 500 _{- 34 }^{+ 32 }$    &  $ 85 _{- 4 }^{+ 4 }$    &  0  &0 \\
 &  &  & & $...$  & $...$ & $...$  & $0$ & $0$  & $...$  & $...$  & $...$ & $...$  & $...$  & $0.3$  & $74.7$  & 0.03 &  66.24 & 14  & $...$  & $...$ &  \cite{Yao:2018jmc} \\
$\Xi'_c(3373)$  & $ \mathbf{\frac{7}{2}^+}$ & $\vert \,2\,,\,0\,,\,0\,,\,0 \,\rangle $ &$^{4}D_{\lambda\lambda,7/2}$&$ 1.4 _{- 0.7 }^{+ 0.8 }$    &  $ 8 _{- 2 }^{+ 2 }$    &  $ 57 _{- 24 }^{+ 23 }$    &  0  &$ 4.7 _{- 2.1 }^{+ 2.6 }$    &  0  &0  &0  &$ 0.1 _{- 0.0 }^{+ 0.1 }$    &  $ 0.1 _{- 0.1 }^{+ 0.1 }$    &  0  &$ 0.8 _{- 0.4 }^{+ 0.6 }$    &  $ 8 _{- 4 }^{+ 5 }$    &  $ 22 _{- 8 }^{+ 11 }$    &  $ 869 _{- 91 }^{+ 92 }$    &  0  &0 \\
 &  &  & & $...$  & $...$ & $...$  & $0$ & $0$  & $...$  & $...$  & $...$ & $...$  & $...$  & $<0.01$  & $2.71$  & 0.22 &  7.26  & 33.7  & $...$  & $...$ &  \cite{Yao:2018jmc} \\
&  &  & & 0 & 0.6 & 9.3  & $0$ & $0.4$  & $...$  & $...$  & $...$  & $...$  & $...$  & $0.1$  & $...$&  $1.1$  & $0.1$  & $219.6$  & $...$ & $...$ & \cite{Peng:2024pyl} \\
$\Xi'_c(3250)$  & $ \mathbf{\frac{1}{2}^+}$ & $\vert \,0\,,\,0\,,\,1\,,\,0 \,\rangle $ &$^{2}S_{1/2}$&$ 2.5 _{- 1.2 }^{+ 1.5 }$    &  $ 36 _{- 12 }^{+ 13 }$    &  $ 3.4 _{- 1.0 }^{+ 1.2 }$    &  $ 5 _{- 2 }^{+ 3 }$    &  $ 12 _{- 5 }^{+ 7 }$    &  0  &0  &$ 0.1 _{- 0.0 }^{+ 0.1 }$    &  0  &0  &$ 555 _{- 81 }^{+ 88 }$    &  0  &$ 3.2 _{- 3.2 }^{+ 5.8 }$    &  $ 0.1 _{- 0.0 }^{+ 0.1 }$    &  $ 1.9 _{- 1.0 }^{+ 1.4 }$    &  0  &0 \\
&  &  & & 1.2 & 7.4 & 0  & $0.1$ & $0.1$  & $...$  & $...$  & $...$  & $...$  & $...$  & $27.5$  & $...$ &  10.6  & $77.8$  & $0.1$  & $...$ & $...$ & \cite{Peng:2024pyl} \\
$\Xi'_c(3319)$  & $ \mathbf{\frac{3}{2}^+}$ & $\vert \,0\,,\,0\,,\,1\,,\,0 \,\rangle $ &$^{4}S_{3/2}$&$ 3.3 _{- 1.6 }^{+ 2.0 }$    &  $ 17 _{- 4 }^{+ 5 }$    &  $ 26 _{- 11 }^{+ 12 }$    &  $ 8 _{- 3 }^{+ 4 }$    &  $ 20 _{- 8 }^{+ 10 }$    &  0  &0  &0  &0  &0  &$ 13 _{- 6 }^{+ 6 }$    &  $ 174 _{- 17 }^{+ 17 }$    &  $ 32 _{- 14 }^{+ 16 }$    &  $ 210 _{- 18 }^{+ 20 }$    &  $ 3.3 _{- 0.6 }^{+ 2.5 }$    &  0  &$ 0.1 _{- 0.0 }^{+ 0.1 }$   \\
&  &  & & 1.3 & 7.6 & 5.4  & $...$ & $...$  & $...$  & $...$  & $...$  & $...$  & $...$  & 8.7  & $...$ &  38.1  & 11.3  & $95.2$  & $...$ & $...$ & \cite{Peng:2024pyl} \\
$\Xi'_c(3544)$  & $ \mathbf{\frac{1}{2}^+}$ & $\vert \,0\,,\,0\,,\,0\,,\,1 \,\rangle $ &$^{2}S_{1/2}$&$ 6 _{- 2 }^{+ 2 }$    &  $ 123 _{- 26 }^{+ 24 }$    &  $ 13 _{- 3 }^{+ 3 }$    &  $ 3.0 _{- 1.8 }^{+ 2.5 }$    &  $ 7 _{- 4 }^{+ 6 }$    &  $ 62 _{- 22 }^{+ 23 }$    &  0  &$ 10 _{- 5 }^{+ 7 }$    &  $ 0.1 _{- 0.0 }^{+ 0.0 }$    &  $ 2.5 _{- 1.2 }^{+ 1.5 }$    &  $ 10 _{- 6 }^{+ 10 }$    &  0  &$ 97 _{- 47 }^{+ 63 }$    &  $ 0.1 _{- 0.1 }^{+ 0.1 }$    &  $ 2.5 _{- 1.4 }^{+ 2.1 }$    &  $ 124 _{- 36 }^{+ 40 }$    &  $ 316 _{- 99 }^{+ 114 }$   \\
$\Xi'_c(3613)$  & $ \mathbf{\frac{3}{2}^+}$ & $\vert \,0\,,\,0\,,\,0\,,\,1 \,\rangle $ &$^{4}S_{3/2}$&$ 7 _{- 2 }^{+ 2 }$    &  $ 36 _{- 7 }^{+ 6 }$    &  $ 109 _{- 24 }^{+ 21 }$    &  $ 4.8 _{- 2.8 }^{+ 3.7 }$    &  $ 12 _{- 7 }^{+ 10 }$    &  $ 4.1 _{- 1.5 }^{+ 1.6 }$    &  $ 11 _{- 3 }^{+ 4 }$    &  $ 12 _{- 4 }^{+ 5 }$    &  $ 14 _{- 4 }^{+ 4 }$    &  $ 7 _{- 4 }^{+ 5 }$    &  $ 14 _{- 7 }^{+ 9 }$    &  $ 1.3 _{- 0.2 }^{+ 0.1 }$    &  $ 33 _{- 16 }^{+ 22 }$    &  $ 5 _{- 2 }^{+ 2 }$    &  $ 55 _{- 29 }^{+ 38 }$    &  $ 170 _{- 46 }^{+ 43 }$    &  $ 450 _{- 130 }^{+ 130 }$   \\
$\Xi'_c(3370)$  & $ \mathbf{\frac{3}{2}^+}$ & $\vert \,1\,,\,1\,,\,0\,,\,0 \,\rangle $ &$^{2}D_{\lambda\rho,3/2}$&$ 7 _{- 2 }^{+ 2 }$    &  $ 40 _{- 12 }^{+ 13 }$    &  $ 14 _{- 5 }^{+ 5 }$    &  $ 9 _{- 2 }^{+ 2 }$    &  $ 2.5 _{- 0.8 }^{+ 1.0 }$    &  $ 0.3 _{- 0.1 }^{+ 0.2 }$    &  $ 0.6 _{- 0.4 }^{+ 0.5 }$    &  $ 0.2 _{- 0.1 }^{+ 0.2 }$    &  $ 0.2 _{- 0.1 }^{+ 0.1 }$    &  0  &$ 6 _{- 2 }^{+ 2 }$    &  $ 16 _{- 5 }^{+ 7 }$    &  $ 6 _{- 1 }^{+ 2 }$    &  $ 4.6 _{- 1.6 }^{+ 2.2 }$    &  $ 0.3 _{- 0.1 }^{+ 0.1 }$    &  $ 246 _{- 13 }^{+ 12 }$    &  $ 38 _{- 3 }^{+ 4 }$   \\
$\Xi'_c(3415)$  & $ \mathbf{\frac{5}{2}^+}$ & $\vert \,1\,,\,1\,,\,0\,,\,0 \,\rangle $ &$^{2}D_{\lambda\rho,5/2}$&$ 7 _{- 2 }^{+ 2 }$    &  $ 48 _{- 14 }^{+ 15 }$    &  $ 18 _{- 6 }^{+ 6 }$    &  $ 0.7 _{- 0.3 }^{+ 0.4 }$    &  $ 10 _{- 3 }^{+ 3 }$    &  $ 0.7 _{- 0.3 }^{+ 0.4 }$    &  $ 0.1 _{- 0.1 }^{+ 0.1 }$    &  $ 0.5 _{- 0.2 }^{+ 0.3 }$    &  $ 0.3 _{- 0.2 }^{+ 0.3 }$    &  $ 0.2 _{- 0.1 }^{+ 0.2 }$    &  $ 16 _{- 4 }^{+ 4 }$    &  $ 2.7 _{- 1.1 }^{+ 1.3 }$    &  $ 11 _{- 2 }^{+ 3 }$    &  $ 9 _{- 3 }^{+ 3 }$    &  $ 5 _{- 2 }^{+ 2 }$    &  $ 3.6 _{- 1.3 }^{+ 1.5 }$    &  $ 306 _{- 18 }^{+ 16 }$   \\
$\Xi'_c(3379)$  & $ \mathbf{\frac{1}{2}^-}$ & $\vert \,1\,,\,1\,,\,0\,,\,0 \,\rangle $ &$^{2}P_{\lambda\rho,1/2}$&$ 0.4 _{- 0.1 }^{+ 0.2 }$    &  0  &0  &$ 10 _{- 3 }^{+ 3 }$    &  $ 0.9 _{- 0.2 }^{+ 0.2 }$    &  0  &$ 0.8 _{- 0.4 }^{+ 0.6 }$    &  0  &$ 0.5 _{- 0.3 }^{+ 0.4 }$    &  $ 0.1 _{- 0.0 }^{+ 0.1 }$    &  0  &$ 21 _{- 7 }^{+ 9 }$    &  0  &$ 14 _{- 4 }^{+ 5 }$    &  $ 2.2 _{- 0.8 }^{+ 1.1 }$    &  $ 259 _{- 23 }^{+ 20 }$    &  $ 34 _{- 4 }^{+ 4 }$   \\
$\Xi'_c(3406)$  & $ \mathbf{\frac{3}{2}^-}$ & $\vert \,1\,,\,1\,,\,0\,,\,0 \,\rangle $ &$^{2}P_{\lambda\rho,3/2}$&$ 0.4 _{- 0.1 }^{+ 0.1 }$    &  0  &0  &$ 2.5 _{- 0.6 }^{+ 0.7 }$    &  $ 9 _{- 2 }^{+ 2 }$    &  $ 1.0 _{- 0.5 }^{+ 0.6 }$    &  $ 0.3 _{- 0.2 }^{+ 0.3 }$    &  $ 0.4 _{- 0.2 }^{+ 0.3 }$    &  $ 0.4 _{- 0.2 }^{+ 0.3 }$    &  $ 0.1 _{- 0.1 }^{+ 0.1 }$    &  $ 23 _{- 7 }^{+ 7 }$    &  $ 9 _{- 3 }^{+ 3 }$    &  $ 10 _{- 3 }^{+ 3 }$    &  $ 10 _{- 3 }^{+ 4 }$    &  $ 2.8 _{- 1.0 }^{+ 1.3 }$    &  $ 75 _{- 7 }^{+ 7 }$    &  $ 225 _{- 19 }^{+ 19 }$   \\
$\Xi'_c(3397)$  & $ \mathbf{\frac{1}{2}^+}$ & $\vert \,1\,,\,1\,,\,0\,,\,0 \,\rangle $ &$^{2}S_{\lambda\rho,1/2}$&0  &0  &0  &$ 10 _{- 3 }^{+ 3 }$    &  $ 4.8 _{- 1.5 }^{+ 1.7 }$    &  $ 0.3 _{- 0.1 }^{+ 0.2 }$    &  $ 0.1 _{- 0.1 }^{+ 0.1 }$    &  $ 0.6 _{- 0.3 }^{+ 0.3 }$    &  0  &$ 0.1 _{- 0.1 }^{+ 0.1 }$    &  $ 7 _{- 1 }^{+ 1 }$    &  $ 3.0 _{- 0.7 }^{+ 0.7 }$    &  $ 13 _{- 2 }^{+ 2 }$    &  $ 0.5 _{- 0.1 }^{+ 0.1 }$    &  $ 3.4 _{- 1.0 }^{+ 1.1 }$    &  $ 180 _{- 21 }^{+ 23 }$    &  $ 71 _{- 9 }^{+ 8 }$   \\
$\Xi'_c(3517)$  & $ \mathbf{\frac{3}{2}^+}$ & $\vert \,0\,,\,2\,,\,0\,,\,0 \,\rangle $ &$^{2}D_{\rho\rho,3/2}$&$ 2.4 _{- 0.9 }^{+ 1.0 }$    &  $ 24 _{- 3 }^{+ 3 }$    &  $ 4.7 _{- 1.2 }^{+ 1.2 }$    &  $ 0.2 _{- 0.1 }^{+ 0.1 }$    &  $ 0.3 _{- 0.2 }^{+ 0.2 }$    &  $ 12 _{- 2 }^{+ 3 }$    &  $ 1.3 _{- 0.7 }^{+ 0.9 }$    &  $ 1.5 _{- 0.4 }^{+ 0.5 }$    &  $ 0.4 _{- 0.2 }^{+ 0.3 }$    &  0  &$ 14 _{- 3 }^{+ 3 }$    &  $ 0.1 _{- 0.1 }^{+ 0.1 }$    &  $ 8 _{- 2 }^{+ 2 }$    &  0  &$ 0.1 _{- 0.1 }^{+ 0.1 }$    &  $ 15 _{- 5 }^{+ 6 }$    &  $ 15 _{- 6 }^{+ 6 }$   \\
$\Xi'_c(3563)$  & $ \mathbf{\frac{5}{2}^+}$ & $\vert \,0\,,\,2\,,\,0\,,\,0 \,\rangle $ &$^{2}D_{\rho\rho,5/2}$&$ 2.6 _{- 0.9 }^{+ 1.0 }$    &  $ 211 _{- 17 }^{+ 14 }$    &  $ 5 _{- 1 }^{+ 1 }$    &  $ 0.2 _{- 0.1 }^{+ 0.2 }$    &  $ 0.4 _{- 0.2 }^{+ 0.3 }$    &  $ 8 _{- 3 }^{+ 4 }$    &  $ 0.2 _{- 0.1 }^{+ 0.1 }$    &  $ 37 _{- 10 }^{+ 11 }$    &  $ 0.7 _{- 0.3 }^{+ 0.4 }$    &  $ 0.5 _{- 0.2 }^{+ 0.3 }$    &  $ 15 _{- 5 }^{+ 6 }$    &  $ 0.2 _{- 0.1 }^{+ 0.2 }$    &  $ 54 _{- 19 }^{+ 22 }$    &  0  &$ 0.2 _{- 0.1 }^{+ 0.1 }$    &  $ 36 _{- 12 }^{+ 14 }$    &  $ 28 _{- 9 }^{+ 10 }$   \\
$\Xi'_c(3532)$  & $ \mathbf{\frac{1}{2}^+}$ & $\vert \,0\,,\,2\,,\,0\,,\,0 \,\rangle $ &$^{4}D_{\rho\rho,1/2}$&$ 1.2 _{- 0.5 }^{+ 0.5 }$    &  $ 6 _{- 1 }^{+ 1 }$    &  $ 4.3 _{- 3.1 }^{+ 3.6 }$    &  $ 0.1 _{- 0.1 }^{+ 0.1 }$    &  $ 0.2 _{- 0.1 }^{+ 0.2 }$    &  $ 1.0 _{- 0.4 }^{+ 0.6 }$    &  $ 5 _{- 3 }^{+ 3 }$    &  0  &$ 3.6 _{- 1.2 }^{+ 1.4 }$    &  $ 0.3 _{- 0.2 }^{+ 0.2 }$    &  $ 0.2 _{- 0.1 }^{+ 0.2 }$    &  $ 6 _{- 2 }^{+ 1 }$    &  $ 0.4 _{- 0.2 }^{+ 0.3 }$    &  $ 3.7 _{- 1.2 }^{+ 1.2 }$    &  $ 2.2 _{- 0.6 }^{+ 0.5 }$    &  $ 31 _{- 12 }^{+ 15 }$    &  $ 0.1 _{- 0.0 }^{+ 0.0 }$   \\
$\Xi'_c(3559)$  & $ \mathbf{\frac{3}{2}^+}$ & $\vert \,0\,,\,2\,,\,0\,,\,0 \,\rangle $ &$^{4}D_{\rho\rho,3/2}$&$ 2.6 _{- 0.9 }^{+ 1.0 }$    &  $ 13 _{- 3 }^{+ 2 }$    &  $ 29 _{- 3 }^{+ 3 }$    &  $ 0.2 _{- 0.1 }^{+ 0.2 }$    &  $ 0.4 _{- 0.2 }^{+ 0.3 }$    &  $ 2.5 _{- 1.1 }^{+ 1.3 }$    &  $ 13 _{- 4 }^{+ 4 }$    &  $ 0.1 _{- 0.1 }^{+ 0.1 }$    &  $ 2.5 _{- 1.1 }^{+ 1.2 }$    &  $ 0.9 _{- 0.3 }^{+ 0.4 }$    &  $ 0.6 _{- 0.3 }^{+ 0.5 }$    &  $ 7 _{- 2 }^{+ 2 }$    &  $ 0.9 _{- 0.5 }^{+ 0.8 }$    &  $ 0.3 _{- 0.0 }^{+ 0.0 }$    &  $ 5 _{- 2 }^{+ 2 }$    &  $ 82 _{- 29 }^{+ 34 }$    &  $ 4.4 _{- 1.6 }^{+ 1.9 }$   \\
$\Xi'_c(3604)$  & $ \mathbf{\frac{5}{2}^+}$ & $\vert \,0\,,\,2\,,\,0\,,\,0 \,\rangle $ &$^{4}D_{\rho\rho,5/2}$&$ 3.7 _{- 1.2 }^{+ 1.3 }$    &  $ 19 _{- 4 }^{+ 3 }$    &  $ 54 _{- 4 }^{+ 3 }$    &  $ 0.4 _{- 0.3 }^{+ 0.3 }$    &  $ 0.7 _{- 0.4 }^{+ 0.6 }$    &  $ 1.1 _{- 0.4 }^{+ 0.5 }$    &  $ 4.0 _{- 1.8 }^{+ 2.1 }$    &  $ 2.2 _{- 0.9 }^{+ 1.1 }$    &  $ 23 _{- 6 }^{+ 7 }$    &  $ 1.6 _{- 0.5 }^{+ 0.5 }$    &  $ 1.2 _{- 0.6 }^{+ 0.7 }$    &  $ 3.7 _{- 0.7 }^{+ 0.6 }$    &  $ 1.9 _{- 1.0 }^{+ 1.3 }$    &  $ 12 _{- 4 }^{+ 4 }$    &  $ 8 _{- 3 }^{+ 3 }$    &  $ 38 _{- 12 }^{+ 12 }$    &  $ 77 _{- 24 }^{+ 27 }$   \\
$\Xi'_c(3668)$  & $ \mathbf{\frac{7}{2}^+}$ & $\vert \,0\,,\,2\,,\,0\,,\,0 \,\rangle $ &$^{4}D_{\rho\rho,7/2}$&$ 2.5 _{- 0.8 }^{+ 0.8 }$    &  $ 14 _{- 2 }^{+ 2 }$    &  $ 218 _{- 15 }^{+ 13 }$    &  $ 0.4 _{- 0.2 }^{+ 0.3 }$    &  $ 0.7 _{- 0.4 }^{+ 0.5 }$    &  $ 0.1 _{- 0.0 }^{+ 0.0 }$    &  $ 0.6 _{- 0.3 }^{+ 0.4 }$    &  $ 2.9 _{- 1.1 }^{+ 1.3 }$    &  $ 6 _{- 3 }^{+ 3 }$    &  $ 45 _{- 13 }^{+ 14 }$    &  $ 1.2 _{- 0.6 }^{+ 0.7 }$    &  $ 16 _{- 7 }^{+ 8 }$    &  $ 2.0 _{- 1.0 }^{+ 1.3 }$    &  $ 17 _{- 5 }^{+ 6 }$    &  $ 50 _{- 20 }^{+ 23 }$    &  $ 1.1 _{- 0.6 }^{+ 0.8 }$    &  $ 109 _{- 32 }^{+ 34 }$   \\
\hline \hline
\end{tabular}
\endgroup
}
\label{cascades_negative-_EM}
\end{table*}
\end{turnpage}

\begin{turnpage}
\begin{table*}[htp]
\caption{Same as table \ref{sigmas+_EM}, but for $\Xi_{c}^{'+}$ states with isospin $1/2$.}
\begin{center}
\scriptsize{
\begingroup
\setlength{\tabcolsep}{1.75pt} 
\renewcommand{\arraystretch}{1.35} 

\begin{tabular}{c c c c  p{1.0cm}  p{1.0cm}  p{1.0cm}  p{1.0cm}  p{1.0cm}  p{1.0cm}  p{1.0cm}  p{1.0cm}  p{1.0cm}  p{1.0cm}  p{1.0cm}  p{1.0cm}  p{1.0cm}  p{1.0cm}  p{1.0cm}  p{1.0cm}  p{1.0cm}  p{1.0cm}  p{1.0cm}  p{1.0cm}} \hline \hline
&    &    &    & $\Xi_{c}^{+} \gamma$  & $\Xi'^{+}_{c} \gamma$  & $\Xi^{*+}_{c} \gamma$  & $\Xi^{+}_{c} \gamma$  & $\Xi^{+}_{c} \gamma$  & $\Xi^{+}_{c} \gamma$  & $\Xi^{+}_{c} \gamma$  & $\Xi^{+}_{c} \gamma$  & $\Xi^{+}_{c} \gamma$  & $\Xi^{+}_{c} \gamma$  & $\Xi'^{+}_{c} \gamma$  & $\Xi^{*+}_{c} \gamma$  & $\Xi'^{+}_{c} \gamma$  & $\Xi^{*+}_{c} \gamma$  & $\Xi^{*+}_{c} \gamma$  & $\Xi'^{+}_{c} \gamma$  & $\Xi'^{+}_{c} \gamma$ \\
$\mathbf{\Xi'_c(snc)}$  & $\mathbf{J^P}$  & $\vert l_{\lambda}, l_{\rho}, k_{\lambda}, k_{\rho} \rangle$  & $^{2S+1}L_{x,J}$  & $^2S_{1/2}$  & $^2S_{1/2}$  & $^4S_{3/2}$  & $^2P_{\lambda,1/2}$  & $^2P_{\lambda,3/2}$  & $^2P_{\rho,1/2}$  & $^4P_{\rho,1/2}$  & $^2P_{\rho,3/2}$  & $^4P_{\rho,3/2}$  & $^4P_{\rho,5/2}$  & $^2P_{\lambda,1/2}$  & $^4P_{\lambda,1/2}$  & $^2P_{\lambda,3/2}$  & $^4P_{\lambda,3/2}$  & $^4P_{\lambda,5/2}$  & $^2P_{\rho,1/2}$  & $^2P_{\rho,3/2}$  \\ \hline
\hline
 $N=2$  &  &  &  &  &  \\\ 
$\Xi'_c(3223)$  & $ \mathbf{\frac{3}{2}^+}$ & $\vert \,2\,,\,0\,,\,0\,,\,0 \,\rangle $ &$^{2}D_{\lambda\lambda,3/2}$&$ 41 _{- 14 }^{+ 16 }$    &  $ 26 _{- 2 }^{+ 2 }$    &  $ 0.5 _{- 0.3 }^{+ 0.3 }$    &  $ 27 _{- 8 }^{+ 9 }$    &  $ 27 _{- 8 }^{+ 10 }$    &  $ 2.5 _{- 1.0 }^{+ 1.5 }$    &  0  &$ 0.5 _{- 0.2 }^{+ 0.3 }$    &  0  &0  &$ 14 _{- 11 }^{+ 14 }$    &  $ 2.6 _{- 1.1 }^{+ 1.6 }$    &  $ 1.4 _{- 0.7 }^{+ 0.9 }$    &  $ 0.5 _{- 0.2 }^{+ 0.3 }$    &  0  &0  &0 \\
  &  &  & & $...$  & $...$ & $...$  & $7.91$ & $24.4$ & $...$  & $...$  & $...$ & $...$  & $...$  & $0.02$  & $1.36$  & 0.45 &  0.9  & 0.19 & $...$  & $...$ &  \cite{Yao:2018jmc} \\
$\Xi'_c(3268)$  & $ \mathbf{\frac{5}{2}^+}$ & $\vert \,2\,,\,0\,,\,0\,,\,0 \,\rangle $ &$^{2}D_{\lambda\lambda,5/2}$&$ 51 _{- 17 }^{+ 19 }$    &  $ 96 _{- 19 }^{+ 22 }$    &  $ 0.6 _{- 0.4 }^{+ 0.4 }$    &  $ 68 _{- 20 }^{+ 22 }$    &  $ 52 _{- 16 }^{+ 17 }$    &  $ 1.1 _{- 0.5 }^{+ 0.6 }$    &  0  &$ 4.6 _{- 2.0 }^{+ 2.9 }$    &  0  &0  &$ 16 _{- 6 }^{+ 8 }$    &  $ 0.4 _{- 0.2 }^{+ 0.2 }$    &  $ 60 _{- 34 }^{+ 37 }$    &  $ 1.5 _{- 0.5 }^{+ 0.7 }$    &  $ 0.8 _{- 0.4 }^{+ 0.5 }$    &  0  &0 \\
 &  &  & & $...$  & $...$ & $...$  & $12.4$ & $17.5$ & $...$  & $...$  & $...$ & $...$  & $...$  & $0.32$  & $0.08$  & 1.98 &  0.4  & 0.98 & $...$  & $...$ &  \cite{Yao:2018jmc} \\
$\Xi'_c(3238)$  & $ \mathbf{\frac{1}{2}^+}$ & $\vert \,2\,,\,0\,,\,0\,,\,0 \,\rangle $ &$^{4}D_{\lambda\lambda,1/2}$&$ 22 _{- 8 }^{+ 10 }$    &  $ 0.8 _{- 0.5 }^{+ 0.6 }$    &  $ 12 _{- 2 }^{+ 1 }$    &  $ 57 _{- 21 }^{+ 28 }$    &  $ 0.3 _{- 0.1 }^{+ 0.0 }$    &  0  &$ 0.9 _{- 0.6 }^{+ 1.0 }$    &  0  &$ 0.5 _{- 0.3 }^{+ 0.6 }$    &  0  &$ 3.2 _{- 1.4 }^{+ 1.9 }$    &  $ 9 _{- 8 }^{+ 12 }$    &  0  &$ 3.3 _{- 2.1 }^{+ 2.5 }$    &  $ 0.2 _{- 0.1 }^{+ 0.2 }$    &  0  &0 \\
 &  &  & & $...$  & $...$ & $...$  & $9.33$ & $7.49$ & $...$  & $...$  & $...$ & $...$  & $...$  & $0.49$  & $<0.01$  & 0.13 &  0.07  & 0.02 & $...$  & $...$ &  \cite{Yao:2018jmc} \\
&  &  & & 4.4 & 0.1 & 38.1  & $16.1$ & $7.6$  & $...$  & $...$  & $...$  & $...$  & $...$  & 0.6  & $...$ &  0.3  & 0.7  & $0.1$  & $...$ & $...$ & \cite{Peng:2024pyl} \\
$\Xi'_c(3265)$  & $ \mathbf{\frac{3}{2}^+}$ & $\vert \,2\,,\,0\,,\,0\,,\,0 \,\rangle $ &$^{4}D_{\lambda\lambda,3/2}$&$ 50 _{- 18 }^{+ 21 }$    &  $ 1.9 _{- 1.1 }^{+ 1.4 }$    &  $ 15 _{- 2 }^{+ 2 }$    &  $ 156 _{- 51 }^{+ 63 }$    &  $ 8 _{- 3 }^{+ 3 }$    &  0  &$ 1.3 _{- 0.8 }^{+ 1.1 }$    &  0  &$ 0.2 _{- 0.1 }^{+ 0.2 }$    &  $ 0.2 _{- 0.1 }^{+ 0.2 }$    &  $ 10 _{- 3 }^{+ 5 }$    &  $ 17 _{- 10 }^{+ 12 }$    &  $ 0.5 _{- 0.2 }^{+ 0.2 }$    &  $ 4.0 _{- 3.4 }^{+ 4.8 }$    &  $ 0.6 _{- 0.4 }^{+ 0.5 }$    &  0  &0 \\
 &  &  & & $...$  & $...$ & $...$  & $34.2$ & $2.41$ & $...$  & $...$  & $...$ & $...$  & $...$  & $0.47$  & $0.54$  & 0.04 &  0.04  & 0.04 & $...$  & $...$ &  \cite{Yao:2018jmc} \\
$\Xi'_c(3310)$  & $ \mathbf{\frac{5}{2}^+}$ & $\vert \,2\,,\,0\,,\,0\,,\,0 \,\rangle $ &$^{4}D_{\lambda\lambda,5/2}$&$ 82 _{- 28 }^{+ 31 }$    &  $ 3.3 _{- 1.9 }^{+ 2.2 }$    &  $ 30 _{- 4 }^{+ 5 }$    &  $ 75 _{- 22 }^{+ 24 }$    &  $ 149 _{- 44 }^{+ 50 }$    &  0  &$ 1.5 _{- 0.7 }^{+ 1.0 }$    &  0  &$ 1.5 _{- 0.7 }^{+ 1.1 }$    &  $ 0.4 _{- 0.2 }^{+ 0.4 }$    &  $ 6 _{- 2 }^{+ 2 }$    &  $ 6 _{- 3 }^{+ 4 }$    &  $ 10 _{- 3 }^{+ 4 }$    &  $ 32 _{- 20 }^{+ 22 }$    &  $ 2.0 _{- 1.6 }^{+ 2.3 }$    &  0  &0 \\
 &  &  & & $...$  & $...$ & $...$  & $12.4$ & $22$ & $...$  & $...$  & $...$ & $...$  & $...$  & $0.16$  & $0.05$  & 0.4 &  0.73  & 0.05 & $...$  & $...$ &  \cite{Yao:2018jmc} \\
$\Xi'_c(3373)$  & $ \mathbf{\frac{7}{2}^+}$ & $\vert \,2\,,\,0\,,\,0\,,\,0 \,\rangle $ &$^{4}D_{\lambda\lambda,7/2}$&$ 65 _{- 22 }^{+ 22 }$    &  $ 2.7 _{- 1.5 }^{+ 1.8 }$    &  $ 115 _{- 23 }^{+ 26 }$    &  $ 1.6 _{- 0.9 }^{+ 1.3 }$    &  $ 224 _{- 64 }^{+ 74 }$    &  $ 0.1 _{- 0.0 }^{+ 0.1 }$    &  $ 0.6 _{- 0.4 }^{+ 0.6 }$    &  $ 0.1 _{- 0.1 }^{+ 0.1 }$    &  $ 3.2 _{- 1.6 }^{+ 2.2 }$    &  $ 4.9 _{- 2.8 }^{+ 4.7 }$    &  0  &$ 0.4 _{- 0.3 }^{+ 0.4 }$    &  $ 18 _{- 6 }^{+ 7 }$    &  $ 10 _{- 5 }^{+ 6 }$    &  $ 67 _{- 40 }^{+ 44 }$    &  0  &0 \\
 &  &  & & $...$  & $...$ & $...$  & $0.05$ & $19.5$ & $...$  & $...$  & $...$ & $...$  & $...$  & $<0.01$  & $<0.01$  & 0.35 &  0.02  & 1.25 & $...$  & $...$ &  \cite{Yao:2018jmc} \\
&  &  & & 6.5 & 0.1 & 14  & $0$ & $25.2$  & $...$  & $...$  & $...$  & $...$  & $...$  & 0  & $...$ &  0.7  & 0.5  & $4.2$  & $...$ & $...$ & \cite{Peng:2024pyl} \\
$\Xi'_c(3250)$  & $ \mathbf{\frac{1}{2}^+}$ & $\vert \,0\,,\,0\,,\,1\,,\,0 \,\rangle $ &$^{2}S_{1/2}$&$ 116 _{- 40 }^{+ 43 }$    &  $ 23 _{- 10 }^{+ 12 }$    &  $ 1.4 _{- 0.8 }^{+ 1.0 }$    &  $ 242 _{- 65 }^{+ 74 }$    &  $ 591 _{- 167 }^{+ 189 }$    &  $ 0.8 _{- 0.2 }^{+ 0.3 }$    &  0  &$ 2.7 _{- 1.4 }^{+ 2.2 }$    &  0  &0  &$ 120 _{- 52 }^{+ 60 }$    &  $ 0.1 _{- 0.0 }^{+ 0.0 }$    &  $ 44 _{- 17 }^{+ 18 }$    &  $ 0.2 _{- 0.1 }^{+ 0.1 }$    &  $ 3.6 _{- 1.6 }^{+ 2.2 }$    &  0  &0 \\
&  &  & & 46.9 & 6.2 & 1.1  & $4.4$ & $6.9$  & $...$  & $...$  & $...$  & $...$  & $...$  & 0.3  & $...$ &  0.2  & 0.2  & $0.1$  & $...$ & $...$ & \cite{Peng:2024pyl} \\
$\Xi'_c(3319)$  & $ \mathbf{\frac{3}{2}^+}$ & $\vert \,0\,,\,0\,,\,1\,,\,0 \,\rangle $ &$^{4}S_{3/2}$&$ 156 _{- 55 }^{+ 55 }$    &  $ 6 _{- 4 }^{+ 4 }$    &  $ 26 _{- 10 }^{+ 12 }$    &  $ 357 _{- 91 }^{+ 95 }$    &  $ 920 _{- 243 }^{+ 260 }$    &  $ 0.5 _{- 0.3 }^{+ 0.4 }$    &  $ 1.4 _{- 0.7 }^{+ 1.0 }$    &  $ 0.8 _{- 0.5 }^{+ 0.8 }$    &  $ 1.6 _{- 0.8 }^{+ 1.2 }$    &  $ 1.2 _{- 0.8 }^{+ 1.6 }$    &  $ 30 _{- 8 }^{+ 9 }$    &  $ 14 _{- 9 }^{+ 10 }$    &  $ 70 _{- 20 }^{+ 24 }$    &  $ 17 _{- 10 }^{+ 12 }$    &  $ 12 _{- 7 }^{+ 9 }$    &  0  &0 \\
&  &  & & 74.9 & 1.4 & 6  & $5.9$ & $9.9$  & $...$  & $...$  & $...$  & $...$  & $...$  & 0.2  & $...$ &  0.5  & 0.2  & $0.2$  & $...$ & $...$ & \cite{Peng:2024pyl} \\
$\Xi'_c(3544)$  & $ \mathbf{\frac{1}{2}^+}$ & $\vert \,0\,,\,0\,,\,0\,,\,1 \,\rangle $ &$^{2}S_{1/2}$&$ 296 _{- 52 }^{+ 45 }$    &  $ 90 _{- 25 }^{+ 26 }$    &  $ 9 _{- 3 }^{+ 3 }$    &  $ 140 _{- 63 }^{+ 80 }$    &  $ 338 _{- 161 }^{+ 206 }$    &  $ 2950 _{- 458 }^{+ 427 }$    &  $ 1.1 _{- 0.3 }^{+ 0.3 }$    &  $ 490 _{- 192 }^{+ 203 }$    &  $ 4.2 _{- 1.2 }^{+ 1.3 }$    &  $ 117 _{- 40 }^{+ 48 }$    &  $ 7 _{- 5 }^{+ 7 }$    &  0  &$ 71 _{- 37 }^{+ 52 }$    &  $ 0.1 _{- 0.1 }^{+ 0.1 }$    &  $ 1.8 _{- 1.1 }^{+ 1.8 }$    &  $ 90 _{- 34 }^{+ 36 }$    &  $ 230 _{- 87 }^{+ 100 }$   \\
$\Xi'_c(3613)$  & $ \mathbf{\frac{3}{2}^+}$ & $\vert \,0\,,\,0\,,\,0\,,\,1 \,\rangle $ &$^{4}S_{3/2}$&$ 327 _{- 46 }^{+ 38 }$    &  $ 26 _{- 7 }^{+ 7 }$    &  $ 79 _{- 24 }^{+ 24 }$    &  $ 225 _{- 102 }^{+ 120 }$    &  $ 559 _{- 260 }^{+ 317 }$    &  $ 194 _{- 38 }^{+ 31 }$    &  $ 531 _{- 55 }^{+ 48 }$    &  $ 545 _{- 122 }^{+ 105 }$    &  $ 655 _{- 70 }^{+ 60 }$    &  $ 341 _{- 150 }^{+ 176 }$    &  $ 10 _{- 6 }^{+ 7 }$    &  $ 0.9 _{- 0.2 }^{+ 0.2 }$    &  $ 24 _{- 13 }^{+ 18 }$    &  $ 3.7 _{- 1.5 }^{+ 1.9 }$    &  $ 40 _{- 23 }^{+ 33 }$    &  $ 123 _{- 41 }^{+ 43 }$    &  $ 328 _{- 117 }^{+ 124 }$   \\
$\Xi'_c(3370)$  & $ \mathbf{\frac{3}{2}^+}$ & $\vert \,1\,,\,1\,,\,0\,,\,0 \,\rangle $ &$^{2}D_{\lambda\rho,3/2}$&$ 334 _{- 35 }^{+ 30 }$    &  $ 29 _{- 11 }^{+ 12 }$    &  $ 10 _{- 4 }^{+ 5 }$    &  $ 429 _{- 32 }^{+ 32 }$    &  $ 121 _{- 17 }^{+ 18 }$    &  $ 12 _{- 4 }^{+ 5 }$    &  $ 30 _{- 14 }^{+ 18 }$    &  $ 12 _{- 4 }^{+ 5 }$    &  $ 8 _{- 4 }^{+ 6 }$    &  $ 0.5 _{- 0.3 }^{+ 0.4 }$    &  $ 4.5 _{- 1.3 }^{+ 1.7 }$    &  $ 12 _{- 5 }^{+ 6 }$    &  $ 4.3 _{- 1.3 }^{+ 1.5 }$    &  $ 3.3 _{- 1.3 }^{+ 1.8 }$    &  $ 0.2 _{- 0.1 }^{+ 0.1 }$    &  $ 11 _{- 7 }^{+ 9 }$    &  $ 4.0 _{- 2.2 }^{+ 2.4 }$   \\
$\Xi'_c(3415)$  & $ \mathbf{\frac{5}{2}^+}$ & $\vert \,1\,,\,1\,,\,0\,,\,0 \,\rangle $ &$^{2}D_{\lambda\rho,5/2}$&$ 323 _{- 28 }^{+ 26 }$    &  $ 35 _{- 13 }^{+ 14 }$    &  $ 13 _{- 5 }^{+ 6 }$    &  $ 32 _{- 11 }^{+ 12 }$    &  $ 504 _{- 31 }^{+ 34 }$    &  $ 32 _{- 11 }^{+ 14 }$    &  $ 4.9 _{- 2.3 }^{+ 3.4 }$    &  $ 22 _{- 8 }^{+ 9 }$    &  $ 16 _{- 7 }^{+ 10 }$    &  $ 10 _{- 4 }^{+ 6 }$    &  $ 12 _{- 4 }^{+ 4 }$    &  $ 2.0 _{- 0.8 }^{+ 1.2 }$    &  $ 8 _{- 2 }^{+ 3 }$    &  $ 6 _{- 2 }^{+ 3 }$    &  $ 3.9 _{- 1.5 }^{+ 1.8 }$    &  $ 0.7 _{- 0.4 }^{+ 0.6 }$    &  $ 12 _{- 8 }^{+ 9 }$   \\
$\Xi'_c(3379)$  & $ \mathbf{\frac{1}{2}^-}$ & $\vert \,1\,,\,1\,,\,0\,,\,0 \,\rangle $ &$^{2}P_{\lambda\rho,1/2}$&$ 18 _{- 4 }^{+ 3 }$    &  0  &0  &$ 497 _{- 28 }^{+ 30 }$    &  $ 44 _{- 4 }^{+ 4 }$    &  0  &$ 38 _{- 16 }^{+ 22 }$    &  0  &$ 23 _{- 11 }^{+ 14 }$    &  $ 3.5 _{- 1.8 }^{+ 2.8 }$    &  0  &$ 15 _{- 6 }^{+ 8 }$    &  0  &$ 10 _{- 4 }^{+ 5 }$    &  $ 1.6 _{- 0.7 }^{+ 0.9 }$    &  $ 17 _{- 11 }^{+ 12 }$    &  $ 2.9 _{- 1.8 }^{+ 2.1 }$   \\
$\Xi'_c(3406)$  & $ \mathbf{\frac{3}{2}^-}$ & $\vert \,1\,,\,1\,,\,0\,,\,0 \,\rangle $ &$^{2}P_{\lambda\rho,3/2}$&$ 17 _{- 4 }^{+ 3 }$    &  0  &0  &$ 120 _{- 6 }^{+ 7 }$    &  $ 412 _{- 22 }^{+ 24 }$    &  $ 45 _{- 16 }^{+ 20 }$    &  $ 16 _{- 8 }^{+ 10 }$    &  $ 18 _{- 7 }^{+ 9 }$    &  $ 18 _{- 8 }^{+ 10 }$    &  $ 4.8 _{- 2.4 }^{+ 3.5 }$    &  $ 17 _{- 6 }^{+ 7 }$    &  $ 6 _{- 2 }^{+ 3 }$    &  $ 7 _{- 2 }^{+ 3 }$    &  $ 7 _{- 3 }^{+ 3 }$    &  $ 2.1 _{- 0.8 }^{+ 1.1 }$    &  $ 2.7 _{- 2.1 }^{+ 2.6 }$    &  $ 18 _{- 10 }^{+ 12 }$   \\
$\Xi'_c(3397)$  & $ \mathbf{\frac{1}{2}^+}$ & $\vert \,1\,,\,1\,,\,0\,,\,0 \,\rangle $ &$^{2}S_{\lambda\rho,1/2}$&0  &0  &0  &$ 462 _{- 55 }^{+ 52 }$    &  $ 231 _{- 29 }^{+ 28 }$    &  $ 15 _{- 5 }^{+ 5 }$    &  $ 6 _{- 2 }^{+ 3 }$    &  $ 26 _{- 9 }^{+ 10 }$    &  $ 1.0 _{- 0.4 }^{+ 0.5 }$    &  $ 6 _{- 3 }^{+ 4 }$    &  $ 5 _{- 1 }^{+ 1 }$    &  $ 2.2 _{- 0.6 }^{+ 0.6 }$    &  $ 9 _{- 2 }^{+ 2 }$    &  $ 0.4 _{- 0.1 }^{+ 0.1 }$    &  $ 2.5 _{- 0.8 }^{+ 1.0 }$    &  $ 14 _{- 9 }^{+ 10 }$    &  $ 11 _{- 5 }^{+ 6 }$   \\
$\Xi'_c(3517)$  & $ \mathbf{\frac{3}{2}^+}$ & $\vert \,0\,,\,2\,,\,0\,,\,0 \,\rangle $ &$^{2}D_{\rho\rho,3/2}$&$ 113 _{- 21 }^{+ 19 }$    &  $ 17 _{- 2 }^{+ 2 }$    &  $ 3.4 _{- 1.1 }^{+ 1.2 }$    &  $ 8 _{- 4 }^{+ 5 }$    &  $ 13 _{- 6 }^{+ 8 }$    &  $ 562 _{- 90 }^{+ 79 }$    &  $ 63 _{- 22 }^{+ 27 }$    &  $ 72 _{- 6 }^{+ 8 }$    &  $ 18 _{- 7 }^{+ 9 }$    &  $ 1.5 _{- 0.6 }^{+ 0.7 }$    &  $ 10 _{- 3 }^{+ 3 }$    &  $ 0.1 _{- 0.1 }^{+ 0.1 }$    &  $ 6 _{- 2 }^{+ 2 }$    &  0  &$ 0.1 _{- 0.0 }^{+ 0.1 }$    &  $ 11 _{- 4 }^{+ 5 }$    &  $ 11 _{- 5 }^{+ 6 }$   \\
$\Xi'_c(3563)$  & $ \mathbf{\frac{5}{2}^+}$ & $\vert \,0\,,\,2\,,\,0\,,\,0 \,\rangle $ &$^{2}D_{\rho\rho,5/2}$&$ 122 _{- 20 }^{+ 17 }$    &  $ 153 _{- 25 }^{+ 24 }$    &  $ 3.9 _{- 1.2 }^{+ 1.2 }$    &  $ 11 _{- 5 }^{+ 6 }$    &  $ 18 _{- 8 }^{+ 10 }$    &  $ 353 _{- 100 }^{+ 102 }$    &  $ 8 _{- 3 }^{+ 4 }$    &  $ 1786 _{- 119 }^{+ 122 }$    &  $ 33 _{- 11 }^{+ 12 }$    &  $ 23 _{- 9 }^{+ 9 }$    &  $ 11 _{- 5 }^{+ 6 }$    &  $ 0.2 _{- 0.1 }^{+ 0.1 }$    &  $ 40 _{- 17 }^{+ 19 }$    &  0  &$ 0.1 _{- 0.1 }^{+ 0.1 }$    &  $ 26 _{- 10 }^{+ 12 }$    &  $ 20 _{- 8 }^{+ 9 }$   \\
$\Xi'_c(3532)$  & $ \mathbf{\frac{1}{2}^+}$ & $\vert \,0\,,\,2\,,\,0\,,\,0 \,\rangle $ &$^{4}D_{\rho\rho,1/2}$&$ 58 _{- 11 }^{+ 10 }$    &  $ 4.3 _{- 1.3 }^{+ 1.4 }$    &  $ 3.1 _{- 2.2 }^{+ 2.4 }$    &  $ 4.4 _{- 2.3 }^{+ 3.4 }$    &  $ 7 _{- 4 }^{+ 6 }$    &  $ 46 _{- 15 }^{+ 16 }$    &  $ 244 _{- 142 }^{+ 162 }$    &  0  &$ 169 _{- 28 }^{+ 27 }$    &  $ 15 _{- 6 }^{+ 6 }$    &  $ 0.2 _{- 0.1 }^{+ 0.2 }$    &  $ 4.5 _{- 1.4 }^{+ 1.4 }$    &  $ 0.3 _{- 0.2 }^{+ 0.3 }$    &  $ 2.7 _{- 1.0 }^{+ 1.2 }$    &  $ 1.6 _{- 0.6 }^{+ 0.5 }$    &  $ 22 _{- 10 }^{+ 13 }$    &  $ 0.1 _{- 0.0 }^{+ 0.0 }$   \\
$\Xi'_c(3559)$  & $ \mathbf{\frac{3}{2}^+}$ & $\vert \,0\,,\,2\,,\,0\,,\,0 \,\rangle $ &$^{4}D_{\rho\rho,3/2}$&$ 121 _{- 21 }^{+ 19 }$    &  $ 9 _{- 3 }^{+ 3 }$    &  $ 21 _{- 4 }^{+ 4 }$    &  $ 11 _{- 5 }^{+ 7 }$    &  $ 18 _{- 9 }^{+ 12 }$    &  $ 118 _{- 33 }^{+ 37 }$    &  $ 620 _{- 55 }^{+ 53 }$    &  $ 7 _{- 2 }^{+ 2 }$    &  $ 124 _{- 60 }^{+ 61 }$    &  $ 40 _{- 8 }^{+ 8 }$    &  $ 0.4 _{- 0.3 }^{+ 0.4 }$    &  $ 5 _{- 2 }^{+ 2 }$    &  $ 0.7 _{- 0.4 }^{+ 0.6 }$    &  $ 0.2 _{- 0.0 }^{+ 0.0 }$    &  $ 3.8 _{- 1.7 }^{+ 2.1 }$    &  $ 60 _{- 24 }^{+ 30 }$    &  $ 3.2 _{- 1.4 }^{+ 1.7 }$   \\
$\Xi'_c(3604)$  & $ \mathbf{\frac{5}{2}^+}$ & $\vert \,0\,,\,2\,,\,0\,,\,0 \,\rangle $ &$^{4}D_{\rho\rho,5/2}$&$ 176 _{- 27 }^{+ 21 }$    &  $ 14 _{- 4 }^{+ 4 }$    &  $ 39 _{- 6 }^{+ 6 }$    &  $ 20 _{- 9 }^{+ 11 }$    &  $ 34 _{- 16 }^{+ 19 }$    &  $ 50 _{- 13 }^{+ 12 }$    &  $ 189 _{- 53 }^{+ 54 }$    &  $ 105 _{- 28 }^{+ 29 }$    &  $ 1114 _{- 76 }^{+ 83 }$    &  $ 78 _{- 24 }^{+ 25 }$    &  $ 0.9 _{- 0.5 }^{+ 0.6 }$    &  $ 2.7 _{- 0.8 }^{+ 0.7 }$    &  $ 1.4 _{- 0.8 }^{+ 1.0 }$    &  $ 9 _{- 4 }^{+ 4 }$    &  $ 6 _{- 2 }^{+ 3 }$    &  $ 28 _{- 10 }^{+ 11 }$    &  $ 56 _{- 21 }^{+ 26 }$   \\
$\Xi'_c(3668)$  & $ \mathbf{\frac{7}{2}^+}$ & $\vert \,0\,,\,2\,,\,0\,,\,0 \,\rangle $ &$^{4}D_{\rho\rho,7/2}$&$ 118 _{- 14 }^{+ 11 }$    &  $ 10 _{- 2 }^{+ 2 }$    &  $ 158 _{- 24 }^{+ 23 }$    &  $ 18 _{- 8 }^{+ 9 }$    &  $ 31 _{- 14 }^{+ 17 }$    &  $ 2.4 _{- 1.2 }^{+ 1.5 }$    &  $ 26 _{- 12 }^{+ 14 }$    &  $ 135 _{- 33 }^{+ 30 }$    &  $ 287 _{- 87 }^{+ 89 }$    &  $ 2140 _{- 172 }^{+ 165 }$    &  $ 0.9 _{- 0.5 }^{+ 0.7 }$    &  $ 12 _{- 6 }^{+ 7 }$    &  $ 1.5 _{- 0.8 }^{+ 1.1 }$    &  $ 12 _{- 5 }^{+ 5 }$    &  $ 37 _{- 16 }^{+ 20 }$    &  $ 0.8 _{- 0.5 }^{+ 0.7 }$    &  $ 79 _{- 29 }^{+ 32 }$   \\
\hline \hline
\end{tabular}

\endgroup
}
\end{center}
\label{cascades-_EM}
\end{table*}
\end{turnpage}

\begin{table*}[htp]
\caption{Same as table \ref{sigmas+_EM}, but for $\Omega_{c}$ states with isospin 0.}
\begin{center}
\scriptsize{
\begingroup
\setlength{\tabcolsep}{1.75pt} 
\renewcommand{\arraystretch}{1.35} 

\begin{tabular}{c c c c  p{1.0cm}  p{1.0cm}  p{1.0cm}  p{1.0cm}  p{1.0cm}  p{1.0cm}  p{1.0cm}  p{1.0cm}  p{1.0cm}  p{1.0cm}  p{1.0cm}  p{1.0cm}} \hline \hline
&    &    &    & $\Omega_{c} \gamma$  & $\Omega^{*}_{c} \gamma$  & $\Omega_{c} \gamma$  & $\Omega_{c} \gamma$  & $\Omega_{c} \gamma$  & $\Omega_{c} \gamma$  & $\Omega_{c} \gamma$  & $\Omega_{c} \gamma$  & $\Omega_{c} \gamma$ \\
$\mathbf{\Omega_c(ssc)}$  & $\mathbf{J^P}$  & $\vert l_{\lambda}, l_{\rho}, k_{\lambda}, k_{\rho} \rangle$  & $^{2S+1}L_{x,J}$  & $^2S_{1/2}$  & $^4S_{3/2}$  & $^2P_{\lambda,1/2}$  & $^4P_{\lambda,1/2}$  & $^2P_{\lambda,3/2}$  & $^4P_{\lambda,3/2}$  & $^4P_{\lambda,5/2}$  & $^2P_{\rho,1/2}$  & $^2P_{\rho,3/2}$  \\ 
\hline
 $N=2$  &  &  &  &  &  \\ 
$\Omega_c(3315)$  & $ \mathbf{\frac{3}{2}^+}$ & $\vert \,2\,,\,0\,,\,0\,,\,0 \,\rangle $ &$^{2}D_{\lambda\lambda,3/2}$&$ 1.9 _{- 1.0 }^{+ 1.0 }$    &  $ 0.4 _{- 0.1 }^{+ 0.1 }$    &  $ 384 _{- 29 }^{+ 29 }$    &  $ 0.3 _{- 0.1 }^{+ 0.1 }$    &  $ 44 _{- 3 }^{+ 4 }$    &  $ 0.1 _{- 0.0 }^{+ 0.0 }$    &  0  &0  &0 \\
 &  &  & & $...$  & $...$ & $91.8$  & $0.13$  & 18.4 &  0.2  & 0.03 & $...$  & $...$ &  \cite{Yao:2018jmc} \\
$\Omega_c(3360)$  & $ \mathbf{\frac{5}{2}^+}$ & $\vert \,2\,,\,0\,,\,0\,,\,0 \,\rangle $ &$^{2}D_{\lambda\lambda,5/2}$&$ 20 _{- 8 }^{+ 8 }$    &  $ 0.6 _{- 0.1 }^{+ 0.1 }$    &  $ 7 _{- 2 }^{+ 2 }$    &  $ 0.1 _{- 0.0 }^{+ 0.0 }$    &  $ 564 _{- 47 }^{+ 42 }$    &  $ 0.2 _{- 0.1 }^{+ 0.1 }$    &  $ 0.1 _{- 0.0 }^{+ 0.1 }$    &  0  &0 \\
 &  &  & & $...$  & $...$ & $9.79$  & $0.008$  & 39.7 &  0.09  & 0.14 & $...$  & $...$ &  \cite{Yao:2018jmc} \\
$\Omega_c(3330)$  & $ \mathbf{\frac{1}{2}^+}$ & $\vert \,2\,,\,0\,,\,0\,,\,0 \,\rangle $ &$^{4}D_{\lambda\lambda,1/2}$&$ 0.8 _{- 0.3 }^{+ 0.3 }$    &  $ 0.8 _{- 0.5 }^{+ 0.5 }$    &  $ 0.4 _{- 0.2 }^{+ 0.2 }$    &  $ 315 _{- 22 }^{+ 21 }$    &  0  &$ 24 _{- 4 }^{+ 4 }$    &  $ 0.1 _{- 0.1 }^{+ 0.1 }$    &  0  &0 \\
 &  &  & & $...$  & $...$ & $0.03$  & $70.3$  & 0.04 &  27.5  & 2.44 & $...$  & $...$ &  \cite{Yao:2018jmc} \\
 &  &  & & 0.2  & 11.4 & $187.7$ & $109.7$  & $44.4$  & $10.4$  & $0.2$ & ... & $...$ & \cite{Peng:2024pyl} \\
$\Omega_c(3357)$  & $ \mathbf{\frac{3}{2}^+}$ & $\vert \,2\,,\,0\,,\,0\,,\,0 \,\rangle $ &$^{4}D_{\lambda\lambda,3/2}$&$ 2.0 _{- 0.5 }^{+ 0.6 }$    &  $ 2.0 _{- 1.0 }^{+ 0.9 }$    &  $ 1.3 _{- 0.7 }^{+ 0.8 }$    &  $ 206 _{- 18 }^{+ 17 }$    &  $ 0.1 _{- 0.0 }^{+ 0.0 }$    &  $ 142 _{- 10 }^{+ 9 }$    &  $ 4.8 _{- 0.7 }^{+ 0.7 }$    &  0  &0 \\
 &  &  & & $...$  & $...$ & $0.49$  & $34.7$  & 0.02 &  54.1  & 7.95 & $...$  & $...$ &  \cite{Yao:2018jmc} \\
$\Omega_c(3402)$  & $ \mathbf{\frac{5}{2}^+}$ & $\vert \,2\,,\,0\,,\,0\,,\,0 \,\rangle $ &$^{4}D_{\lambda\lambda,5/2}$&$ 3.7 _{- 0.7 }^{+ 0.9 }$    &  $ 4.3 _{- 2.0 }^{+ 1.9 }$    &  $ 0.8 _{- 0.4 }^{+ 0.4 }$    &  $ 6 _{- 2 }^{+ 2 }$    &  $ 1.4 _{- 0.7 }^{+ 0.7 }$    &  $ 378 _{- 28 }^{+ 27 }$    &  $ 71 _{- 5 }^{+ 5 }$    &  0  &0 \\
 &  &  & & $...$  & $...$ & $0.19$  & $5.56$  & 0.17 &  39.8  & 19.6 & $...$  & $...$ &  \cite{Yao:2018jmc} \\
$\Omega_c(3466)$  & $ \mathbf{\frac{7}{2}^+}$ & $\vert \,2\,,\,0\,,\,0\,,\,0 \,\rangle $ &$^{4}D_{\lambda\lambda,7/2}$&$ 3.6 _{- 0.8 }^{+ 0.9 }$    &  $ 20 _{- 9 }^{+ 9 }$    &  0  &$ 0.3 _{- 0.1 }^{+ 0.2 }$    &  $ 2.5 _{- 1.3 }^{+ 1.4 }$    &  $ 10 _{- 3 }^{+ 4 }$    &  $ 637 _{- 75 }^{+ 70 }$    &  0  &0 \\
 &  &  & & $...$  & $...$ & $0$  & $1.07$  & 0.13 &  5.62  & 23.1 & $...$  & $...$ &  \cite{Yao:2018jmc} \\
 &  &  & & 0.4  & 3.9 & $0.1$  & $0.1$  & $0.7$  & $0$  & $220.5$ & ... & $...$ & \cite{Peng:2024pyl} \\
$\Omega_c(3342)$  & $ \mathbf{\frac{1}{2}^+}$ & $\vert \,0\,,\,0\,,\,1\,,\,0 \,\rangle $ &$^{2}S_{1/2}$&$ 13 _{- 3 }^{+ 4 }$    &  $ 1.2 _{- 0.3 }^{+ 0.3 }$    &  $ 344 _{- 40 }^{+ 38 }$    &  0  &$ 16 _{- 6 }^{+ 7 }$    &  0  &$ 0.4 _{- 0.2 }^{+ 0.3 }$    &  0  &0 \\
 &  &  & & 3  & 0.2 & $10$  & $24$  & $8.6$  & $51.5$  & $0$ & ... & $...$ & \cite{Peng:2024pyl} \\
$\Omega_c(3411)$  & $ \mathbf{\frac{3}{2}^+}$ & $\vert \,0\,,\,0\,,\,1\,,\,0 \,\rangle $ &$^{4}S_{3/2}$&$ 7 _{- 2 }^{+ 2 }$    &  $ 8 _{- 3 }^{+ 3 }$    &  $ 4.0 _{- 2.0 }^{+ 1.9 }$    &  $ 125 _{- 12 }^{+ 12 }$    &  $ 10 _{- 4 }^{+ 5 }$    &  $ 153 _{- 13 }^{+ 12 }$    &  $ 5 _{- 2 }^{+ 3 }$    &  0  &0 \\
 &  &  & & 5.4  & 1.9 & $13.7$  & $8$  & $34.8$  & $8.5$  & $70.2$ & ... & $...$ & \cite{Peng:2024pyl} \\
$\Omega_c(3585)$  & $ \mathbf{\frac{1}{2}^+}$ & $\vert \,0\,,\,0\,,\,0\,,\,1 \,\rangle $ &$^{2}S_{1/2}$&$ 57 _{- 11 }^{+ 12 }$    &  $ 5 _{- 1 }^{+ 1 }$    &  $ 0.9 _{- 0.6 }^{+ 1.0 }$    &  0  &$ 21 _{- 9 }^{+ 11 }$    &  0  &$ 0.4 _{- 0.2 }^{+ 0.3 }$    &  $ 55 _{- 14 }^{+ 16 }$    &  $ 132 _{- 37 }^{+ 42 }$   \\
$\Omega_c(3654)$  & $ \mathbf{\frac{3}{2}^+}$ & $\vert \,0\,,\,0\,,\,0\,,\,1 \,\rangle $ &$^{4}S_{3/2}$&$ 18 _{- 4 }^{+ 3 }$    &  $ 51 _{- 11 }^{+ 10 }$    &  $ 3.2 _{- 1.5 }^{+ 1.9 }$    &  $ 0.8 _{- 0.1 }^{+ 0.1 }$    &  $ 7 _{- 3 }^{+ 4 }$    &  $ 1.8 _{- 0.5 }^{+ 0.6 }$    &  $ 11 _{- 5 }^{+ 7 }$    &  $ 85 _{- 19 }^{+ 20 }$    &  $ 214 _{- 53 }^{+ 56 }$   \\
$\Omega_c(3437)$  & $ \mathbf{\frac{3}{2}^+}$ & $\vert \,1\,,\,1\,,\,0\,,\,0 \,\rangle $ &$^{2}D_{\lambda\rho,3/2}$&$ 14 _{- 4 }^{+ 4 }$    &  $ 4.7 _{- 1.4 }^{+ 1.6 }$    &  $ 2.6 _{- 0.7 }^{+ 0.7 }$    &  $ 6 _{- 2 }^{+ 2 }$    &  $ 2.6 _{- 0.6 }^{+ 0.7 }$    &  $ 1.5 _{- 0.5 }^{+ 0.5 }$    &  $ 0.1 _{- 0.0 }^{+ 0.0 }$    &  $ 194 _{- 14 }^{+ 13 }$    &  $ 28 _{- 2 }^{+ 2 }$   \\
$\Omega_c(3482)$  & $ \mathbf{\frac{5}{2}^+}$ & $\vert \,1\,,\,1\,,\,0\,,\,0 \,\rangle $ &$^{2}D_{\lambda\rho,5/2}$&$ 19 _{- 5 }^{+ 5 }$    &  $ 6 _{- 2 }^{+ 2 }$    &  $ 7 _{- 2 }^{+ 2 }$    &  $ 0.9 _{- 0.3 }^{+ 0.4 }$    &  $ 5 _{- 1 }^{+ 1 }$    &  $ 3.4 _{- 0.9 }^{+ 1.2 }$    &  $ 2.0 _{- 0.6 }^{+ 0.8 }$    &  $ 1.8 _{- 0.6 }^{+ 0.7 }$    &  $ 245 _{- 19 }^{+ 19 }$   \\
$\Omega_c(3446)$  & $ \mathbf{\frac{1}{2}^-}$ & $\vert \,1\,,\,1\,,\,0\,,\,0 \,\rangle $ &$^{2}P_{\lambda\rho,1/2}$&0  &0  &0  &$ 8 _{- 3 }^{+ 3 }$    &  0  &$ 4.6 _{- 1.3 }^{+ 1.5 }$    &  $ 0.7 _{- 0.2 }^{+ 0.3 }$    &  $ 192 _{- 15 }^{+ 16 }$    &  $ 24 _{- 2 }^{+ 2 }$   \\
$\Omega_c(3473)$  & $ \mathbf{\frac{3}{2}^-}$ & $\vert \,1\,,\,1\,,\,0\,,\,0 \,\rangle $ &$^{2}P_{\lambda\rho,3/2}$&0  &0  &$ 10 _{- 2 }^{+ 3 }$    &  $ 3.3 _{- 1.1 }^{+ 1.3 }$    &  $ 3.8 _{- 0.9 }^{+ 1.1 }$    &  $ 3.6 _{- 0.9 }^{+ 1.1 }$    &  $ 0.9 _{- 0.3 }^{+ 0.4 }$    &  $ 57 _{- 6 }^{+ 5 }$    &  $ 165 _{- 14 }^{+ 13 }$   \\
$\Omega_c(3464)$  & $ \mathbf{\frac{1}{2}^+}$ & $\vert \,1\,,\,1\,,\,0\,,\,0 \,\rangle $ &$^{2}S_{\lambda\rho,1/2}$&0  &0  &$ 3.7 _{- 0.7 }^{+ 0.7 }$    &  $ 1.4 _{- 0.4 }^{+ 0.4 }$    &  $ 6 _{- 1 }^{+ 1 }$    &  $ 0.2 _{- 0.1 }^{+ 0.1 }$    &  $ 1.3 _{- 0.4 }^{+ 0.5 }$    &  $ 127 _{- 13 }^{+ 13 }$    &  $ 49 _{- 5 }^{+ 4 }$   \\
$\Omega_c(3558)$  & $ \mathbf{\frac{3}{2}^+}$ & $\vert \,0\,,\,2\,,\,0\,,\,0 \,\rangle $ &$^{2}D_{\rho\rho,3/2}$&$ 24 _{- 2 }^{+ 2 }$    &  $ 1.8 _{- 0.5 }^{+ 0.5 }$    &  $ 6 _{- 2 }^{+ 2 }$    &  0  &$ 3.4 _{- 0.9 }^{+ 0.9 }$    &  0  &0  &$ 6 _{- 2 }^{+ 2 }$    &  $ 6 _{- 2 }^{+ 2 }$   \\
$\Omega_c(3603)$ & $ \mathbf{\frac{5}{2}^+}$ & $\vert \,0\,,\,2\,,\,0\,,\,0 \,\rangle $ &$^{2}D_{\rho\rho,5/2}$&$ 141 _{- 12 }^{+ 11 }$    &  $ 2.3 _{- 0.5 }^{+ 0.5 }$    &  $ 5 _{- 2 }^{+ 2 }$    &  0  &$ 18 _{- 6 }^{+ 7 }$    &  0  &0  &$ 15 _{- 4 }^{+ 5 }$    &  $ 12 _{- 3 }^{+ 4 }$   \\
$\Omega_c(3573)$  & $ \mathbf{\frac{1}{2}^+}$ & $\vert \,0\,,\,2\,,\,0\,,\,0 \,\rangle $ &$^{4}D_{\rho\rho,1/2}$&$ 2.7 _{- 0.7 }^{+ 0.7 }$    &  $ 10 _{- 3 }^{+ 3 }$    &  0  &$ 2.8 _{- 0.9 }^{+ 0.9 }$    &  $ 0.1 _{- 0.0 }^{+ 0.1 }$    &  $ 1.4 _{- 0.5 }^{+ 0.5 }$    &  $ 1.0 _{- 0.3 }^{+ 0.3 }$    &  $ 12 _{- 5 }^{+ 5 }$    &  $ 0.1 _{- 0.0 }^{+ 0.0 }$   \\
$\Omega_c(3600)$  & $ \mathbf{\frac{3}{2}^+}$ & $\vert \,0\,,\,2\,,\,0\,,\,0 \,\rangle $ &$^{4}D_{\rho\rho,3/2}$&$ 6 _{- 2 }^{+ 1 }$    &  $ 20 _{- 1 }^{+ 1 }$    &  $ 0.1 _{- 0.1 }^{+ 0.1 }$    &  $ 3.0 _{- 0.9 }^{+ 0.9 }$    &  $ 0.2 _{- 0.1 }^{+ 0.1 }$    &  $ 0.1 _{- 0.0 }^{+ 0.0 }$    &  $ 1.7 _{- 0.6 }^{+ 0.7 }$    &  $ 35 _{- 12 }^{+ 13 }$    &  $ 1.8 _{- 0.6 }^{+ 0.7 }$   \\
$\Omega_c(3645)$  & $ \mathbf{\frac{5}{2}^+}$ & $\vert \,0\,,\,2\,,\,0\,,\,0 \,\rangle $ &$^{4}D_{\rho\rho,5/2}$&$ 10 _{- 2 }^{+ 2 }$    &  $ 39 _{- 2 }^{+ 2 }$    &  $ 0.3 _{- 0.1 }^{+ 0.1 }$    &  $ 2.0 _{- 0.4 }^{+ 0.4 }$    &  $ 0.4 _{- 0.2 }^{+ 0.2 }$    &  $ 4.5 _{- 1.4 }^{+ 1.4 }$    &  $ 2.9 _{- 1.0 }^{+ 1.0 }$    &  $ 17 _{- 5 }^{+ 5 }$    &  $ 34 _{- 9 }^{+ 11 }$   \\
$\Omega_c(3708)$  & $ \mathbf{\frac{7}{2}^+}$ & $\vert \,0\,,\,2\,,\,0\,,\,0 \,\rangle $ &$^{4}D_{\rho\rho,7/2}$&$ 8 _{- 1 }^{+ 1 }$    &  $ 150 _{- 13 }^{+ 12 }$    &  $ 0.3 _{- 0.1 }^{+ 0.2 }$    &  $ 5 _{- 2 }^{+ 2 }$    &  $ 0.5 _{- 0.2 }^{+ 0.3 }$    &  $ 7 _{- 2 }^{+ 2 }$    &  $ 17 _{- 6 }^{+ 7 }$    &  $ 0.3 _{- 0.1 }^{+ 0.2 }$    &  $ 53 _{- 14 }^{+ 15 }$   \\
\hline \hline
\end{tabular}
\endgroup
}
\end{center}
\label{omegas_EM_2shell}
\end{table*}


This section reports the electromagnetic decay widths of the singly charmed baryons $\Sigma_c$, $\Xi'_c$, and $\Omega_c$ in Tables~\ref{sigmasEM}-\ref{omegas_EM_2shell}, computed using the formalism of Ref.~\cite{Garcia-Tecocoatzi:2023btk,Alejandra_MScThesis}. For the energy band $N=0$, we consider $S$-wave to $S$-wave transitions, which are presented in Tables\ref{sigmasEM}–\ref{omegasEM} for the $\Sigma_c$, $\Xi'_c$, and $\Omega_c$ baryons, respectively. For the energy band $N=1$, we evaluate $P$-wave to $S$-wave transitions, also shown in Tables~\ref{sigmasEM}–\ref{omegasEM}. 


Tables~\ref{sigmasEM}--\ref{omegasEM} summarize our numerical results for the electromagnetic decay widths of excited baryons belonging to the flavor multiplet  $\mathcal{F} = {\mathbf{6}}_{\rm F}$. We report the baryon names along with their predicted masses, taken from Ref.~\cite{Garcia-Tecocoatzi:2022zrf}. Additionally, we provide the total angular momentum and parity $\mathbf{J^{P}}$, the internal configuration of the baryon given by $\left| l_{\lambda}, l_{\rho}, k_{\lambda}, k_{\rho} \right\rangle$, and the corresponding spectroscopic notation $^{2S+1}L_{x,J}$, where the subscript $x$ indicates the type of orbital excitation. In these tables, $x = \lambda$ or $\rho$, with $\lambda$ referring to a single $\lambda$ mode excitation and $\rho$ to a single $\rho$ mode excitation. 
Starting from the fifth column, we present our predictions. Each column corresponds to a specific decay channel, with the final state baryons indicated at the top. The second row shows the spectroscopic notation $^{2S+1}L_{x,J}$ of the final states. Zero values correspond to decays that are either kinematically forbidden or too suppressed to be displayed at this scale. Our predictions are compared with results from previous studies such as LC-QCD~\cite{Luo:2025pzb} and CQM~\cite{Wang:2017kfr,Ortiz-Pacheco:2023kjn}. We do not include the results from Refs.~\cite{Cheng:1992xi,Banuls:1999br,Jiang:2015xqa,Wang:2018cre,Zhu:1998ih,Wang:2010xfj,Wang:2009cd,Aliev:2009jt,Aliev:2011bm,Aliev:2014bma,Aliev:2016xvq,Bernotas:2013eia,Shah:2016nxi,Gandhi:2019xfw,Gandhi:2019bju,Yang:2019tst,Kim:2021xpp,Chow:1995nw,Gamermann:2010ga,Zhu:2000py,Bijker:2020tns,Ivanov:1998wj,Ivanov:1999bk,Tawfiq:1999cf}, as only a limited number of channels were considered in these studies. 

Finally, for $N=2$, we compute decays of $S$-, $P$-, and $D$-wave states to $S$-, and $P$-wave final states, which are presented in Tables~\ref{sigmas+_EM}–\ref{omegas_EM_2shell} for the $\Sigma_c$, $\Xi'_c$, and $\Omega_c$ baryons, respectively. Tables~\ref{sigmas+_EM}--\ref{omegas_EM_2shell} are organized according to the electric charge of the baryons. In these tables, the subscript $x$ takes additional values: $\lambda$, $\lambda\lambda$, $\rho$, $\rho\rho$, and $\lambda\rho$. Here, $\lambda\lambda$ and $\rho\rho$ indicate double excitations in their respective modes, while $\lambda\rho$ represents a mixed excitation involving both $\lambda$- and $\rho$-mode components. Our predictions are compared with some of the results in Refs.~\cite{Yao:2018jmc, Peng:2024pyl}. However, because those works use different basis states, a full one-to-one comparison is not possible.

The experimental decay widths include contributions from strong, electromagnetic, and weak interactions, with the strong decays generally providing the dominant component. Electromagnetic decay widths are relatively small in comparison, yet they provide important information for experimentalists. Specifically, electromagnetic decay widths enable the estimation of branching ratios, which are key observables that can be measured in particle accelerator experiments. The measurement of these branching ratios can provide valuable input for the identification and assignment of singly charmed baryon states.
Electromagnetic decay widths are particularly valuable in cases where strong decays are forbidden. 
Moreover, electromagnetic decay widths may be useful in the assignments of resonances when they have the same mass and strong decay widths.

\subsection{Assignment of the $\Omega_c (3327)$ state}
\label{Omegac_assignment}

\begin{table}[h!tp]
\caption{Partial strong decay widths (in MeV) for the $\Omega_c (3315)$ and $\Omega_c (3330)$, as from Ref.~\cite{Garcia-Tecocoatzi:2022zrf}. 
The first column denotes the baryon name along with its theoretical mass. The second column provides the spectroscopic notation $^{2S+1}L_{x,J}$ associated with each state. Starting from the third column, each column corresponds to a strong decay channel, and the decay products are indicated at the top of the column. }
\centering
\begin{tabular}{cc ccccc}
\hline\hline
$\mathbf{\Omega_c (ssc)}$ & $^{2S+1}L_{x,J}$ & $\Xi_c K$ & $\Xi'_c K$ & $\Xi^{*}_c K$ & $\Omega_c \eta$ & $\Xi_8 D$  \\
\hline
$\Omega_c (3315)$ & $^{2}D_{\lambda\lambda,3/2}$ & 1.9 & 1.8 & 2.3 & 0.3 & 4.3 \\
$\Omega_c (3330)$ & $^{4}D_{\lambda\lambda,1/2}$ & 0.2 & 0.2 & 3.3 & 0.1 & 12.3 \\
\hline\hline
\end{tabular}
\label{partial_st_dw_omega}
\end{table}

One useful application of our results   is the identification of recent $\Omega_{c}(3327)^{0}$, observed in the $\Xi_{c}^{+}K^{-}$ final state by LHCb~\cite{LHCb:2023sxp}
\begin{eqnarray}
M_{\Omega_c(3327)} &=& 3327.1 \pm 1.2 ^{+0.1}_{-1.3} \pm 0.2 \, \text{MeV}, \nonumber \\
\Gamma_{\Omega_c(3327)} &=& 20 \pm 5 ^{+13}_{-1.0} \, \text{MeV}.
\end{eqnarray}
Although the mass and width of this excited state have been measured with the highest precision to date, its spin quantum numbers remain undetermined. Ref. \cite{Garcia-Tecocoatzi:2022zrf} predicted two states in this energy range, the $\lambda$-mode $\Omega_c$ $^2D_{\lambda\lambda,3/2}$ state, which has a mass of 3315 MeV and a strong decay width of 11 MeV, while the $\lambda$-mode $\Omega_c$ $^4D_{\lambda\lambda,1/2}$ state has a mass of 3330 MeV and a strong decay width of 16 MeV. Thus, the $\Omega_{c}(3327)^{0}$ can be one of these two states. Moreover, if we analyze the strong decay channels, we can see that the branching ratios are comparable, i.e. the differences in the branching ratios are up to two orders of magnitude for the same decay channels when the initial baryon is the $\Omega_c (3315)$ or the $\Omega_c (3330)$ (see Table~\ref{partial_st_dw_omega}). Thus, even the study of strong decay channels and the measurement of the branching ratios in those channels won't allow to distinguish which of the two states corresponds to the $\Omega_{c}(3327)^{0}$. Then, a possible way to make a conclusive assignment is to study its radiative decay channels. Some useful channels might be the following: 
\begin{eqnarray}
&& \Gamma_{em} [\Omega_c (3315)   \to \Omega_c (3008)  \gamma ] = 384_{-29 }^{+29 } \, \text{keV} , \\
&& \Gamma_{em} [\Omega_c (3330)   \to \Omega_c (3008)  \gamma ] = 0.4_{-0.2 }^{+0.2 } \, \text{keV} ,  \\
&& \Gamma_{em} [\Omega_c (3315)   \to \Omega_c (3050)  \gamma ] =  0.3 _{- 0.1 }^{+ 0.1 } \, \text{keV} ,  \\
&& \Gamma_{em} [\Omega_c (3330)   \to \Omega_c (3050)  \gamma ] = 315_{-22 }^{+21 } \, \text{keV} .
\end{eqnarray}
With these decay widths, we can get several branching ratios. If we analyze, for example the following cases
\begin{eqnarray}
&&\frac{\Gamma_{em} [\Omega_c (3315)   \to \Omega_c (3008)  \gamma]}{\Gamma_{em} [\Omega_c (3315)   \to \Omega_c (3050)  \gamma]} = 1280^{+437}_{-437}\, , \\
&&\frac{\Gamma_{em} [\Omega_c (3330)   \to \Omega_c (3008)  \gamma]}{\Gamma_{em} [\Omega_c (3330)   \to \Omega_c (3050)  \gamma]} = 0.0013^{+0.0006}_{-0.0006}\, ,
\end{eqnarray} 
we observe that the differences in the branching ratios are of six orders of magnitude for the same decay
channels when the initial baryon is the $\Omega_c (3315)$ or the $\Omega_c (3330)$. Thus, this can be a good channel to make the identification of the recently observed $\Omega_{c}(3327)^{0}$ baryon.

\subsection{Assignment of the $\Xi_c(2923)^+$ baryon}

Following the recent discovery made by the LHCb Collaboration~\cite{LHCb:2025mge} of the $\Xi'_c(2923)^{+}$ baryon, we discuss our assignment based on the theoretical results of the charmed baryons reported in Ref.~\cite{Garcia-Tecocoatzi:2022zrf}. The $\Xi_c(2923)^{+}$ is identified as the $P_\lambda$-wave excitation with quantum numbers $J^P=1/2^{-}$ with spin $S= 3/2$ that belongs to the flavor sextet ($\Xi'_c(1P)$ states). The mass and decay width reported in Ref.~\cite{LHCb:2025mge} are   $2922.85$ MeV and $5.3$ MeV, respectively, being compatible with the theoretical values from Ref.~\cite{Garcia-Tecocoatzi:2022zrf}. This state is consistent with being the isospin partner of the previously observed $\Xi'_c(2923)^{0}$. We also reported the following electromagnetic decay widths for: 

\begin{itemize}
    \item $\Xi_c(2923)^{+}$

    \begin{enumerate}

  \item $\Gamma_{em}[\Xi'_c(2923)^{+} \to \Xi_c^{+} \, \gamma ] = 73 _{- 23 }^{+ 26 } $ keV,

\item $\Gamma_{em}[\Xi'_c(2923)^{+} \to \Xi_c'^{+}  \, \gamma ]  = 49 _{- 25 }^{+ 28 } $ keV,

\item $\Gamma_{em}[\Xi'_c(2923)^{+} \to \Xi_c^{*+}  \, \gamma  ] = 0.8 _{- 0.3 }^{+ 0.3 } $ keV,

    \end{enumerate}

\item $\Xi'_c(2923)^{0}$
\begin{enumerate}
    \item $\Gamma_{em}[\Xi'_c(2923)^{0} \to \Xi_c^{0}  \, \gamma ] = 1.6 _{- 0.7 }^{+ 0.9 } $ keV,

\item $\Gamma_{em}[\Xi'_c(2923)^{0} \to \Xi_c'^{0}  \, \gamma ]  = 449 _{- 36 }^{+ 36 } $ keV,

\item $\Gamma_{em}[\Xi'_c(2923)^{0} \to \Xi_c^{*0}  \, \gamma ] = 0.4 _{- 0.2 }^{+ 0.2 } $ keV ,

\end{enumerate}
\end{itemize}

For the charged state $\Xi_c(2923)^{+}$,
 we observe that our values for the $\Xi_c^{+}\gamma$ and $\Xi'_c{}^{+}\gamma$ channels are approximately 30\% larger than those reported in Ref.~\cite{Ortiz-Pacheco:2023kjn} (see Table~\ref{cascadesprimeEM}), while for the $\Xi_c^{+}\gamma$ channel our result is about 200\% larger. Compared with Ref.~\cite{Wang:2017kfr}, our value for the $\Xi_c^{+}\gamma$ channel is roughly 30\% higher, for the $\Xi'_c{}^{+}\gamma$ channel approximately 300\% higher, and for the $\Xi_c^{+}\gamma$ channel about 50\% of the value reported in Ref.~\cite{Wang:2017kfr}. Similarly, for $\Xi_c(2923)^{0}$, our values for the channels
$\Xi_c^{0}\gamma$, $\Xi_c'^{0}\gamma$, and $\Xi_c^{*0}\gamma$ are about $30\%$ larger than those reported in Ref.~\cite{Ortiz-Pacheco:2023kjn}. When comparing with Ref.~\cite{Wang:2017kfr}, our prediction for the $\Xi_c'^{0}\gamma$ channel is also about $30\%$ higher, while for the $\Xi_c^{*0}\gamma$ channel our result is roughly $50\%$ of theirs. For the $\Xi_c^{0}\gamma$ channel, Ref.~\cite{Wang:2017kfr} reports a value of zero.

\subsection{Identification of $\Xi_c$ $^2D_{\rho\lambda,3/2}$ and $\Xi'_c$ $^4D_{\lambda\lambda,3/2}$}
Another important example is the $\Xi_c$ $^2D_{\rho\lambda,3/2}$ mixed state, which has a mass of 3265 MeV and a strong decay width of 54 MeV, while the $\Xi'_c$ $^4D_{\lambda\lambda,3/2}$ state has a mass of 3265 MeV and a strong decay width of 53 MeV. Consequently, if experimentalists detect a state with this mass and total decay width, they will not be able to determine whether it is a $\Xi_c$ or $\Xi'_c$ state based on that information alone. Moreover, as in the case of the $\Omega_c$ states discussed in subsection~\ref{Omegac_assignment}, both $\Xi_c$ and $\Xi'_c$ decay into the same strong decay channels, and their branching ratios are comparable (see Table~\ref{partial_st_dw_cascade}). Thus, a possible way to achieve a conclusive assignment is by studying their radiative decay channels. Some potentially useful channels include the following: 
\begin{eqnarray}
&& \Gamma_{em} [\Xi_c (3265)^{0}   \to \Xi_c^{' 0} \gamma ] = 2.6^{+0.4}_{-0.4} \, \text{keV} ,  \\
&&  \Gamma_{em} [\Xi'_c (3265)^{0}   \to \Xi_c^{' 0} \gamma ] = 4.9^{+1.9}_{-1.5} \, \text{keV} , \\
&& \Gamma_{em} [\Xi_c (3265)^{0}   \to \Xi_c^{* 0} \gamma ] = 0.14^{+0.04}_{-0.06} \, \text{keV} ,  \\
&&  \Gamma_{em} [\Xi'_c (3265)^{0}   \to \Xi_c^{* 0} \gamma ] = 6^{+3}_{-3} \, \text{keV} , \\
&& \Gamma_{em} [\Xi_c (3265)^{+}   \to \Xi_c^{'+} \gamma ] = 130^{+5}_{-6} \, \text{keV} ,  \\
&&  \Gamma_{em} [\Xi'_c (3265)^{+}   \to \Xi_c^{'+} \gamma ] = 1.9^{+1.4}_{-1.1} \, \text{keV} , \\
&& \Gamma_{em} [\Xi_c (3265)^{+}   \to \Xi_c^{* +} \gamma ] = 6.5^{+1.4}_{-1.5} \, \text{keV} , \\
&&  \Gamma_{em} [\Xi'_c (3265)^{+}  \to \Xi_c^{* +} \gamma ] = 15^{+2}_{-2} \, \text{keV} .
\end{eqnarray}
With these decay widths we can get several branching ratios. If we analyze, for example the following cases
\begin{eqnarray}
&&\frac{\Gamma_{em} [\Xi_c (3265)   \to \Xi_c^{'+} \gamma]}{\Gamma_{em} [\Xi_c (3265)   \to \Xi_c^{* +} \gamma]} = 20^{+4}_{-5} \, , \\
&&\frac{\Gamma_{em} [\Xi'_c (3265)   \to \Xi_c^{'+} \gamma]}{\Gamma_{em} [\Xi'_c (3265)   \to \Xi_c^{* +} \gamma]} = 0.13^{+0.09}_{-0.07} \, .
\end{eqnarray} 

\begin{table*}[htp]
\caption{ Same as Table \ref{partial_st_dw_omega}, but for $\Xi_c$ and $\Xi'_c$ states. The strong partial decay widths are taken from Ref.~\cite{Garcia-Tecocoatzi:2022zrf}.} 
\centering
\begin{tabular}{cc cccccccccccc}
\hline\hline
$\mathbf{\Xi^{(\prime)}_c(snc)}$ & $^{2S+1}L_{x,J}$ & $\Lambda_c K$ & $\Xi_c \pi$ & $\Xi'_c \pi$ & $\Xi^{*}_c \pi$ & $\Sigma_c K$ & $\Sigma^{*}_c K$ & $\Xi_c \eta$  & $\Lambda^{*}_c K$ & $\Xi_c \rho$ & $\Xi^{*}_c \eta$ & $\Xi_c \omega$ \\ 
\hline
$\Xi_c(3265) $ & $^{2}D_{\lambda\rho,3/2}$ & 1.3 & 1.7 & 2.5 & 10.6 & 6.3 & 21.8 & 0.3 & 6.9 & 1.6 & 0.3 & 0.4 \\
$\Xi'_c(3265) $ & $^{4}D_{\lambda\lambda,3/2}$ & 0.7 & 0.8 & 0.2 & 2.1 & 0.5 & 4.6 & 0.1 & 1.8 & 0.4 & 0.1 & 0.1\\ 
\hline\hline
\end{tabular}

\label{partial_st_dw_cascade}
\end{table*}
We observe that the differences in the branching ratios for different electromagnetic decay channels are at least one order of magnitude when the initial baryon is a $\Xi_c$ or a $\Xi'_c$. Hence the measurement of these electromagnetic decay widths or branching ratios will significantly help distinguish between states.

In this sense, our results may help make a correct assignment of states similar to the examples described above. The electromagnetic decay width values for the second shell $\Xi_c$ baryon states are preliminary results taken from~\cite{Charmantitriplet}, which is currently a work in progress.

\section{Conclusions}
\label{Conclusions}

In summary, we calculate the electromagnetic decay widths of the $\Sigma_c$, $\Xi'_c$ and $\Omega_c$ charmed baryons belonging to the flavor ${\bf {6}}_{\rm F}$-plet. The electromagnetic transitions are calculated from the ground and $P$-wave states to ground states, as well as from the second shell states to both the ground and $P$-wave final states. 

Up to now, there have been no experimental measurements of the electromagnetic decay widths of singly charmed baryons belonging to the ${\bf {6}}_{\rm F}$-plet; therefore, our results can only be compared with other theoretical works.

The EMDs of $P$-wave states have been investigated in Refs.~\cite{Chow:1995nw,Gamermann:2010ga,Zhu:2000py,Luo:2025pzb,Bijker:2020tns,Ortiz-Pacheco:2023kjn,Wang:2017kfr,Peng:2024pyl,Ivanov:1998wj,Ivanov:1999bk,Tawfiq:1999cf}. The EMDs of second shell singly charmed baryons were studied in Refs.~\cite{Yao:2018jmc,Peng:2024pyl}, although in both cases only a subset of second shell states was considered.

It is worth mentioning that this is the first time that the electromagnetic decays of $D_\rho$-wave states, $\rho \lambda$ mixed states, and $\rho$-mode radial excited states in the charm sector are calculated.

Electromagnetic decay widths can help with the assignments when states have the same mass and strong decay widths, like the cases of the examples discussed in Section~\ref{Results}. 
Specifically, they can help in the assignment of the $\Omega_c (3327)$ state, since the branching ratios between the strong decay channels are comparable.
Moreover, we also make the assignment of the recently
discovered $\Xi'_c(2923)^{+}$ baryon, which is consistent with being the isospin partner of the $\Xi'_c(2923)^{0}$, and  we provide its electromagnetic decay widths.

In this work, we have accounted for the propagation of parameter uncertainties using a Monte Carlo bootstrap method. The inclusion of uncertainties is essential, but it is often missed in related research works.

The study of electromagnetic decay widths of the $\Lambda_c$ and $\Xi_c$ belonging to the flavor ${\bf {\bar 3}}_{\rm F}$-plet will be published separately \cite{Charmantitriplet}. 


\section*{Acknowledgements}
C.A. V.-A. is supported by the Secretar\'ia de Ciencia, Humanidades, Tecnolog\'ia e Innovaci\'on (Secihti) Investigadoras e Investigadores por M\'exico project 749 and SNII 58928. 

\clearpage



\end{document}